%% file: main.tex
\documentclass[conference]{IEEEtran}
\IEEEoverridecommandlockouts
\usepackage{mathptmx} 

\usepackage{enumitem}
\usepackage{flushend}
\usepackage{balance}
\usepackage{booktabs}
\usepackage{todonotes}
\usepackage{epsfig}
\usepackage{graphicx}
\usepackage{amsmath}
\usepackage{amssymb}
\usepackage{lipsum}
\usepackage{subfig}
\usepackage{color}

\usepackage[normalem]{ulem}
\usepackage{bigstrut}
\usepackage{multirow}
\usepackage{siunitx}
\usepackage{ulem}
\usepackage{listings}
\usepackage{eso-pic}
\usepackage{tikz}
\usepackage{xspace}
\usepackage{fancyhdr}
\usepackage{mathtools}
\usepackage{soul}
\usepackage[leftcaption]{sidecap}

\usepackage[bookmarks=true,breaklinks=true,letterpaper=true,colorlinks,linkcolor=blue,citecolor=blue,urlcolor=black]{hyperref}
\def\BibTeX{{\rm B\kern-.05em{\sc i\kern-.025em b}\kern-.08em
    T\kern-.1667em\lower.7ex\hbox{E}\kern-.125emX}}

\title{Eudoxus: Characterizing and Accelerating Localization in Autonomous Machines \\
{\large Industry Track Paper}
}

\author{\IEEEauthorblockN{Yiming Gan}
\IEEEauthorblockA{
\textit{University of Rochester}\\
ygan10@ur.rochester.edu}
\and
\IEEEauthorblockN{Bo Yu}
\IEEEauthorblockA{\textit{PerceptIn} \\
bo.yu@perceptin.io}
\and
\IEEEauthorblockN{Boyuan Tian $^\dagger$\thanks{$^\dagger$ Now at UIUC.}}
\IEEEauthorblockA{
\textit{University of Rochester}\\
btian2@ur.rochester.edu}
\and
\IEEEauthorblockN{Leimeng Xu}
\IEEEauthorblockA{\textit{PerceptIn} \\
leimeng.xu@perceptin.io}
\and
\IEEEauthorblockN{Wei Hu}
\IEEEauthorblockA{\textit{PerceptIn} \\
wei.hu@perceptin.io}
\and
\IEEEauthorblockN{Shaoshan Liu$^*$ \thanks{* Corresponding author}}
\IEEEauthorblockA{\textit{PerceptIn} \\
shaoshan.liu@perceptin.io}
\and
\IEEEauthorblockN{Qiang Liu}
\IEEEauthorblockA{
\textit{Tianjin University}\\
qiangliu@tju.edu.cn}
\and
\IEEEauthorblockN{Yanjun Zhang}
\IEEEauthorblockA{
\textit{Beijing Institute of Technology}\\
zhangyj@bit.edu.cn}
\and
\IEEEauthorblockN{Jie Tang}
\IEEEauthorblockA{
\textit{SCUT}\\
cstangjie@scut.edu.cn}
\and
\IEEEauthorblockN{Yuhao Zhu}
\IEEEauthorblockA{
\textit{University of Rochester}\\
yzhu@rochester.edu}
}

\input{macros}
\graphicspath{{figs/}}

\begin{document}
\maketitle
\pagestyle{plain}


\input{abst}
\input{intro}

\input{background}
\input{motivation}

\input{algo}

\input{fe}

\input{be}

\input{eval}

\input{conc}


\bibliographystyle{IEEEtranS}
\bibliography{refs}

\end{document}

%% file: macros.tex
\ifx\figurename\undefined \def\figurename{Figure}\fi
\renewcommand{\figurename}{Fig.}
\renewcommand{\paragraph}[1]{\textbf{#1}}

\newcommand{\Sect}[1]{Sec.~\ref{#1}}
\newcommand{\Fig}[1]{Fig.~\ref{#1}}
\newcommand{\Tbl}[1]{Tbl.~\ref{#1}}
\newcommand{\Equ}[1]{Equ.~\ref{#1}}

\newcommand{\proj}{\textsc{Eudoxus}\xspace}

\newcommand{\sys}[1]{\underline{\textsc{#1}}}

\newcommand{\no}[1]{#1}
\renewcommand{\no}[1]{}
\newcommand{\RNum}[1]{\uppercase\expandafter{\romannumeral #1\relax}}


%% file: abst.tex
\begin{abstract}

We develop and commercialize autonomous machines, such as logistic robots and self-driving cars, around the globe. A critical challenge to our---and any---autonomous machine is accurate and efficient localization under resource constraints, which has fueled specialized localization accelerators recently. Prior acceleration efforts are point solutions in that they each specialize for a specific localization algorithm. In real-world commercial deployments, however, autonomous machines routinely operate under different environments and no single localization algorithm fits all the environments. Simply stacking together point solutions not only leads to cost and power budget overrun, but also results in an overly complicated software stack.



This paper demonstrates our new software-hardware co-designed framework for autonomous machine localization, which adapts to different operating scenarios by fusing fundamental algorithmic primitives. Through characterizing the software framework, we identify ideal acceleration candidates that contribute significantly to the end-to-end latency and/or latency variation. We show how to co-design a hardware accelerator to systematically exploit the parallelisms, locality, and common building blocks inherent in the localization framework. We build, deploy, and evaluate an FPGA prototype on our next-generation self-driving cars. To demonstrate the flexibility of our framework, we also instantiate another FPGA prototype targeting drones, which represent mobile autonomous machines. We achieve about 2$\times$ speedup and 4$\times$ energy reduction compared to widely-deployed, optimized implementations on general-purpose platforms.

\end{abstract}

%% file: intro.tex
\section{Introduction}
\label{sec:intro}

Over the past four years, we have developed and commercialized autonomous machines, such as self-driving cars and mobile logistic robots, in the U.S., Japan, and countries in Europe~\cite{yu2020building}. Throughout our development and deployment process, \textit{localization} is a critical challenge. Under tight resource constraints, the localization task must precisely calculate the position and orientation of the machine itself in a given map of the environment. Efficient localization is a prerequisite to motion planning, navigation, and stabilization~\cite{kelly2013mobile, dudek2010computational}.



While literature is rich with accelerator designs for specific localization algorithms~\cite{li2019879gops, suleiman2019navion, zhang2017visual, liu2019eslam, yoon2010graphics}, prior efforts are mostly \textit{point solutions} in that they each specialize for a particular localization algorithm such as Visual-Inertial Odometry (VIO)~\cite{mourikis2007multi, sun2018robust} and Simultaneous Localization and Mapping (SLAM)~\cite{rublee2011orb, mur2017orb}. However, each algorithm suits only a particular operating scenario, whereas in commercial deployments autonomous machines routinely operate under different scenarios. For instance, our self-driving cars must travel from a place visited before to a new, unmapped place while going through both indoor and outdoor environments.

Using a combination of data from our commercial autonomous vehicles and standard benchmarks, we demonstrate that no single localization algorithm fits all scenarios (\Sect{sec:mot}). For instance, in outdoor environments, which usually provide stable GPS signals, the compute-light VIO algorithm coupled with the GPS signals achieves the best accuracy and performance. In contrast, in unknown, unmapped indoor environments, the SLAM algorithm delivers the best accuracy.

We present \proj, a software-hardware system that adapts to different operating scenarios while providing real-time localization under tight energy budgets. Our localization algorithm framework adapts to different operating scenarios by unifying core primitives in existing algorithms (\Sect{sec:algo:framework}), and thus provides a desirable software baseline to identify general acceleration opportunities. Our algorithm exploits the inherent two-phase design of existing localizations algorithms, which share a similar visual frontend but have different 
optimization backends (e.g., probabilistic estimation vs. constrained optimization) that suit different operating scenarios.

We comprehensively characterize the localization framework to identify acceleration candidates (\Sect{sec:algo:char}). In particular, we focus on kernels that contribute significantly to the overall latency and/or latency \textit{variation}. While most prior work focuses on the overall latency only, latency variation is critical to the predictability and safety of autonomous machines. We show that the worst-case latency could be up to 4$\times$ higher than the best-case latency. Our strategy is to accelerate both high-latency and high-variation kernels in order to reduce both the overall latency and latency variation.


We show that, irrespective of the operating environment, the overall latency is bottlenecked by the vision frontend. We present a principled hardware accelerator for the vision frontend (\Sect{sec:fe}). While much of the prior work focuses on accelerating individual vision tasks (e.g., convolution~\cite{qadeer2013convolution}, stereo matching~\cite{feng2019asv, li20171920}), an autonomous machine's vision frontend necessarily integrates different tasks together. Our design judiciously applies pipelining and resource sharing by exploiting the unique task-level parallelisms, and captures the data locality by provisioning diffe    rent on-chip memory structures to suit different data reuse patterns.

With the frontend accelerated, the backend contributes significantly to the latency variation. Critically, although different operating scenarios invoke different backend kernels, key variation-contributing kernels share fundamental operations that are amenable to matrix blocking. We exploit this algorithmic characteristics to design a flexible and resource-efficient backend architecture (\Sect{sec:be}). The accelerator is coupled with a runtime scheduler, which ensures that reducing latency variation does not hurt the average latency.

We implement \proj on FPGA. We target FPGA for two main reasons. First, FPGA platforms today have mature sensor interfaces (e.g., MIPI Camera Serial Interface~\cite{mipicsi2}) and sensor processing hardware (e.g., Image Signal Processor~\cite{hennessy2017computer}) that greatly ease development and deployment, Second, targeting FPGA accelerates our time to market with low cost.


We demonstrate two FPGA prototypes (\Sect{sec:eval}), one that targets our self-driving car with a high-end Vertex-7 FPGA board, and the other that targets a drone setup using an embedded Zynq Ultrascale+ platform. We show that \proj achieves up to 4$\times$ speedup, 43\% -- 58\% latency variation reduction, as well as 47\% -- 74\% energy reduction.



\no{To our best knowledge, this is the first paper focusing on end-to-end systems and architecture design for general localization in autonomous machines. More specifically:}

This paper makes the following contributions:

\begin{itemize}
	\item We introduce a taxonomy of real-world environment, and quantitatively show that no single localization algorithm fits all operating environments.
	\item We propose a localization algorithm that adapts to different operating scenarios. The new algorithm provides a desirable software target for hardware acceleration.
	\item We present unnormalized, unobscured data collected from our operating environments, which let us identify lucrative acceleration targets and provides motivational data that future work could build upon.
	\item We show how to co-design an accelerator that significantly reduces the latency, energy, and variation of localization by exploiting unique the parallelisms, locality, and common operations in the algorithm.
\end{itemize}

%% file: background.tex
\section{Background and Context}
\label{sec:bck}



\begin{figure}[t]
\centering
\begin{minipage}[t]{0.48\columnwidth}
  \centering
  \includegraphics[width=\columnwidth]{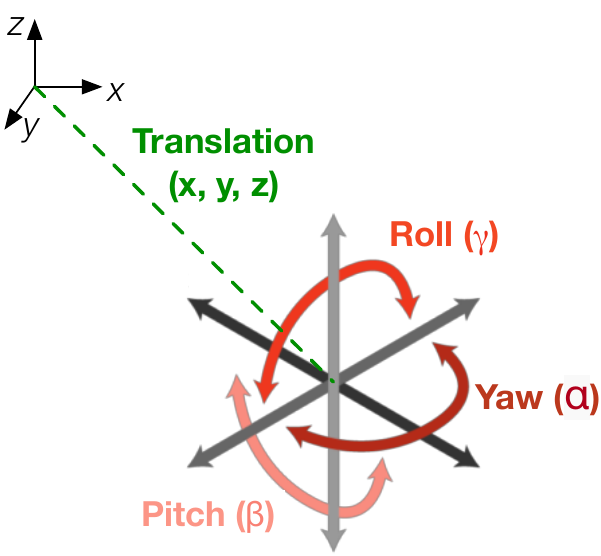}
  \caption{Localization estimates the six degree of freedom (6 DoF) pose of a body ($x, y, z, \alpha, \beta, \gamma$).}
  \label{fig:6dof}
\end{minipage}
\hspace{2pt}
\begin{minipage}[t]{0.48\columnwidth}
  \centering
  \includegraphics[width=\columnwidth]{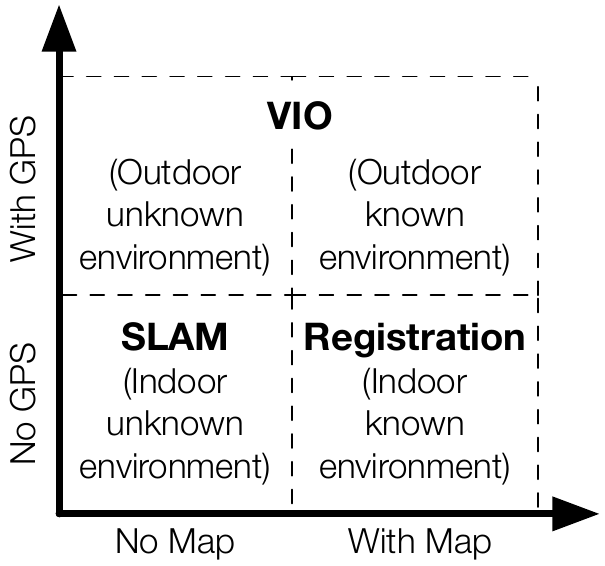}
  \caption{Taxonomy of real-world environments. Each environment prefers a particular localization algorithm.}
  \label{fig:taxonomy}
\end{minipage}
\end{figure}

\paragraph{Localization in Autonomous Machines}Fundamental to autonomous machines is localization, i.e., ego-motion estimation, which calculates the position and orientation of an agent itself in a given frame of reference. Formally, localization generates the six degrees of freedom (DoF) pose shown in~\Fig{fig:6dof}. The six DoF includes the three DoF for the \textit{translational} pose, which specifies the $<x$, $y$, $z>$ position, and the three DoF for the \textit{rotational} pose, which specifies the orientation about three perpendicular axes, i.e., yaw, roll, and pitch.

Localization is fundamental to autonomy. Knowing the translational pose fundamentally enables a robot/self-driving car to plan the path and to navigate. The rotational pose further lets robots and drones stabilize themselves~\cite{kelly2013mobile, dudek2010computational}.

\paragraph{Sensors} Localization is made possible through sensors that interact with the world. Common sensors include cameras, Inertial Measurement Units (IMU), and Global Positioning System (GPS) receivers. An IMU provides the \textit{relative} 6 DoF information by combining a gyroscope and an accelerometer. IMU samples are  noisy~\cite{el2004wavelet}; localization results would quickly drift if relying completely on the IMU. Thus, IMU is usually combined with the more reliable camera sensor in localization. GPS receivers directly provide the 3 translational DoF; they, however, are not used alone because their signals 1) do not provide the 3 rotational DoF, 2) are blocked in an indoor environment, and 3) could be unreliable even outdoor when the multi-path problem occurs~\cite{kos2010effects}.

These sensors are relatively cheap (below \$1,000 combined~\cite{IMUprice,GPSprice,Cameraprice}). We assume that they are available for localization, as in virtually all of today's autonomous machines.


%

%% file: motivation.tex
\section{One Algorithm Does Not Fit All}
\label{sec:mot}

\begin{figure*}[t]
\centering
\subfloat[\small{Indoor unknown environment. SLAM provides the best accuracy.}]
{
  \includegraphics[trim=0 0 0 0, clip, height=1.46in]{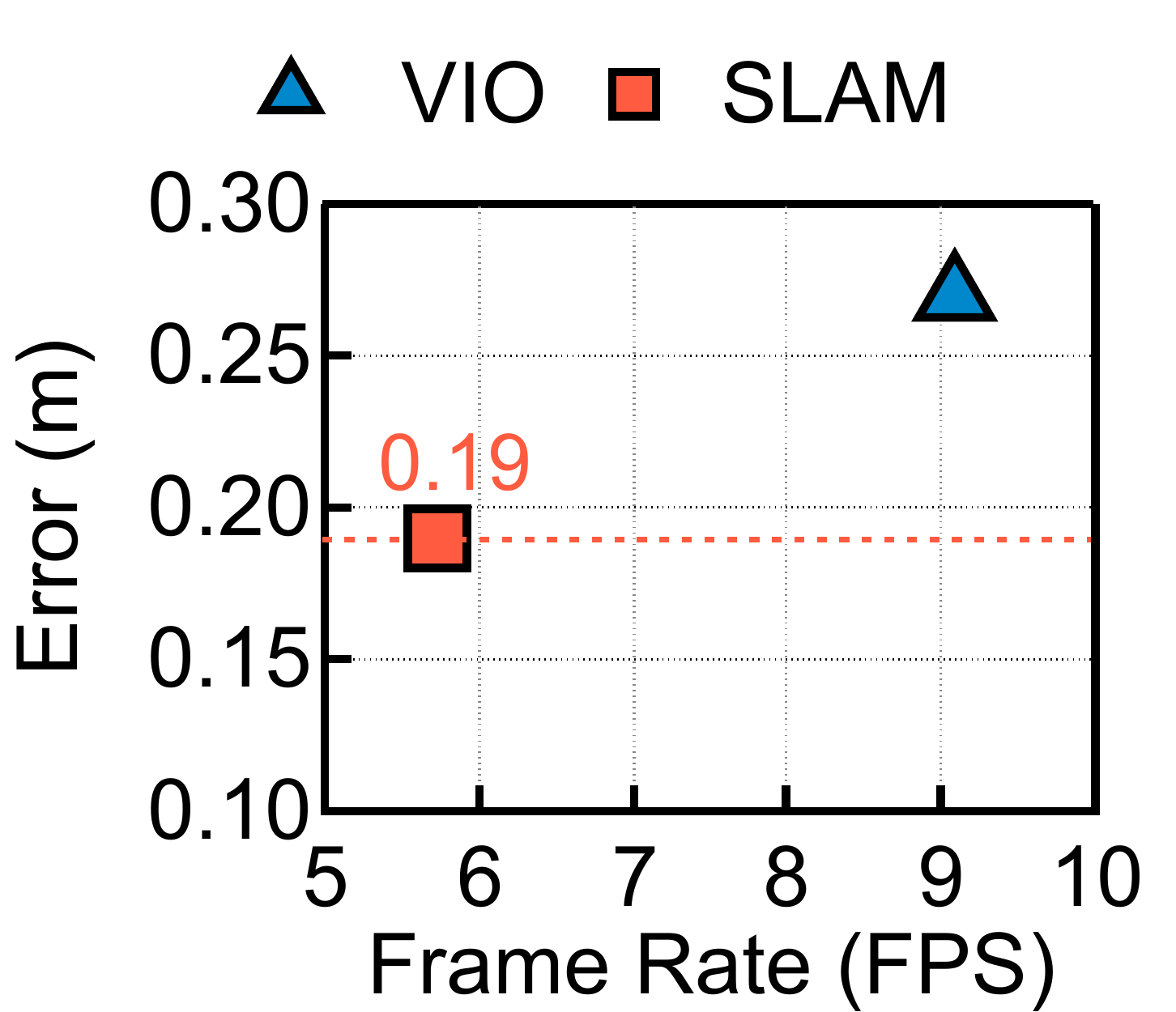}
  \label{fig:indoorwomap}
}
\hspace{2pt}
\subfloat[\small{Indoor known environment. Registration has the best accuracy.}]
{
  \includegraphics[trim=0 0 0 0, clip, height=1.46in]{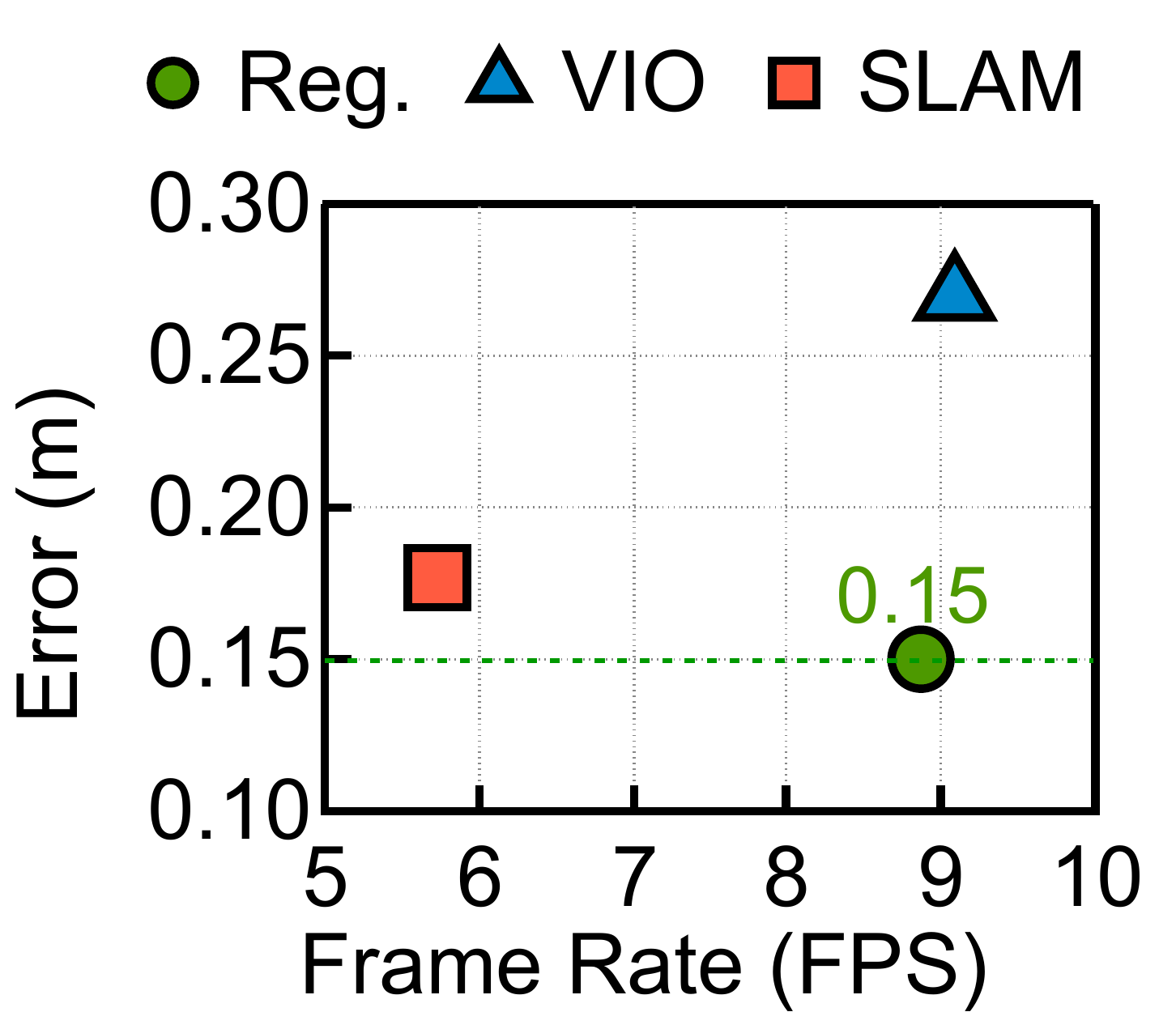}
  \label{fig:indoorwmap}
}
\hspace{2pt}
\subfloat[\small{Outdoor unknown environment. VIO is best in accuracy.}]
{
  \includegraphics[trim=0 0 0 0, clip, height=1.46in]{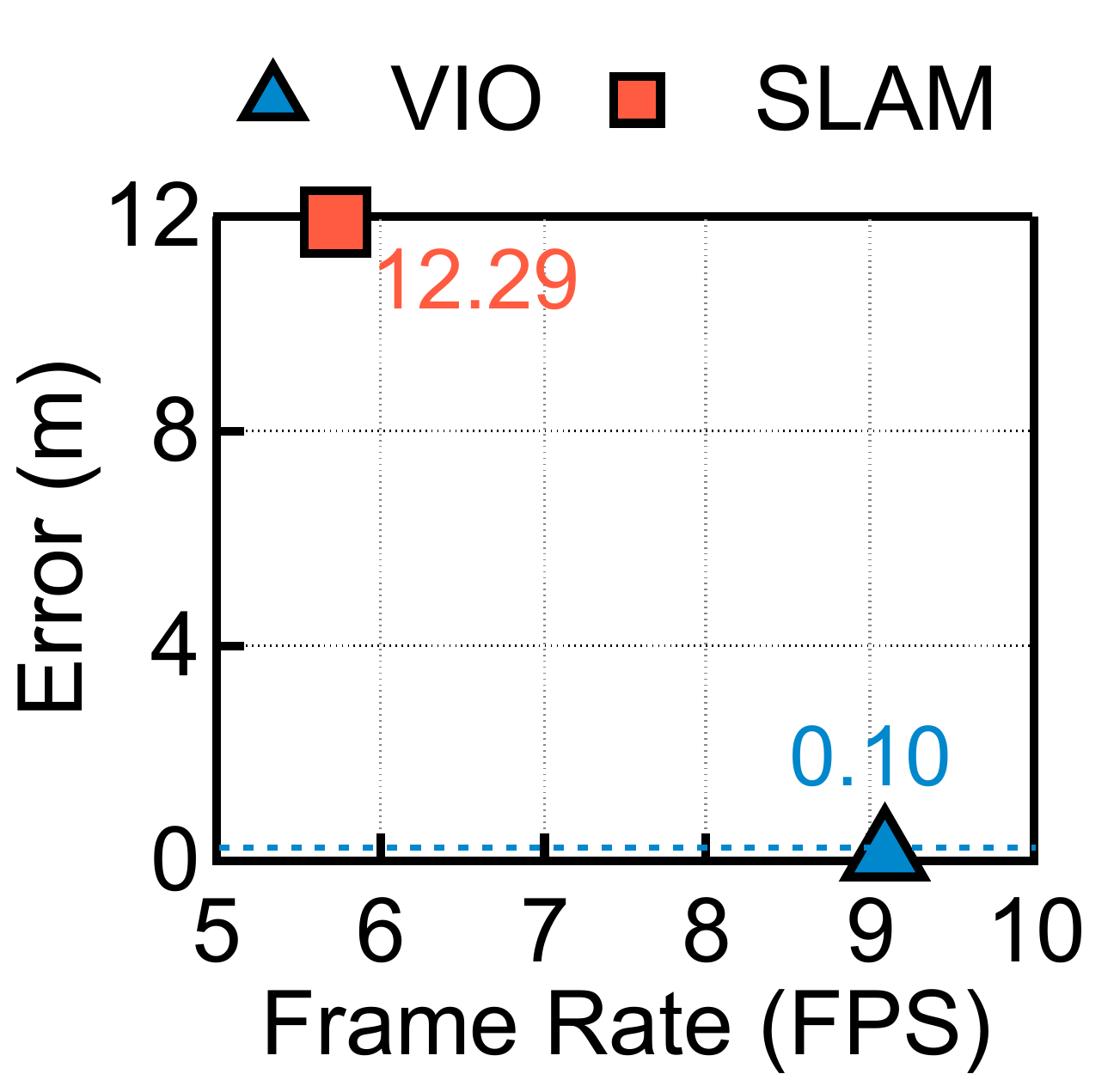}
  \label{fig:outdoorwomap}
}
\hspace{2pt}
\subfloat[\small{Outdoor known environment. VIO is best in accuracy.}]
{
  \includegraphics[trim=0 0 0 0, clip, height=1.46in]{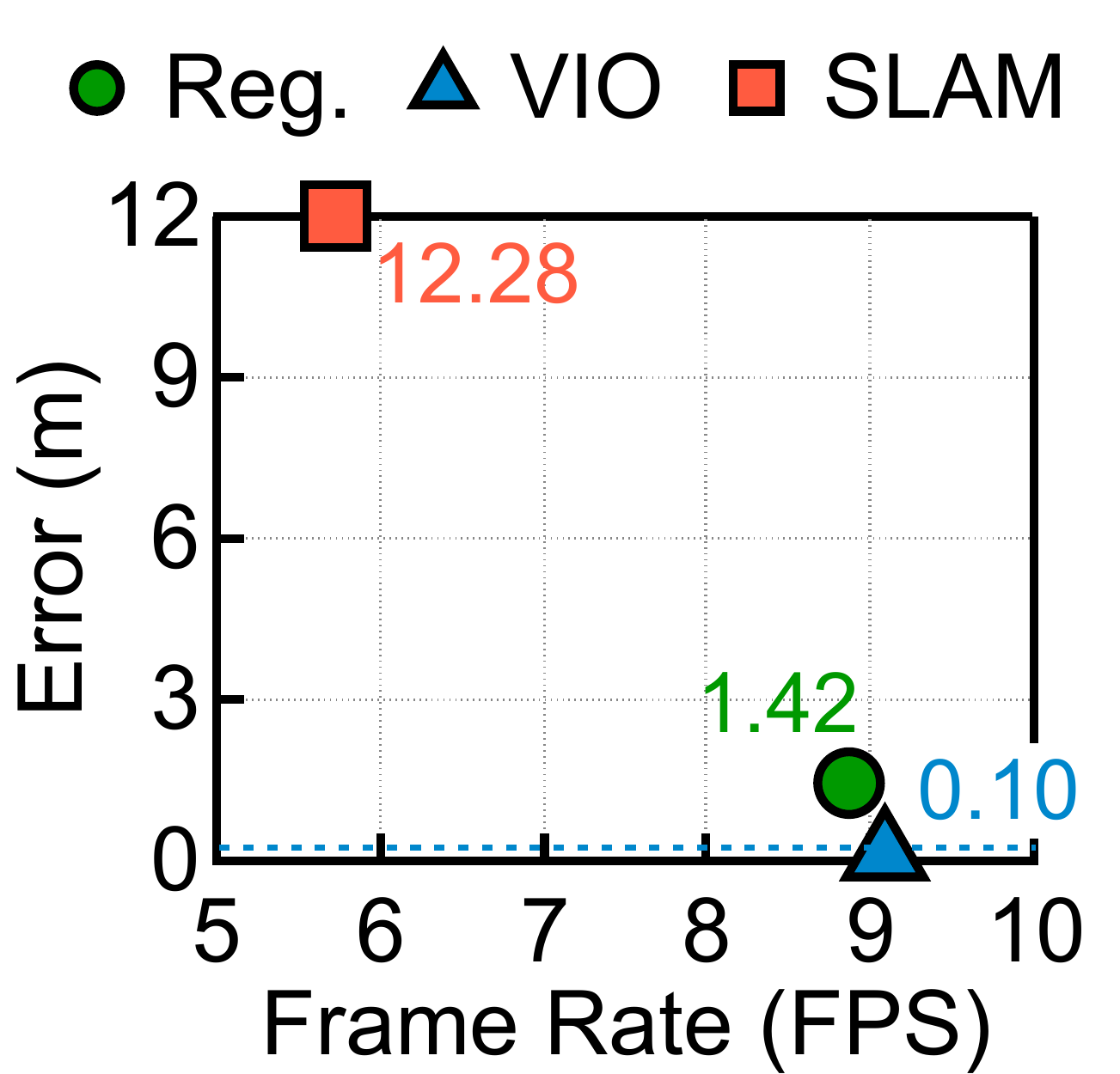}
  \label{fig:outdoorwmap}
}
\caption{Localization error vs. performance trade-off in the four operating scenarios, each requiring a particular localization algorithm to maximize accuracy. \no{We use accuracy as the metric to compare different algorithms here so as to provide the most accurate algorithm for acceleration.} Note that registration does not apply to scenarios without a map. We use a combination of the KITTI Odometry dataset~\mbox{\cite{geiger2012we}} and our in-house dataset to cover all four scenarios. Results are averaged across all frames.}
\label{fig:motivation}
\end{figure*}

Hardware design must target an efficient and broadly applicable software base to begin with. By analyzing three fundamental categories of localization algorithms, we find that 
no one single algorithm applies to all operating scenarios. A flexible localization framework that adapts to different scenarios is needed in practice. While a general-purpose processor easily provides the flexibility, the tight performance requirements call for hardware acceleration.


\paragraph{Operating Environment} Autonomous machines usually have to operate under different scenarios to finish a task. For instance, in one of our commercial deployments, logistics robots transfer cargo between warehouses in an industrial park; during normal operations the robots roughly spend 50\% of the time navigating outdoor and 50\% of the time inside pre-mapped warehouses. When the robots are moved to a different section of the park (to optimize the overall efficiency), the robots would spend a few days mapping new warehouses.


In principle, real-world environments can be classified along two dimensions for localization purposes as shown in \Fig{fig:taxonomy}: the availability of a pre-constructed map and the availability of a direct localization sensor, mainly the GPS. First, an environment's map could be available depending on whether the environment has been previously mapped, i.e., known to the autonomous machine. A map could significantly ease localization because the map provides a reliable frame of reference. Second, depending on whether the environment is indoor or outdoor, GPS could provide absolute positioning information that greatly simplifies localization.


Overall, four scenarios exist in the real-world:

\begin{itemize}
    \setlength\itemsep{-2pt}
	\item $<$No GPS, No Map$>$: indoor unknown environment;
	\item $<$No GPS, With Map$>$: indoor known environment;
	\item $<$With GPS, No Map$>$: outdoor unknown environment.
	\item $<$With GPS, With Map$>$:  outdoor known environment.
\end{itemize}


\paragraph{Localization Algorithms} To understand the most fitting algorithm in each scenario, we analyze three fundamental categories of localization algorithms~\mbox{\cite{kelly2013mobile, budiyono2013towards}} that are: 1) complementary to each other in requirements and capabilities, 2) well-optimized by algorithmic researchers, and 3) widely used in industry. These localization algorithms are:

\begin{itemize}
	\item \textbf{Registration}: It calculates the 6 DoF pose against \textit{a given map}. Given the current camera observation $\mathbf{F}$ and the global map $\mathbf{G}$, registration algorithms calculate the 6 DoF pose that transforms $\mathbf{F}$ to $\mathbf{F}'$ in a way that minimizes the Euclidean distance (i.e., error) between $\mathbf{F}'$ and $\mathbf{G}$. We use the ``bag-of-words'' framework~\cite{galvez2012bags, mur2014fast, dbow2}, which is the backbone of many products such as iRobot~\mbox{\cite{irobot}}.

	\item \textbf{VIO}: A classic way of localization without an explicit map is to formulate localization as a probabilistic nonlinear state estimation problem using Kalman Filter (KF), which effectively calculates the \textit{relative} pose of an autonomous machine with respect to the starting point. Common KF extensions include Extended Kalman Filter (EKF)~\cite{julier2004unscented} and Multi-State Constraint Kalman Filter (MSCKF)~\cite{mourikis2007multi}. Since KF-based methods often appear in a VIO system~\mbox{\cite{li2013high, li2012improving, li2013high, zhang2018pirvs, sun2018robust}}, this paper simply refers to them as VIO.

	~~In our experiments, we use a MSCKF-based framework~\mbox{\cite{sun2018robust, msckfvio}} due to its superior accuracy compared to other state estimation algorithms. For instance on the EuRoC dataset, MSCKF accuracy on average 0.18m error reduction over EKF~\mbox{\cite{sun2018robust}}. MSCKF is also used by many products such as Google ARCore~\mbox{\cite{arcoremsckf}}.

	~~Since VIO calculates the relative trajectory using past observations, its localization errors could accumulate over time~\cite{chen2018ionet}. One effective mitigation is to provide VIO with the absolute positioning information through GPS. When stably available, GPS signals help the VIO algorithm relocalize to correct the localization drift~\cite{dusha2012error}. VIO coupled with GPS is often used for outdoor navigation such as in DJI drones~\mbox{\cite{djiviogps}}.

	\item \textbf{SLAM}: It simultaneously constructs a map while localizing an agent within the map. SLAM avoids the accumulated errors in VIO by constantly remapping the global environment and closing the loop. SLAM is usually formulated as a constrained optimization problem, often through bundle adjustment~\mbox{\cite{hartley2003multiple}}. SLAM algorithms are used in many robots~\mbox{\cite{slamcore}} and AR devices (e.g., Hololens)~\mbox{\cite{mlhololens}}. We use the widely-used VINS-Fusion framework~\cite{qin2017vins, vinsfusion} for experiments.
\end{itemize}

Note that the design space of VIO and SLAM is broader than the specific algorithms we target. For instance, VIO could use factor-graph optimizations~\mbox{\cite{suleiman2019navion}} rather than KF. We use VIO to refer to algorithms using probabilistic state estimation without explicitly constructing a map, and uses SLAM to refer to global optimization-based algorithms that construct a map. It is these fundamental classes of algorithms that we focus on.

\begin{figure*}[t]
\centering
\includegraphics[width=2.1\columnwidth]{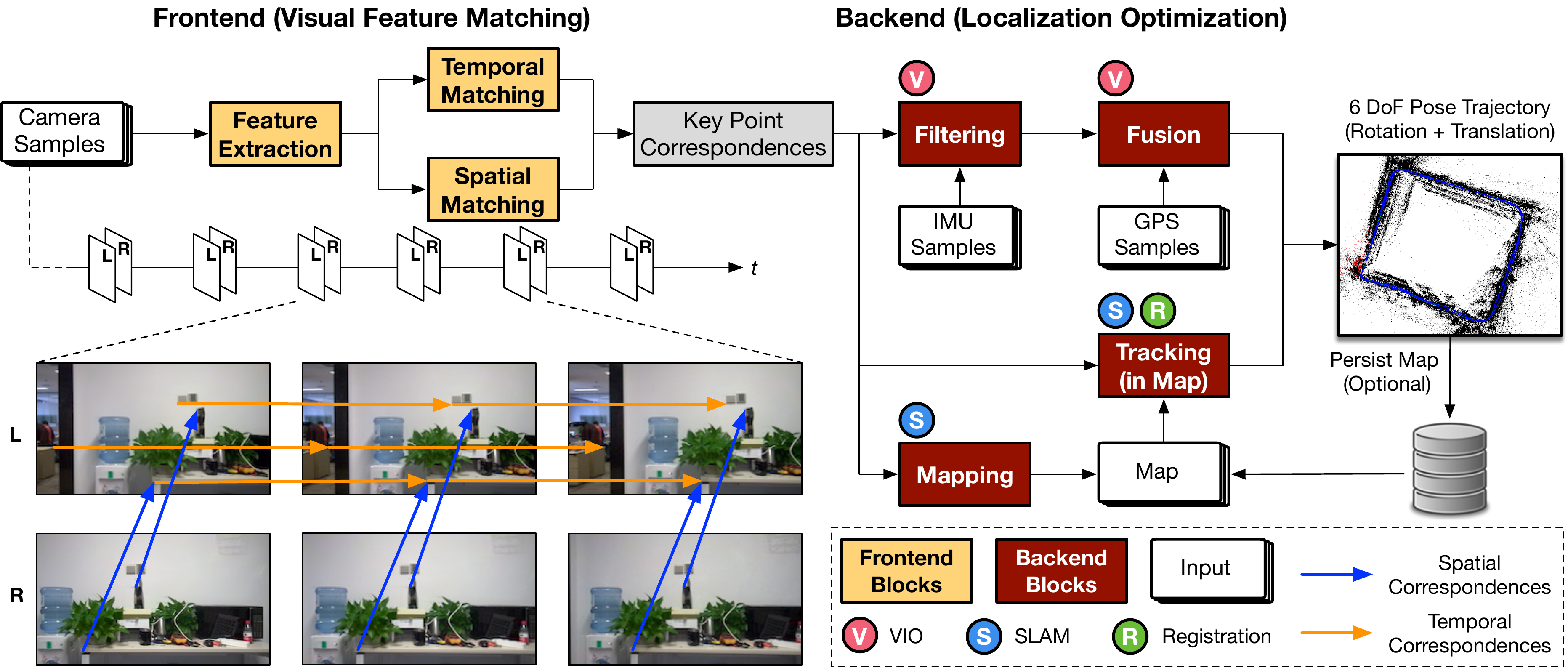}
\caption{Overview of our unified localization algorithm framework that adapts to different operating environments. The goal of our framework is to provide an efficient algorithm target for acceleration. The algorithm consists of a vision frontend and an optimization backend. The frontend establishes feature correspondences and is always activated. The backend operates under three modes, each activated in a particular operating scenario. Essentially, our localization framework fuses the three primitive localization algorithms (i.e. registration, VIO, and SLAM) by sharing common building blocks. This strategy preserves the accuracy under each scenario while minimizing algorithm-level redundancies.}
\label{fig:algo}
\end{figure*}

\paragraph{Accuracy Characterizations} We find that there is no single localization algorithm that fits all. Instead, each operating environment tends to prefer a different localization algorithm to minimize error. \Fig{fig:indoorwomap} through \Fig{fig:outdoorwmap} show the root-mean-square error (RMSE) ($y$-axis) and the average performance ($x$-axis) of the three localization algorithms under the four scenarios above, respectively. Since localization algorithm implementations today target CPUs~\cite{sun2018robust, mur2017orb}, we collect the performance data from a four-core Intel Kaby Lake CPU. \no{The code implementations leverage the multi-threading and SIMD capabilities wherever possible.} See \Sect{sec:eval:exp} for a detailed setup.

In an indoor environment without a map, \Fig{fig:indoorwomap} shows that SLAM delivers much lower error than VIO (0.19 m vs. 0.27 m). The registration algorithm is not applicable in this case as it requires a map. We do not supply the GPS signal to the VIO algorithm here due to the unstable signal reception; supplying unstable GPS signals would worsen the VIO accuracy. The data shows that VIO lacks the relocalization ability to correct accumulated drifts without GPS. 

When an indoor environment has a pre-constructed map, registration achieves higher accuracy while operating at a higher frame rate than SLAM as shown in~\Fig{fig:indoorwmap}. Specifically, the registration algorithm has only a 0.15 meter localization error while operating at 8.9 FPS. VIO almost doubles the error (0.27 m) due to drifts, albeit operating as a slightly higher frame rate (9.1 FPS).


VIO becomes the best algorithm outdoor, both with (\Fig{fig:outdoorwmap}) or without a map (\Fig{fig:outdoorwomap}). VIO achieves the highest accuracy with the help of GPS (0.10 m error) and is the fastest, Pareto-dominating the other two. Even with a pre-constructed map (\Fig{fig:outdoorwmap}), registration still has a much higher error (1.42 m) than VIO; SLAM is the slowest and has a significantly higher error due to difficulties to adapt to changing lightning conditions and drifts. Note that our SLAM error (12.28 m) is lower than prior SLAM work~\mbox{\cite{li2019879gops}} (21 m))


\Fig{fig:taxonomy} summarizes the algorithm affinity of each operating scenario to maximize accuracy: indoor environment with a map prefers registration; indoor environment without a map prefers SLAM; outdoor environment prefers VIO. As an autonomous machine often operates under different scenarios, a localization system must simultaneously support the three algorithms so as to be useful in practice.


\no{\paragraph{Need for Acceleration} Simultaneously supporting the three localization algorithms is challenging. As shown in \Fig{fig:motivation}, the localization algorithms operate below 10 FPS even on a PC machine. The performance is even lower on mobile platforms used in drones and robots (\Sect{sec:eval:sys}).}





%% file: algo.tex
\section{Unified Localization Framework}
\label{sec:algo}

We propose a localization algorithm framework that adapts to different operation environments, providing a desirable software target for hardware acceleration (\Sect{sec:algo:framework}). By characterizing the performance of the new algorithm framework, we identify lucrative acceleration candidates that, when accelerated, would significantly reduce the overall localization latency and latency variation (\Sect{sec:algo:char}).


\subsection{Algorithm Framework}
\label{sec:algo:framework}

We propose a localization framework that flexibly adapts to different operating environments.~\Fig{fig:algo} shows our algorithmic framework. Our strategy is to capture general patterns and to share common building blocks across each of the three primitive algorithms (i.e., registration, VIO, and SLAM). In particular, we find that the three primitive algorithms share the same two-phase design consisting of a \textit{visual feature matching} phase and a localization \textit{optimization} phase. While the optimization technique differs in the three primitive algorithms, the feature matching phase is the same.

\paragraph{Decoupled Framework} Our framework consists of a shared vision frontend, which extracts and matches visual features and is always activated, and an optimization backend, which has three modes---registration, VIO, and SLAM---each triggered under a particular operating scenario (\Fig{fig:taxonomy}). Each mode forms a unique dataflow path by activating a set of blocks in the backend as shown in \Fig{fig:algo}.

\no{Sharing the frontend across different backend modes is driven not only by the algorithmic characteristics, but also by the hardware resource-efficiency. Vision frontend is resource-heavy in localization accelerators. For instance, the frontend area contributes to about 27\% and 53\% of the total chip area in two recent ASIC accelerators~\cite{li2019879gops, suleiman2019navion}. The same is true in FPGA implementations as well~\cite{zhang2017visual, liu2019eslam}. Simply integrating accelerators individually designed for each algorithm would waste chip area or FPGA resources.}

Below we describe each block in the framework. Each block is directly taken from the three individual algorithms (\mbox{\Sect{sec:mot}}), which have been well-optimized and used in many products. While some components have been individually accelerated before~\mbox{\cite{qadeer2013convolution, feng2019asv}}, it is yet known how to provide a unified architecture to efficiently support these components in one system, which is the goal of our hardware design.

\paragraph{Frontend} The visual frontend extracts visual features to find correspondences in consecutive observations, both temporally and spatially, which the backend uses to estimate the pose. In particular, the frontend consists of three blocks.

\begin{itemize}
	\item \textbf{Feature Extraction} The frontend first extracts key feature points, which correspond to salient landmarks in the 3D world. Operating on feature points, as opposed to all image pixels, improves the robustness and compute-efficiency of localization. In particular, key points are detected using the widely-used FAST feature~\cite{rosten2006machine}; each feature point is associated with an ORB descriptor~\cite{rublee2011orb} in preparation for spatial matching later.
	\item \textbf{Stereo Matching} This block matches key points spatially between a pair of stereo images (from the left and right cameras). We use the widely-used blocking matching method~\cite{jakubowski2013block} based on the ORB descriptor of each feature point calculated before~\cite{calonder2010brief, rublee2011orb}.
	\item \textbf{Temporal Matching} This block establishes temporal correspondences by matching the key points of two consecutive images. Instead of searching for matches, this block tracks feature points across frames using the classic Lucas-Kanade optical flow method~\cite{lucas1981iterative}.
\end{itemize}

\paragraph{Backend} The backend calculates the 6 DoF pose from the visual correspondences generated in the frontend. Depending on the operating environment, the backend is dynamically configured to execute in one of the three modes, essentially providing the corresponding primitive algorithm. To that end, the backend consists of the following blocks:

\begin{itemize}
	\item \textbf{Filtering} This block is activated only in the VIO mode. It uses Kalman Filter to integrate a series of measurements observed over time, including the feature correspondences from the frontend and the IMU samples, to estimate the pose. We use MSCKF~\cite{mourikis2007multi}, a Kalman Filter framework that keeps a sliding window of past observations rather than just the most recent past.
	\item \textbf{Fusion} This block is activated only in the VIO mode. It fuses the GPS signals with the pose information generated from the filtering block, essentially correcting the cumulative drift introduced in filtering. We use a loosely-coupled approach~\cite{zhang2018pirvs}, where the GPS positions are integrated through a simple EKF~\cite{julier2004unscented}.
	\item \textbf{Mapping} This block is activated only in the SLAM mode. It uses the feature correspondences from the frontend along with the IMU measurements to calculate the pose and the 3D map. This is done by solving a non-linear optimization problem, which minimizes the projection errors (imposed by the pose estimation) from 2D features to 3D points in the map. The optimization problem is solved using the Levenberg-Marquardt (LM) method~\mbox{\cite{more1978levenberg}}. We target an LM implementation in the Ceres Solver, which is used in products such as Google's Street View~\mbox{\cite{ceresuse}}. In the end, the generated map could be optionally persisted offline and later used in the registration mode.
	\item \textbf{Tracking} This block is activated both in the registration and the SLAM mode. Using the bag-of-words place recognition method~\cite{galvez2012bags, mur2014fast}, this block estimates the pose based on the features in the current frame and a given map. In the registration mode, the map is naturally provided as the input. In the SLAM mode, tracking and mapping are executed in parallel, where the tracking block uses the latest map generated from the mapping block, which continuously updates the map.
\end{itemize}

\begin{figure}[t]
\centering
\begin{minipage}[t]{0.47\columnwidth}
  \centering
  \includegraphics[width=\columnwidth]{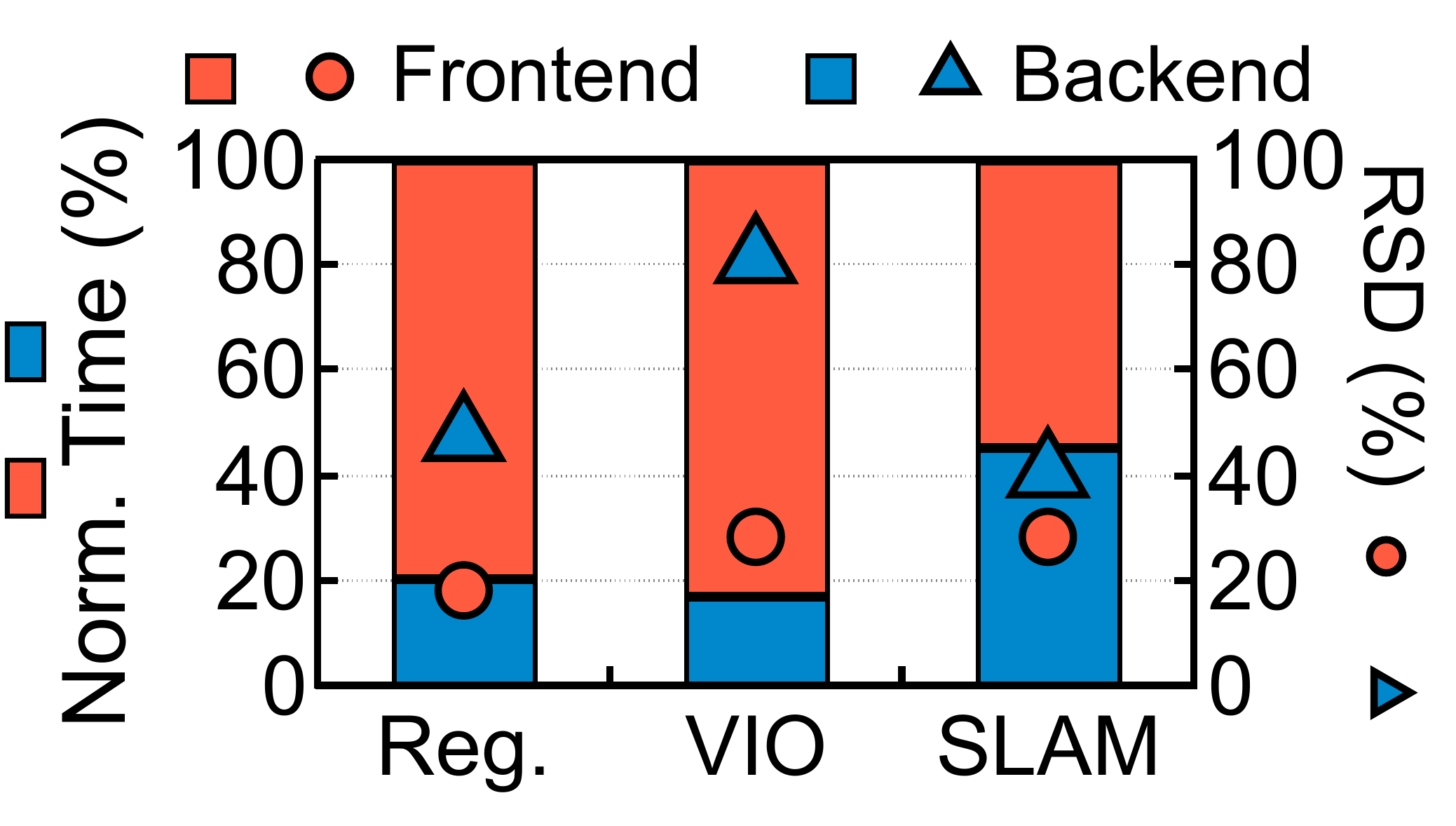}
  \caption{Latency distribution and relative standard deviation (RSD) of frontend and backend in three modes.}
  \label{fig:orb_dist}
\end{minipage}
\hspace{5pt}
\begin{minipage}[t]{0.47\columnwidth}
  \centering
  \includegraphics[width=\columnwidth]{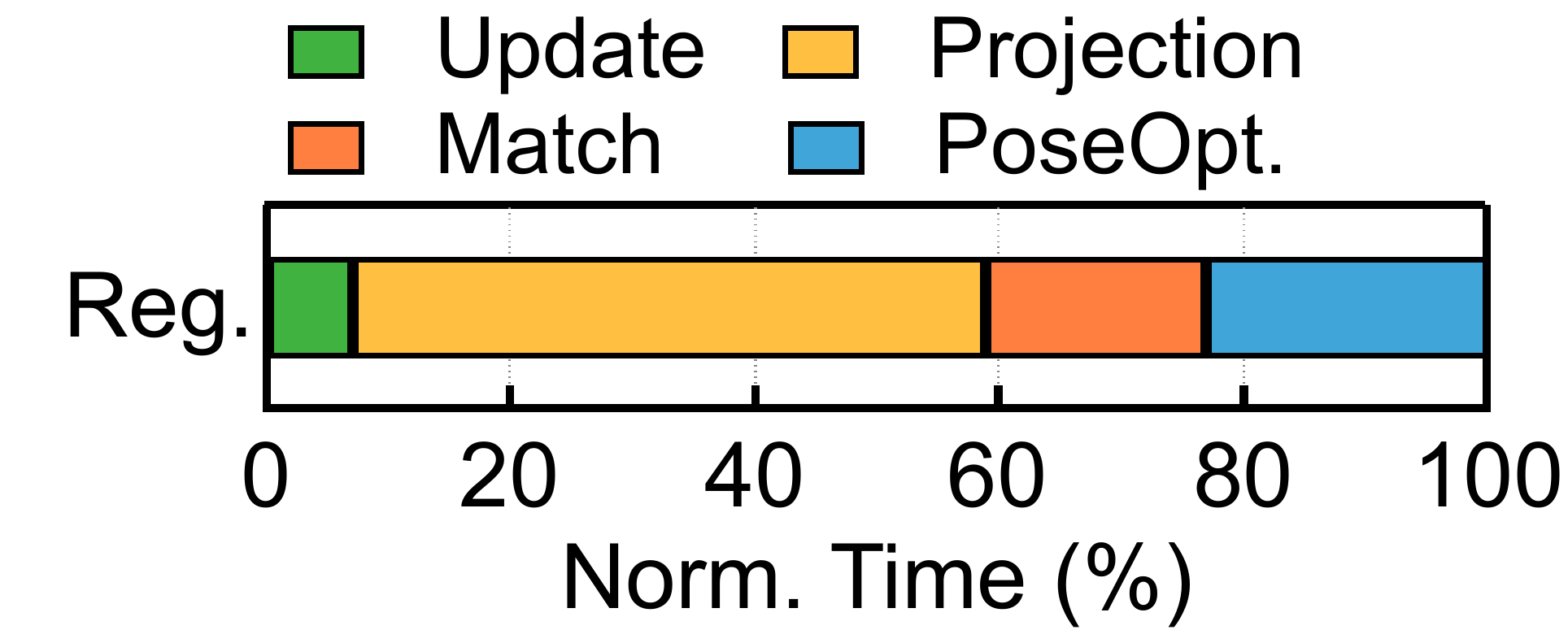}
  \caption{Latency distribution in registration backend.}
  \label{fig:reg_be_time}
\end{minipage}
\\
\vspace{3pt}
\begin{minipage}[t]{0.47\columnwidth}
  \centering
  \includegraphics[width=\columnwidth]{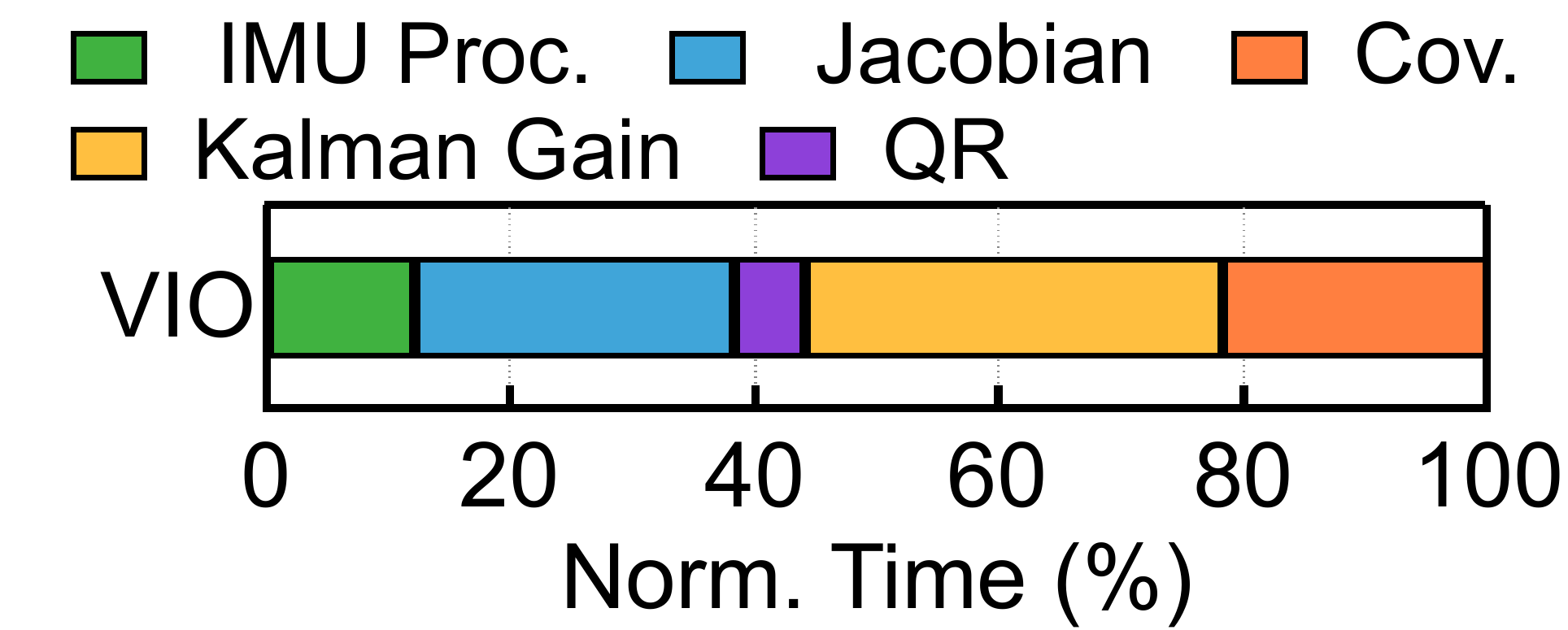}
  \caption{Latency distribution in VIO backend.}
  \label{fig:vio_be_time}
\end{minipage}
\hspace{5pt}
\begin{minipage}[t]{0.47\columnwidth}
  \centering
  \includegraphics[width=\columnwidth]{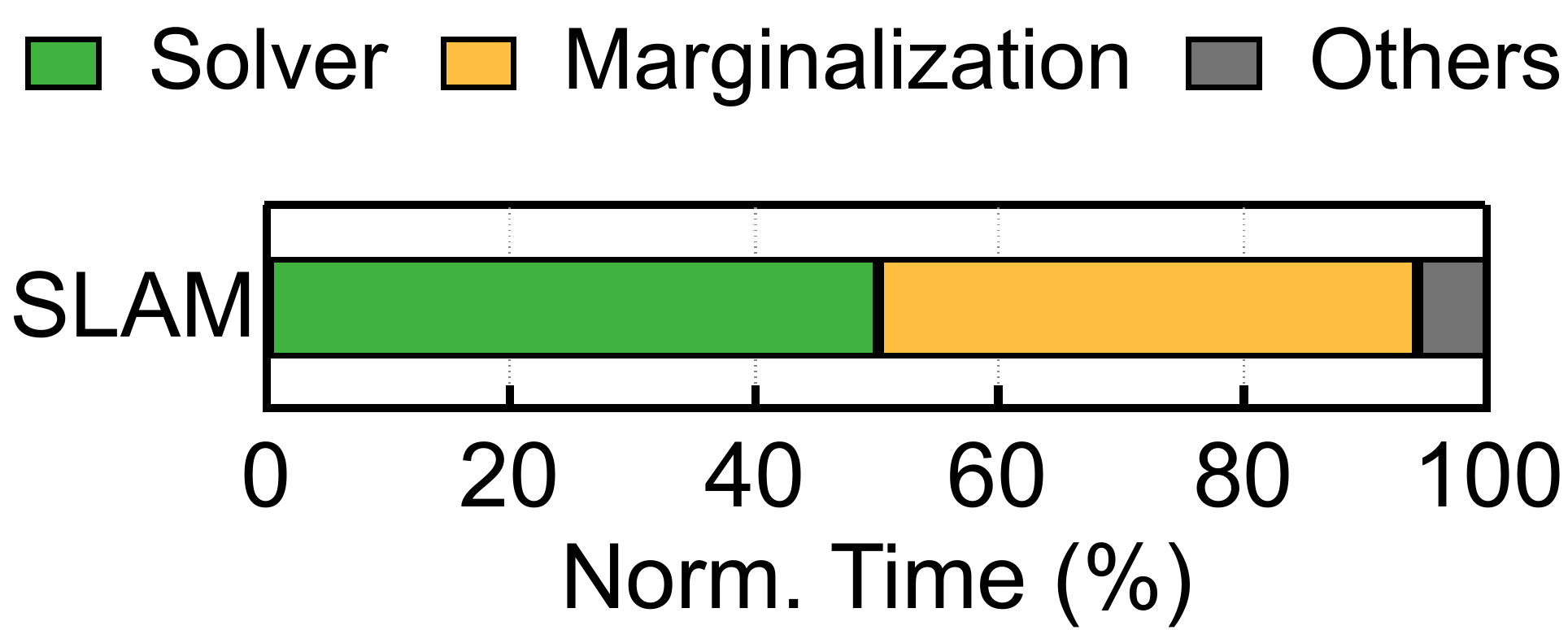}
  \caption{Latency distribution in SLAM backend.}
  \label{fig:slam_be_time}
\end{minipage}
\end{figure}



\no{\paragraph{Mode Switching} Switching between different modes in practice has negligible overhead. The quality of the GPS signal (updated every 100 $ms$) immediately indicates indoor vs. outdoor environment. Once indoor, running registration against a pre-constructed map for one single frame immediately indicates whether the map is useful or not.}

\paragraph{Accuracy and Performance} Our software framework is accurate. On the popular (indoor) drone dataset EuRoC~\mbox{\cite{burri2016euroc}}, our algorithm has a relative trajectory error of 0.28\% (registration) -- 0.42\% (SLAM), on par with prior algorithms, whose errors are within the 0.1\% to 2\% range~\mbox{\cite{fraundorfer2012visual}}. On the (outdoor) KITTI dataset~\mbox{\cite{geiger2012we}}, our algorithm has a negligible error ($<$ 0.01\% error) using VIO+GPS.




Our software framework is about 4\% faster than the dedicated algorithms, because we remove their dependencies on the Robot Operation System (ROS), which is a common framework (libraries, runtime services) on top of Linux for developing robotics applications but is known to incur non-trivial overheads~\cite{profanter2019opc, wei2016rt}. Our framework is thus a clean target for performance characterizations and acceleration.

\subsection{Latency Characterizations}
\label{sec:algo:char}


\no{We identify algorithmic blocks that contribute significantly to the average latency and/or demonstrate high latency variation. These blocks are ideal acceleration candidates. For instance, accelerating a block with a Gaussian latency distribution by $C$ times reduces its standard deviation by $C$ times~\cite{friedman2001elements}.}


\begin{figure}[t]
\centering
\subfloat[\small{Per-frame latency breakdown between frontend and backend.}]
{
  \includegraphics[trim=0 0 0 0, clip, width=0.5\columnwidth]{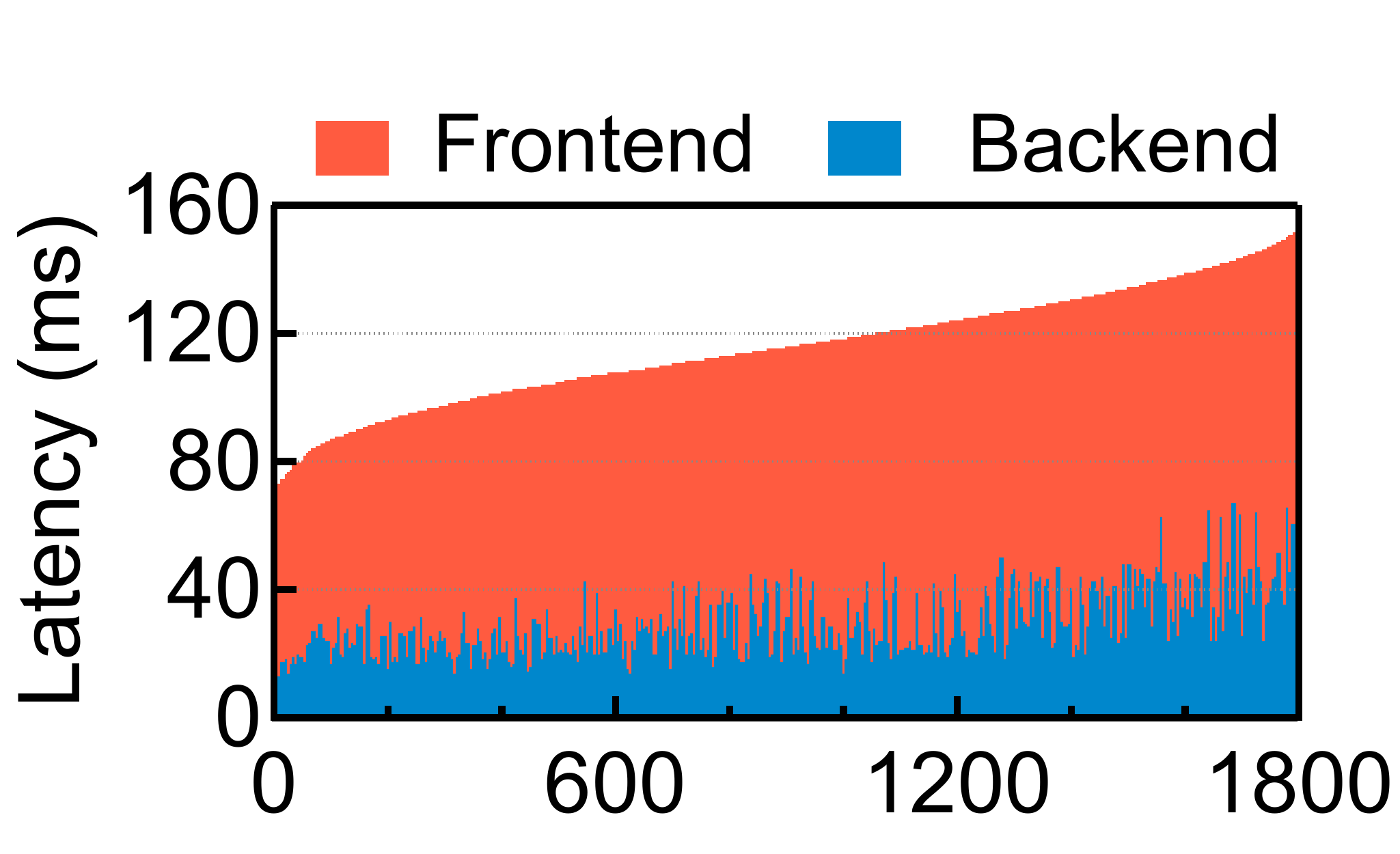}
  \label{fig:reg_variation}
}
\subfloat[\small{Latency breakdown in backend.}]
{
  \includegraphics[trim=0 0 0 0, clip, width=0.5\columnwidth]{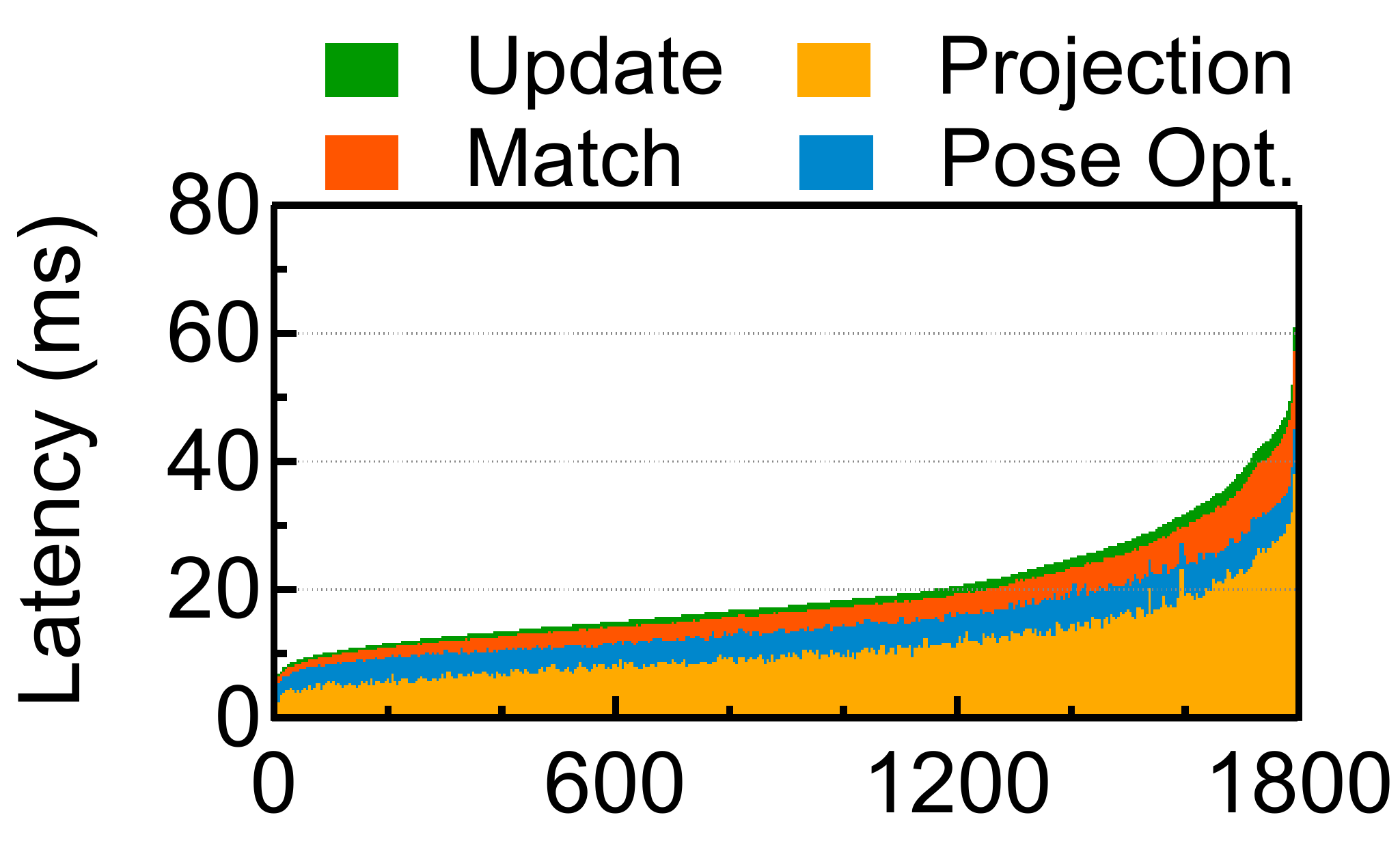}
  \label{fig:reg_be_var}
}
\caption{Latency variation in the registration mode.}
\label{fig:reg_var}
\vspace{-5pt}
\centering
\subfloat[\small{Per-frame latency breakdown between frontend and backend.}]
{
  \includegraphics[trim=0 0 0 0, clip, width=0.5\columnwidth]{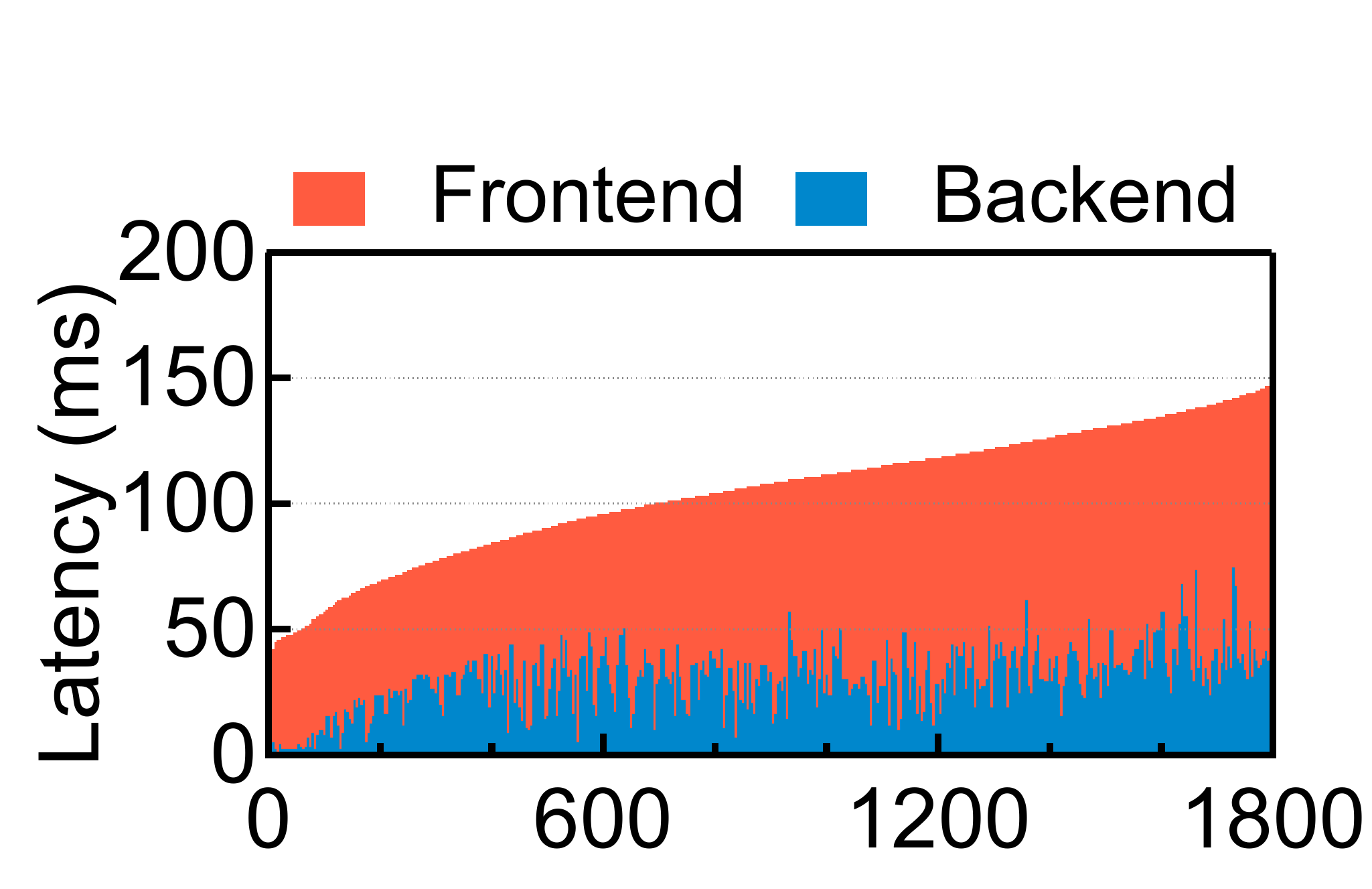}
  \label{fig:vio_variation}
}
\subfloat[\small{Latency breakdown in backend.}]
{
  \includegraphics[trim=0 0 0 0, clip, width=0.5\columnwidth]{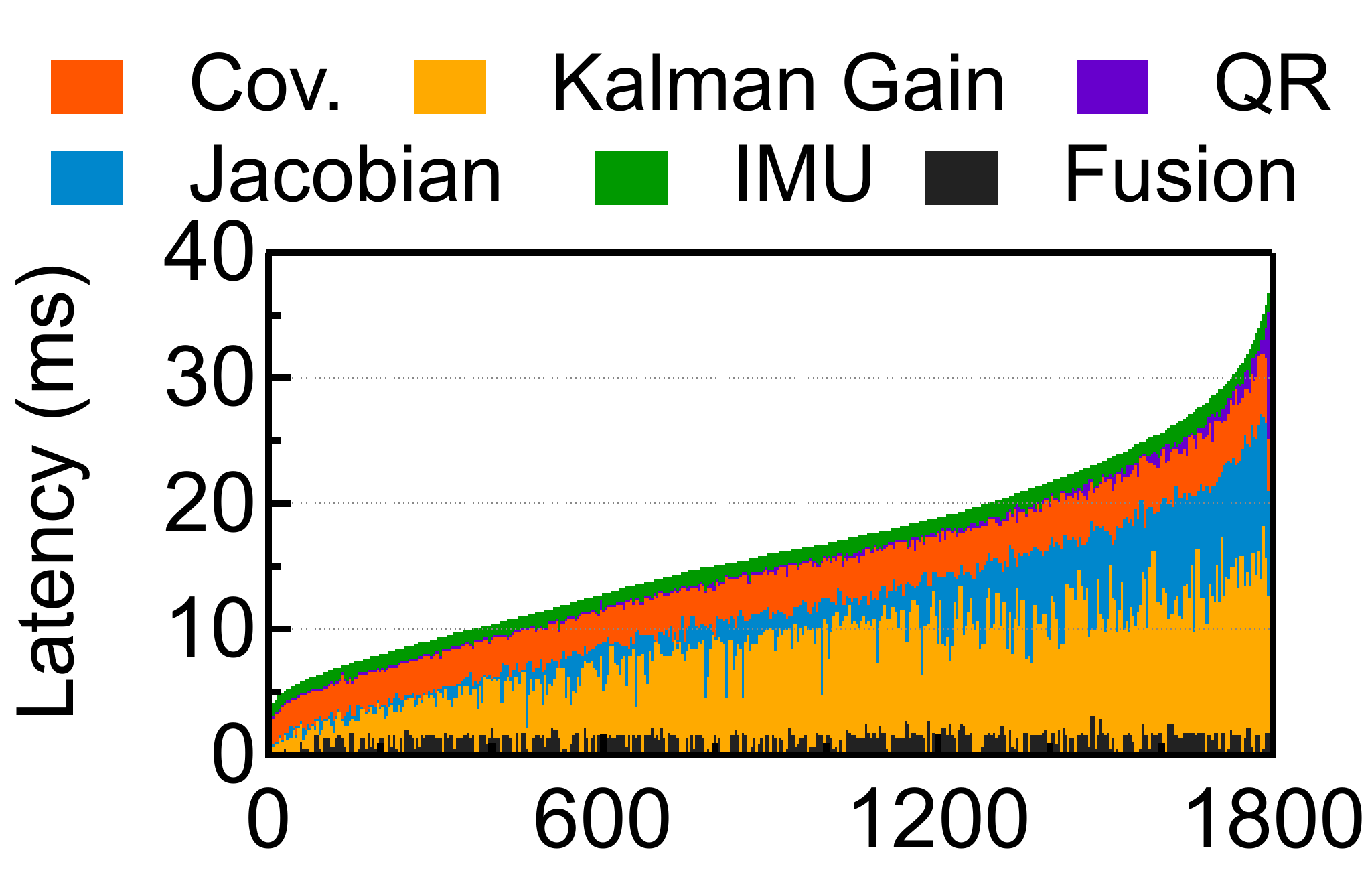}
  \label{fig:vio_be_var}
}
\caption{Latency variation in the VIO mode.}
\label{fig:vio_var}
\vspace{-5pt}
\centering
\subfloat[\small{Per-frame latency breakdown between frontend and backend.}]
{
  \includegraphics[trim=0 0 0 0, clip, width=0.5\columnwidth]{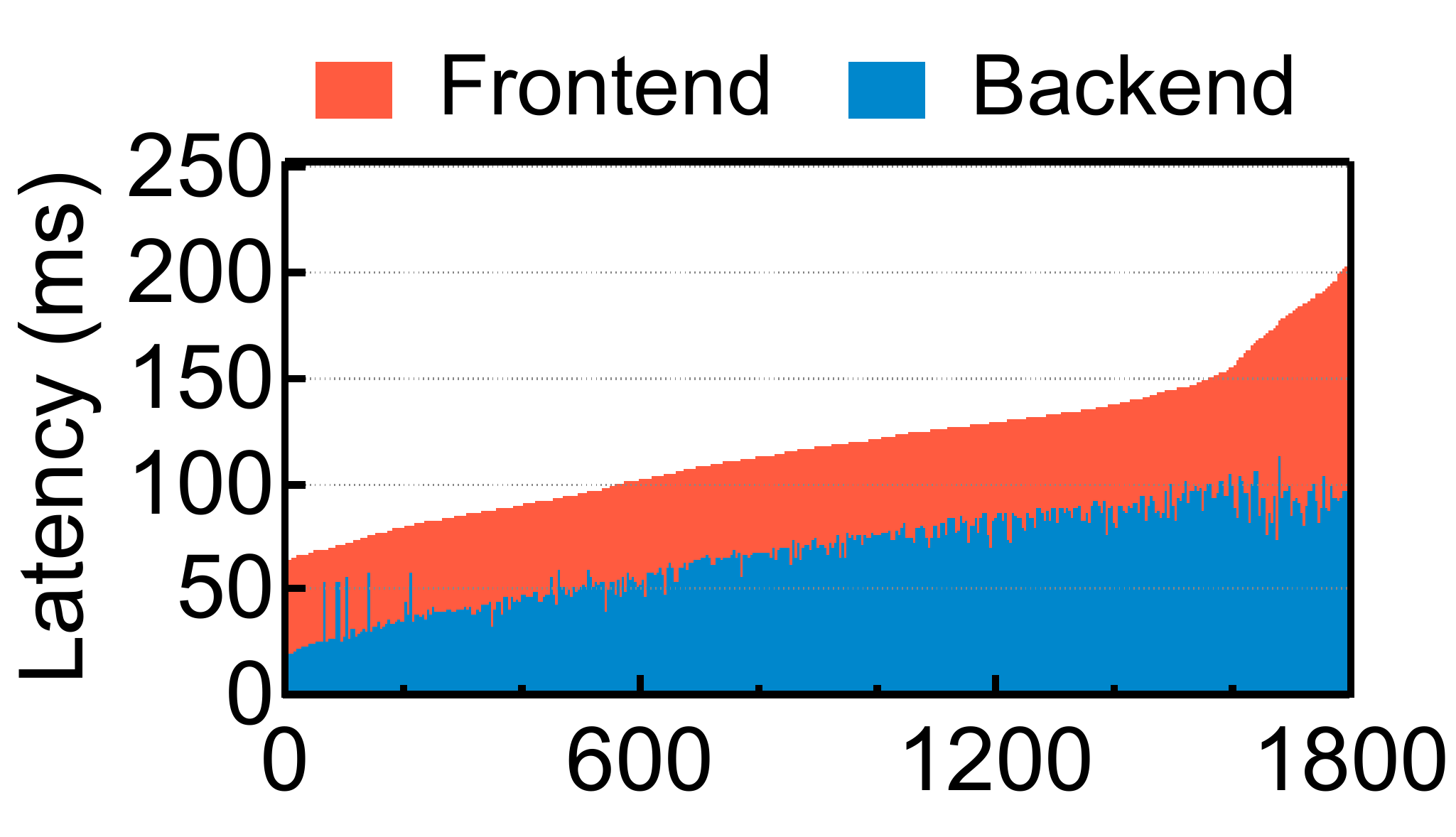}
  \label{fig:slam_variation}
}
\subfloat[\small{Latency breakdown in backend.}]
{
  \includegraphics[trim=0 0 0 0, clip, width=0.5\columnwidth]{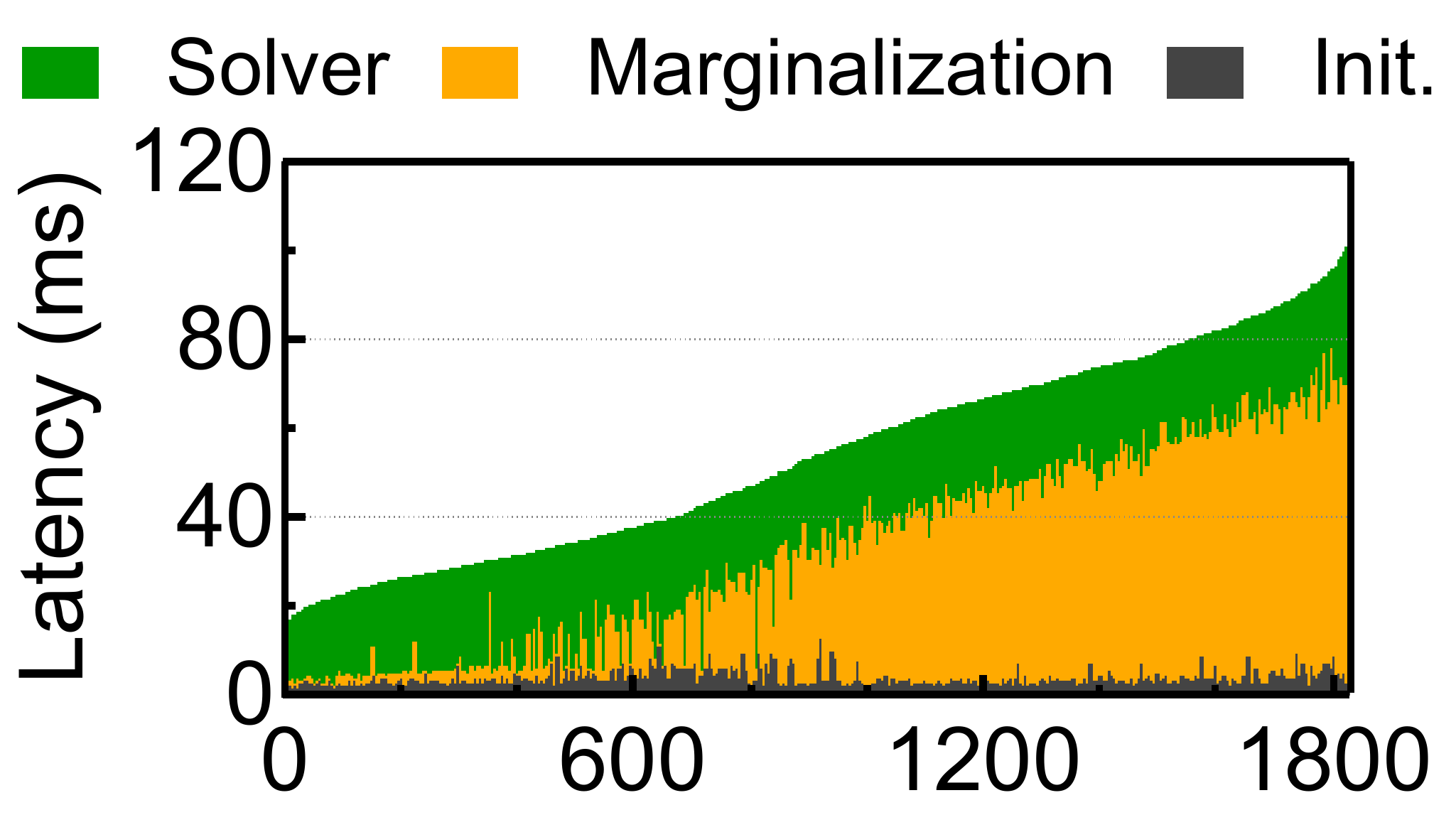}
  \label{fig:slam_be_var}
}
\caption{Latency variation in the SLAM mode.}
\label{fig:slam_var}
\end{figure}

\begin{figure*}[t]
\centering
\includegraphics[width=2.1\columnwidth]{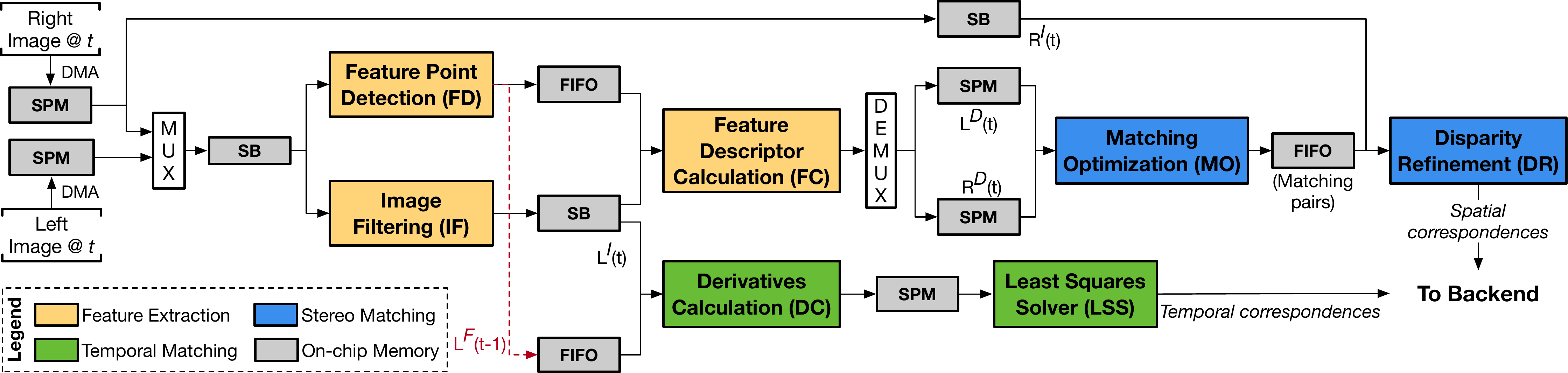}
\caption{Frontend architecture, which consists of three blocks: feature extraction, optical flow, and stereo matching. The left and right camera streams are time-multiplexed in the feature extraction block to save hardware resources. On-chip memories are customized in different ways to suit different types of data reuse: stencil buffer (SB) supports stencil operations, FIFO captures sequential accesses, and scratchpad memory (SPM) supports irregular accesses.}
\label{fig:fe}
\end{figure*}

\paragraph{Latency Distribution} We find that the frontend contributes significantly to the end-to-end latency across all three scenarios. Since the frontend is shared across different backend modes, accelerating the frontend would lead to ``universal'' performance improvement.~\Fig{fig:orb_dist} shows the average latency distribution between the frontend and the backend across the three scenarios. The frontend time varies from 55\% in the SLAM mode to 83\% in the VIO mode. The SLAM backend is the heaviest because it iteratively solves a complex non-linear optimization problem.


While the frontend is a lucrative acceleration target, backend time is non-trivial too, especially after the frontend is accelerated. \Fig{fig:reg_be_time}, \Fig{fig:vio_be_time}, and \Fig{fig:slam_be_time} show the distribution of different blocks within each backend mode. Marginalization in SLAM, computing Kalman gain in VIO, and projection in registration are the biggest contributors to the three backend modes. Interestingly, these three kernels also contribute significantly to the backend variation, which we analyze next.



\paragraph{Latency Variation} Localization has high latency variation under all three modes. ~\Fig{fig:reg_variation},~\Fig{fig:vio_variation}, and~\Fig{fig:slam_variation} show the per-frame latency distribution between the frontend and the backend in the registration, VIO, and SLAM mode, respectively. The data is sorted by the total latency. The longest latency in the SLAM mode is over 4$\times$ longer than the shortest latency. The difference is over 2$\times$ in the registration mode.



Compared to the frontend, the backend exhibits higher variation. We use relative standard deviation (RSD, a.k.a., coefficient of variation), which is defined as the ratio of the standard deviation to the mean of a distribution~\cite{friedman2001elements}, to compare the frontend and backend variation. The right $y$-axis in \Fig{fig:orb_dist} compares the RSDs of the frontend and backend in the three modes. The difference is most significant in the VIO mode where the RSDs of the frontend and the backend are 47.3\% and 81.1\%, respectively.

We further breakdown the latency variation in each backend mode, shown in~\Fig{fig:reg_be_var},~\Fig{fig:vio_be_var}, and~\Fig{fig:slam_be_var}. In each backend there is a single biggest contributor to the variation: camera model projection in registration, computing Kalman gain in VIO, and marginalization in SLAM. They match the overall latency contributors described before.

\no{\paragraph{Summary} The characterization results point out four acceleration candidates. The frontend contributes significantly to the overall latency under all scenarios, which we address in \Sect{sec:fe}. The three backend modes each possess a kernel that contributes heavily to both the backend time and variation. We will show in \Sect{sec:be} that the three kernels share common building blocks, and could be accelerated through a unified backend architecture.}

%% file: fe.tex
\section{Frontend Architecture}
\label{sec:fe}




This section describes our frontend accelerator. After an overview (\Sect{sec:fe:ov}), we discuss two optimizations that 1) improve the performance by exploiting unique task-level parallelisms (\Sect{sec:fe:para}) and 2) reduce on-chip memory usage by capturing data locality at the FPGA synthesis time (\Sect{sec:fe:data}).


\subsection{Overview and Design Principles}
\label{sec:fe:ov}

\Fig{fig:fe} provides an overview of the architecture. The input (left and right) images are streamed through the DMA and double-buffered on-chip. The on-chip memory design mostly allows the frontend to access DRAM only at the beginning and the end of the pipeline as we will discuss later.

The two images go through three \textit{blocks}: feature extraction, spatial matching, and temporal matching. Each block consists of multiple \textit{tasks}. For instance, the feature extraction block consists of three tasks: feature point detection, image filtering, and descriptor calculation.

The feature extraction block is exercised by both the left and right images. The feature points in the left image at time $t-1$ ($L^{F}_{t-1}$) are buffered on-chip, which are combined with the left image at time $t$ ($L^{I}_{t}$) to calculate the temporal correspondence at $t$. Meanwhile, the feature descriptors in both images at $t$ ($L^{D}_{t}$ and $R^{D}_{t}$) are consumed by the stereo matching block to calculate the spatial correspondences at $t$. The temporal and spatial correspondences are about 2 -- 3 KB on average; they are transmitted to the backend.



\no{The key design principle of our frontend architecture is to exploit the data reuse patterns and parallelisms \textit{across tasks}. While much of the prior work focuses on accelerating individual tasks, e.g., convolution~\cite{qadeer2013convolution}, exploiting locality and parallelism across tasks provides opportunities to improve performance while reducing the resource consumption.}




\subsection{Exploiting Task-Level Parallelisms}
\label{sec:fe:para}


%
%
%

\paragraph{Understanding the Parallelisms} \no{\Fig{fig:tlp} shows the detailed task-level dependencies in the frontend. }At the high-level, feature extraction (FE) consumes both the left and right images, which are independent and could be executed in parallel. Stereo matching (SM) must wait until both images finish the Feature Descriptor Calculation (FC) task in the FE block, because SM requires the feature points/descriptors generated from both images. Temporal matching (TM) operates only on the left image and is independent of SM. Thus, TM could start whenever the left image finishes the image filtering (IF) task in FE. The IF task and the feature point detection (FD) task operate in parallel in FE.

TM consists of two serialized tasks: derivatives calculation (DC), whose outputs drive a (linear) least squares solver (LSS). SM consists of a matching optimization (MO) task, which provides initial spatial correspondence by comparing hamming distances between feature descriptors, and a disparity refinement (DR) task which refines the initial correspondences through block matching.


\paragraph{Design} TM latency is usually over 10$\times$ lower than SM latency. Thus, the critical path is FD $\rightarrow$ FC $\rightarrow$ MO $\rightarrow$ DR. The critical path latency is in turn dictated by the SM latency (MO+DR), which is roughly 2 -- 3$\times$ higher than the FE latency (FD+FC). Therefore, we pipeline the critical path between the FE and the SM. Pipelining improves the frontend throughput, which is dictated by the latency of SM.

Interestingly, the FE hardware resource consumption roughly doubles the resource consumption of TM and SM combined, as FE processes raw input images whereas SM and TM process only key points ($<$0.1\% of total pixels). Thus, we time-share the FE hardware between the left and right images. This reduces hardware resource consumption, but does not hurt the throughput since FE is much faster than SM.

\no{\begin{figure}[t]
\centering
\includegraphics[width=\columnwidth]{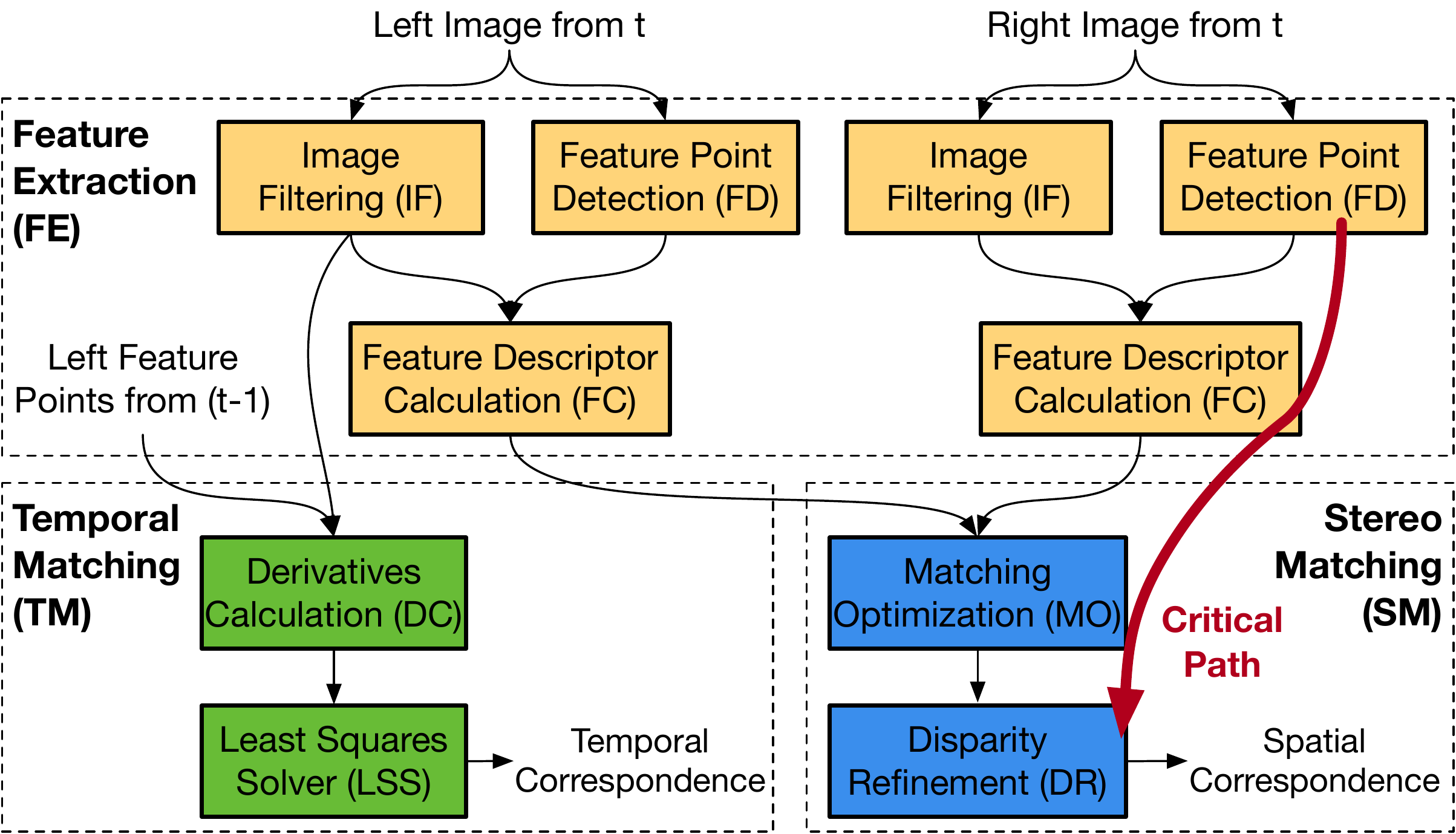}
\caption{Task dependencies in the frontend.}
\label{fig:tlp}
\end{figure}}







\subsection{Capturing Data Locality By Synthesis-Time Specialization}
\label{sec:fe:data}

We carefully design on-chip memory structures to capture three fundamental types of intra-task and inter-task data reuses patterns in the frontend (\Fig{fig:fe}). First, the frontend algorithm has many stencil operations, such as convolution in IF and block-matching in MO. We propose a stencil buffer (SB) design to capture data reuse in stencil operations. Second, many stencil operations read from a list sequentially, for which a FIFO is suitable. For instance, FC operates on feature points detected before one after another. Finally, some stages (e.g., MO) exercise arbitrary memory accesses, for which a generic scratchpad memory (SPM) is more suitable. We use generic FIFO and SPM structures, and describe our unique SB below.


\paragraph{Basic Stencil Buffer Design}  General data reuse in stencil operations, e.g., convolution, has been extensively explored~\cite{chen2014dadiannao, chen2016eyeriss, jouppi2017datacenter}. However, since we target FPGA, our SBs are designed to specialize for \textit{specific stencil sizes} in a given algorithm when we synthesize an FPGA design.

\begin{figure}[t]
\centering
\includegraphics[width=\columnwidth]{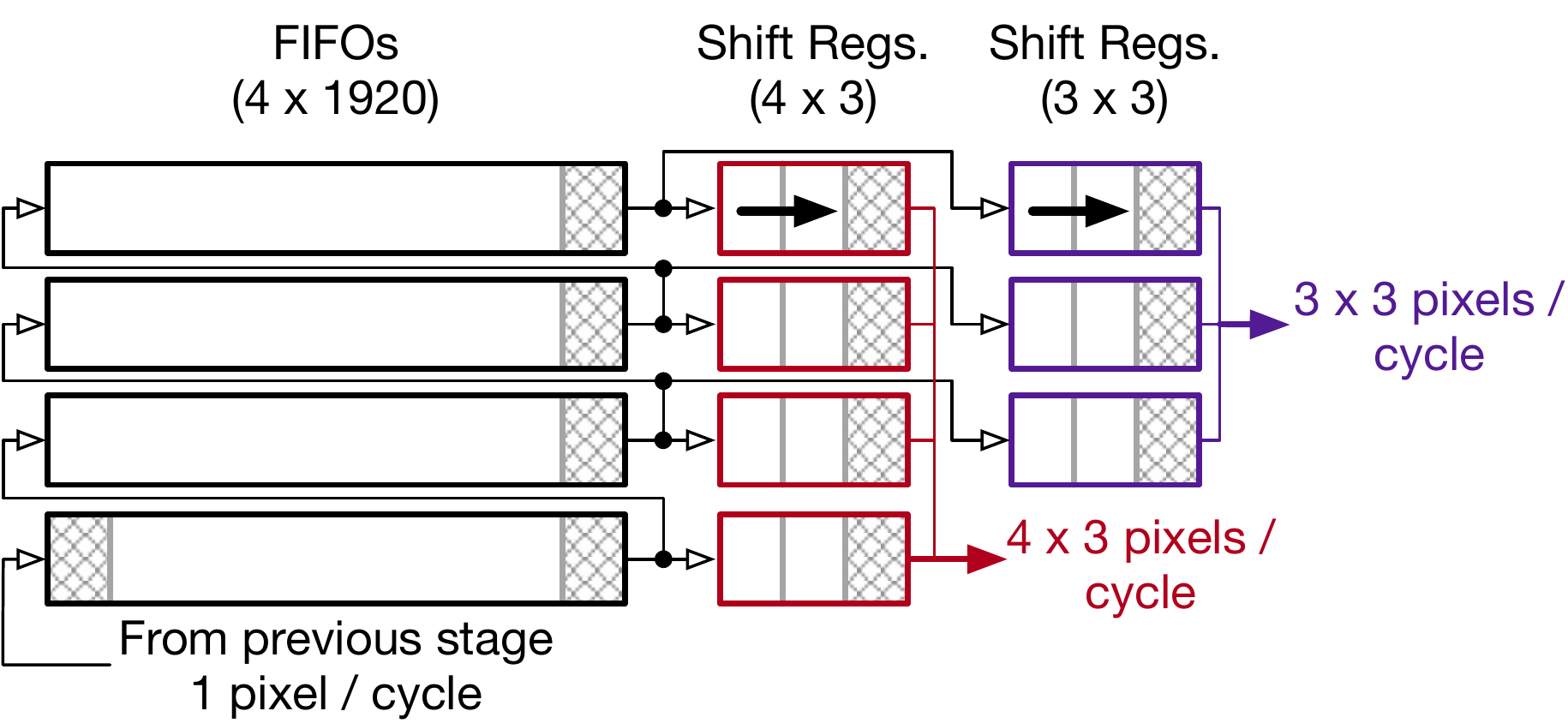}
\caption{Stencil buffer (SB) design. An SB consists of a set of (4 here) sequentially cascaded FIFOs that are connected to one or more (2 here) shift registers. Each cycle, one pixel enters the SB and 4 pixels enter the shift registers.\no{ Each shift register outputs its entirety every cycle for the stencil operation (e.g., convolution).}}
\label{fig:isb}
\vspace{15pt}
%
\centering
\includegraphics[width=\columnwidth]{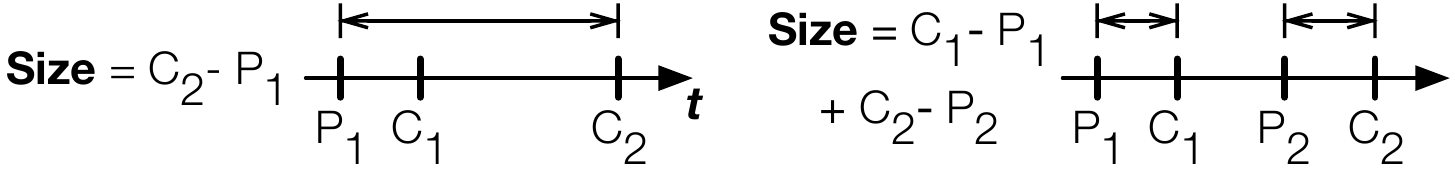}
\caption{SB optimizations. $P_1$ and $P_2$ denote production cycles; $C_1$, and $C_2$ denote consumption cycles. When $P_2 > C_1$, replicating pixels in two SBs requires less memory than sharing one SB at the expense of one extra memory read.}
\label{fig:sb_opt}
\vspace{10pt}
\end{figure}

\Fig{fig:isb} overviews the SB microarchitecture using an example where two stencil operations share the same input. This is similar to our pipeline where FD and IF share the same input image. In this example, the first stencil operation has a size of $4 \times 3$ and the second stencil operation has a size of $3 \times 3$. Thus, the SB buffers 4 lines from the input image. We use 4 cascaded FIFOs, followed by two shift registers, each of which contains the pixels in one stencil window. Each cycle, each FIFO pops one pixel, which is written into both the corresponding shift register(s) and the previous FIFO. The first three FIFOs write to both shift registers, whereas the last FIFO writes to only the first shift register.

\no{Our SB design might initially look similar to line-buffers used in image signal processing~\cite{Hegarty2014darkroom, ragan2013halide}. However, conventional line-buffers allow reading arbitrary number of elements from arbitrary locations, which requires either heavily-ported SRAM/BRAM or a large number of flip-flops, which are area-inefficient. Instead, our SB design uses FIFOs, which are easily implemented as dual-port BRAMs in FPGA.}


\paragraph{Reducing SB Sizes by Redundant DRAM Accesses} Our SB design is reminiscent of the traditional line buffers~\cite{chi2018soda, whatmough2019fixynn, Hegarty2014darkroom, hegarty2016rigel} with a critical difference. The objective of traditional line buffers is to minimize DRAM traffic such that every input element is read from DRAM once. For the localization frontend, doing that would bloat the buffer size, because the consumption and production of a pixel could be millions of cycles apart in a long stencil pipeline. Our design reads pixels multiple times from the DRAM to significantly save on-chip buffer storage. Let us explain below.

Generally, if a pixel is pushed into the SB at cycle $P_1$ and consumed by two stencil operations at cycle $C_1$ and $C_2$, the SB size must be at least $max(C_1, C_2) - P_1$, because a new pixel is pushed into the SB every cycle. Between two consumption cycles, the pixel occupies the SB space without being used. If $|C_1-C_2|$ is large, space waste is significant. This scenario is manifested in our algorithm, where DR reads from the same image that IF and FD read from, but DR is millions of cycles later than IF and FD in the pipeline.

Instead, we use two SBs and read each pixel twice, essentially replicating the pixels in the two SBs. The total SB size would be $(C_1 - P_1) + (C_2 - P_2)$, where $P_1$ and $P_2$ denote the two cycles when the pixel is read. When $P_2 > C_1$, reading pixels multiple times reduces the total SB size. Given the stencil sizes in a particular algorithmic configuration, which is statically known, we first decide whether replicating pixels is beneficial and then calculate the size of each SB, which we specify when synthesizing the FPGA design.


\no{\subsection{Synthesis Specialization}
\label{sec:fe:flex}


Traditional stencil accelerators (e.g., systolic array~\mbox{\cite{kung1982systolic}}) are generic to arbitrary stencil sizes. Our frontend architecture can be regarded as a parameterized soft core customized to different algorithmic parameter choices at \textit{synthesis-time}.}

\no{We expose two algorithmic parameters that affect the sizes of different on-chip memory structures. First, we allow specifying the input image resolution, which varies on different autonomous machines, e.g., drone vs. self-driving car, as we will later evaluate (\Sect{sec:eval:exp}). Input resolution directly affects the sizes of all memory structures. Second, we allow the window size of each stencil operation to be parametrizable, which affects the datapath and SB size. Given a particular algorithmic parameter configuration, we statically calculate the size of each memory structure (SB, FIFO, and SPM), which we then specify when synthesizing for the FPGA.}


\no{As a concrete comparison, specializing the hardware for a particular stencil size has two advantages over a generic systolic array. First, convolutions in our pipeline use one kernel (e.g., Gaussian filtering), for which the PE array in a systolic array would be severely under-utilized; only one column (or row) of the 2D PE array would be occupied in both the output- and weight-stationary data-flows.

Second, our SB is more space efficient than a generic SRAM in a systolic array. For instance, if the stencil window size is 3$\times$3, the SB buffers only 3 image rows while implicitly enabling double-buffering (\mbox{\Fig{fig:isb}}). In contrast, enabling double-buffering in typical systolic array SRAM could require buffering 6 rows (2$\times$ overhead).

Using a recent systolic-array simulator~\mbox{\cite{feng2019asv, asvsim}}, we compare our design against a 16$\times$16 systolic array with the same buffer size performing the same Gaussian filtering. Our design is 14.1$\times$ faster and reduces 83.2\% DRAM accesses.}

\no{Being able to flexibly specify the size of different memory structures depending on the actual algorithm parameters is one key advantage of targeting FPGA over ASIC. In an ASIC design, the on-chip SRAM would need to be dynamically partitioned between FIFO, SB, and SPM at runtime depending on the algorithm parameters. In contrast, targeting FPGA allows us to easily allocate on-chip memory resources at synthesis time while using the appropriate organization most suitable for each data reuse pattern.}


%% file: be.tex
\section{Backend System}
\label{sec:be}

The backend has high latency variation, especially after the frontend is accelerated, and thus is an ideal acceleration target. We first show that the backend kernels share common building blocks, which we exploit to design a flexible and efficient architecture substrate (\Sect{sec:be:design}). Due to the large variation, accelerating the backend is not always beneficial; we exploit the execution behaviors of the backend kernels to design a lightweight runtime scheduler, which ensures that backend acceleration reduces variation without increasing overall latency (\Sect{sec:be:rt}).


\subsection{Backend Architecture}
\label{sec:be:design}

\paragraph{Building Blocks} Recall from \Sect{sec:algo:char} that the each backend mode inherently possesses a kernel that contributes significantly to both the overall latency and the latency variation: \textit{camera model projection} under the registration mode, \textit{computing Kalman gain} under the VIO mode, and \textit{marginalization} under the SLAM mode. Accelerating these kernels reduces the overall latency and latency variation.

While it is possible to spatially instantiate separate hardware logic for each kernel, it would lead to resource waste. This is because the three kernels \textit{share common building blocks}. Fundamentally, each kernel performs matrix operations that manipulate various forms of visual features and IMU states. \Tbl{tbl:beoperator} decomposes each kernel into different matrix primitives.

\begin{table}[t]
\centering
\caption{Latency variation-contributing kernels in the backend are composed of common matrix operations.}
\renewcommand*{\arraystretch}{1}
\renewcommand*{\tabcolsep}{3pt}
\resizebox{\columnwidth}{!}
{
\begin{tabular}{cccccc}
\toprule[0.15em]
\textbf{Building Block}                      & \textbf{Projection} & \textbf{Kalman Gain}              & \textbf{Marginalization}\\
\midrule[0.05em]
Matrix Multiplication & \checkmark & \checkmark & \checkmark\\
Matrix Decomposition  & ~ & \checkmark & \checkmark\\
Matrix Inverse        & ~ & ~ & \checkmark\\
Matrix Transpose      & ~ & \checkmark & \checkmark\\
Fwd./Bwd. Substitution     & ~ & \checkmark & \checkmark\\
\bottomrule[0.15em]
\end{tabular}
}
\label{tbl:beoperator}
\end{table}

For instance, the projection kernel in the registration mode simply multiplies a $3 \times 4$ camera matrix $\mathbf{C}$ with a $4 \times M$ matrix $\mathbf{X}$, where $M$ denotes the number of feature points in the map (each represented by 4D homogeneous coordinates). The Kalman gain $\mathbf{K}$ in VIO is computed by:
\begin{subequations}
\begin{align}
  \mathbf{S} &= \mathbf{H} \times \mathbf{P} \times \mathbf{H}^{T} + \mathbf{R} \label{eq:k1} \\
  \mathbf{S} \times \mathbf{K} &= \mathbf{P} \times \mathbf{H}^{T} \label{eq:k2}
\end{align}
\end{subequations}

\noindent where $\mathbf{H}$ is the Jacobian matrix of the function that maps the true state space into the observed space, $\mathbf{P}$ is the covariance matrix, and $\mathbf{R}$ is an identity noise matrix. Calculating $\mathbf{K}$ requires solving a system of linear equations, which is implemented by matrix ($\mathbf{S}$) decomposition followed by forward/back-substitution. Marginalization combines all five operations; its formulation is omitted here for simplicity.

\paragraph{Design} The backend specializes the hardware for the five matrix operations in~\Tbl{tbl:beoperator}, which the three backend kernels are then mapped to. These matrix operations are low-level enough to allow sharing across different backend kernels but high-level enough to reduce the control flows, which are particularly inefficient to implement on FPGAs.


\label{sec:bedesign}
\begin{figure}[t]
\centering
\includegraphics[width=\columnwidth]{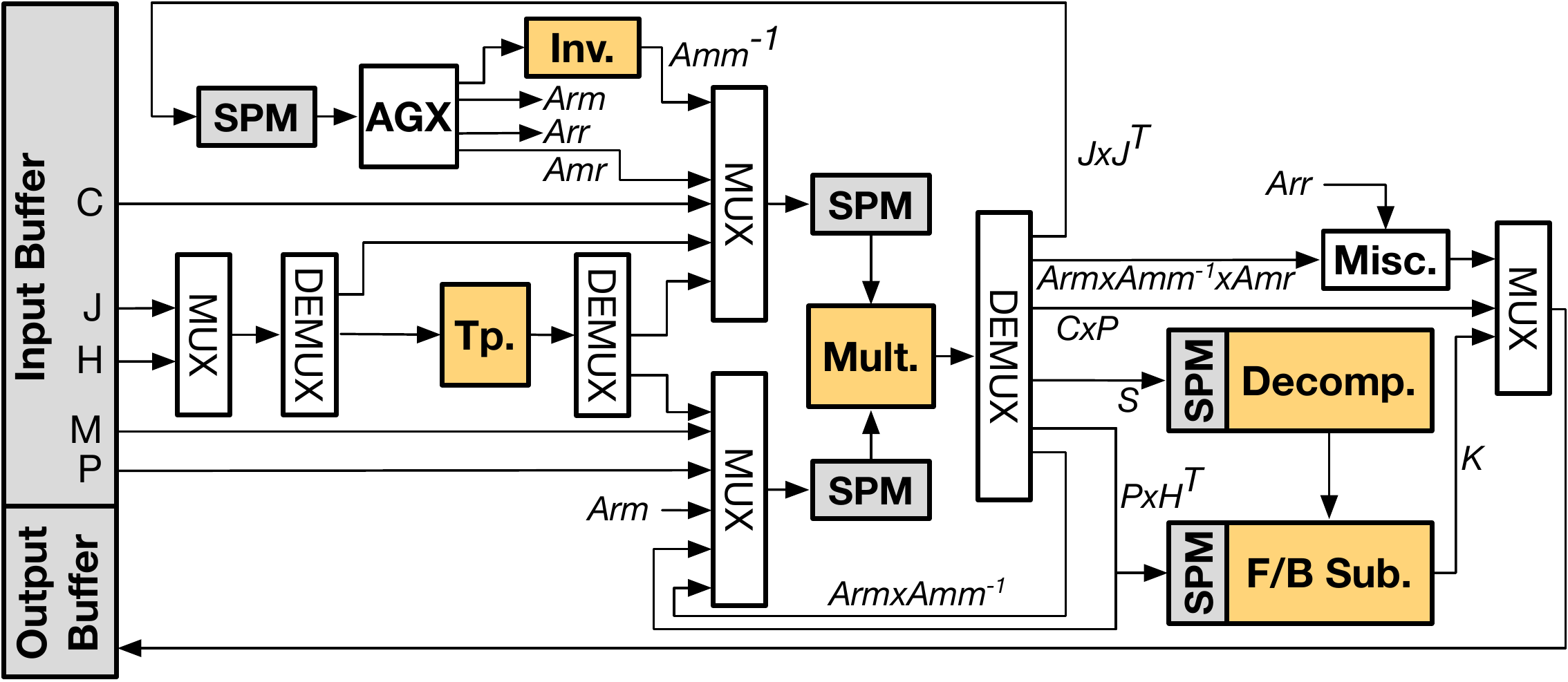}
\caption{Backend architecture. It consists of five basic building blocks for matrix operations: inverse (Inv.), decomposition (Decomp.), transpose (Tp.), multiplication (Mult.), and forward/backward-substitution. ``AGX'' is the address generation logic, and ``Misc.'' contains the rest of the logic not included in any above (e.g., addition).\no{ Scratchpad memories (SPM) are used to stored intermediate data. The input and output is buffered on-chip and DMA-ed from/to the host.}}
\label{fig:bearch}
\end{figure}

\Fig{fig:bearch} shows the backend architecture. The input and output are buffered on-chip and DMA-ed from/to the host. The inputs to each matrix block are stored in the SPMs. The input matrices must be ready before an operation starts. Importantly, the SPMs can not be replaced by SBs (\Fig{fig:isb}) as in the frontend, because these matrix operations, unlike convolution and block matching in the frontend, are not stencil operations.

The architecture accommodates different matrix sizes by exploiting the inherent blocking nature of matrix operations (e.g., multiplication, decomposition), where the output could be computed by iteratively operating on different blocks of the input matrices. Thus, the compute units have to support computations for only a block, although the SPMs need to accommodate the size of the entire input matrices.

\paragraph{Optimization} We exploit unique characteristics inherent in various matrices to further optimize the computation and memory usage. The $\mathbf{S}$ matrix is inherently symmetric (\Equ{eq:k1}). Thus, the computation and storage cost of $\mathbf{S}$ can naturally be reduced by half. In addition, the matrix that requires inversion in marginalization, $A_{mm}$, is a symmetric matrix with a unique blocking structure of $\begin{bsmallmatrix} A & B \\ C & D  \end{bsmallmatrix}$, where $A$ is a diagonal matrix and $D$ is a $6 \times 6$ matrix, where 6 represents the number of degrees of freedom in a pose to be calculated. Therefore, the inversion hardware is specialized for a $6 \times 6$ matrix inversion combined with simple reciprocal structures.


\subsection{Runtime Scheduling}
\label{sec:be:rt}

Offloading backend kernels to the backend accelerator is not always beneficial due to the overhead of data transfer, especially when the size of the matrix involved in a kernel is small. For instance, in many frames the marginalization time is below 1~ms (\Fig{fig:slam_be_var}), in which case offloading marginalization degrades performance.



\begin{figure}[t]
\centering
\subfloat[\small{Projection.}]
{
  \includegraphics[trim=0 0 0 0, clip, width=0.32\columnwidth]{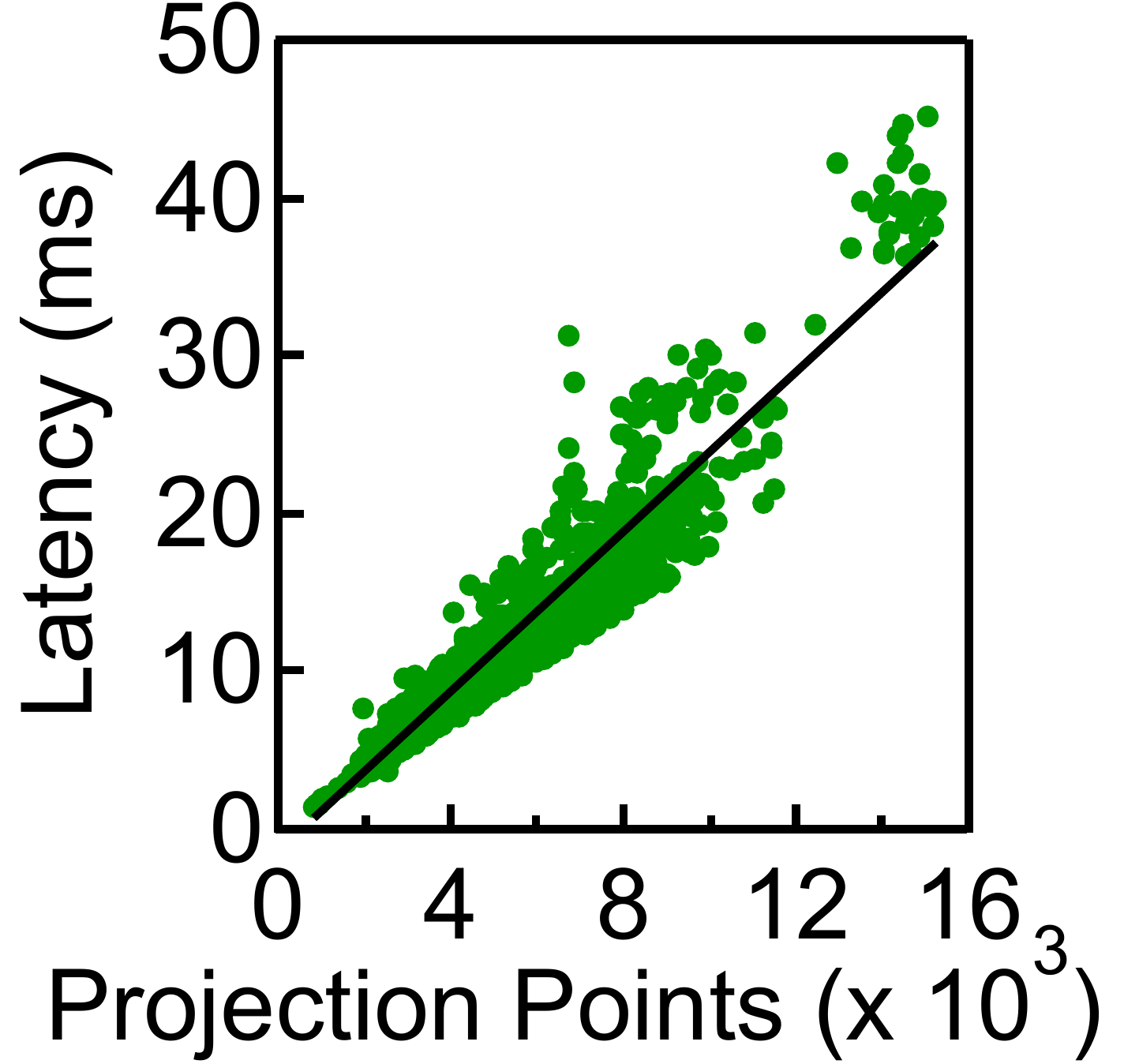}
  \label{fig:regiscause}
}
\subfloat[\small{Kalman gain.}]
{
  \includegraphics[trim=0 0 0 0, clip, width=0.32\columnwidth]{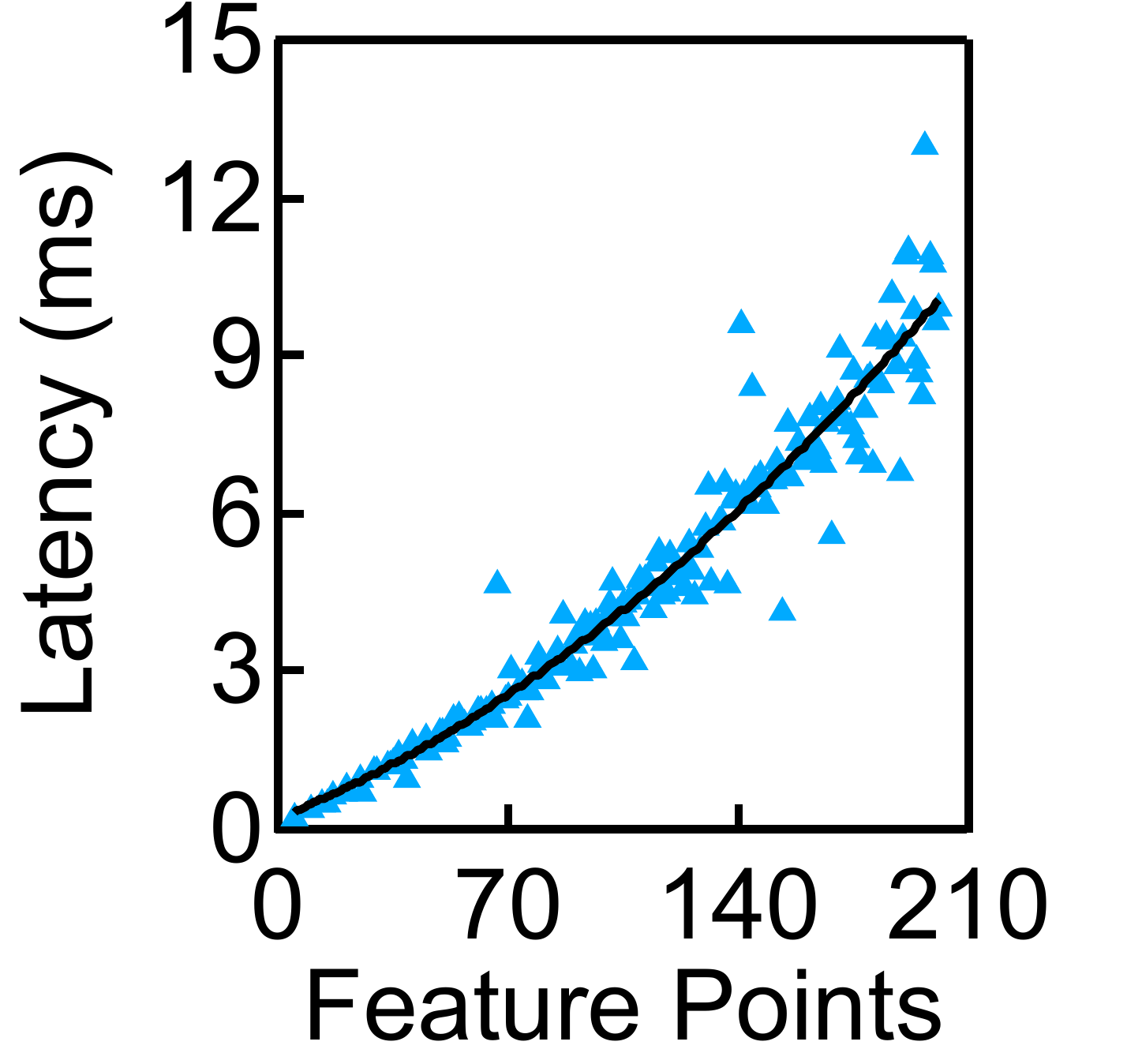}
  \label{fig:vioreason}
}
\subfloat[\small{Marginalization.}]
{
  \includegraphics[trim=0 0 0 0, clip, width=0.32\columnwidth]{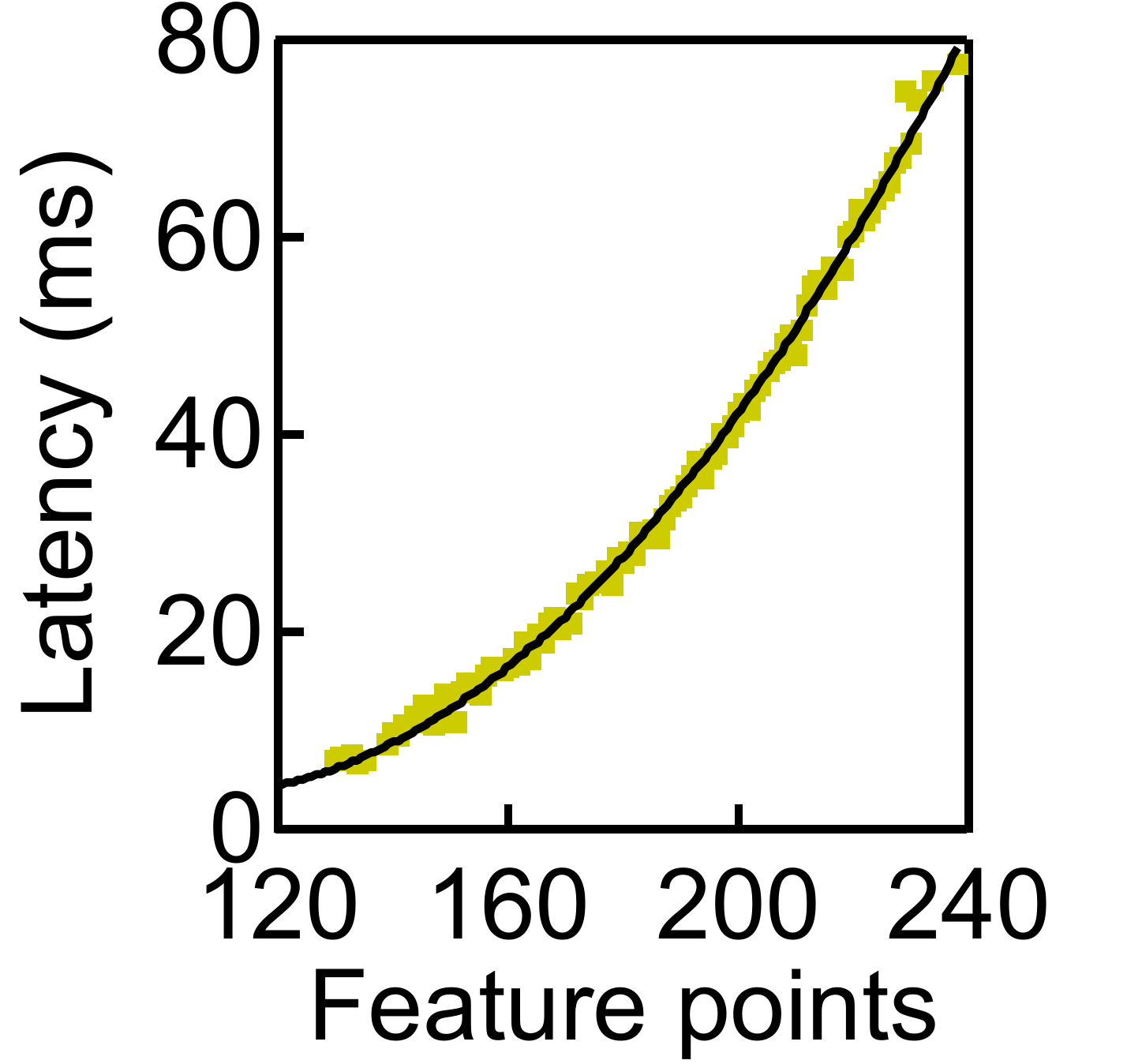}
  \label{fig:slamcause}
}
\caption{The backend kernels' latency is dictated by the size of matrices that they operate on.\no{ Projection time could be estimated using a linear model whereas the other could be estimated by quadratic models.}}
\label{fig:because}
\end{figure}

The latency of each kernel highly correlates with the sizes of the matrices that it operates on, which in turn depends on information calculated in the frontend. Using the same 1,800 frames profiled in \Sect{sec:algo:char}, \Fig{fig:regiscause} shows that the projection latency of a frame ($y$-axis) increases almost linearly with the number of points in the current map ($x$-axis). \Fig{fig:vioreason} shows that the latency of computing the Kalman gain in a frame ($y$-axis) scales with $\mathbf{H}$'s height ($x$-axis), which is dictated by the number of key points in a frame extracted by the frontend. \Fig{fig:slamcause} shows that the marginalization latency increases with the number of feature points, too.


Leveraging this insight, we design a lightweight software scheduler, which offloads the backend only when the latency could be reduced. The scheduler first estimates the CPU execution time using simple regression models constructed offline, as illustrated in~\Fig{fig:because}. In particular, the projection time is fit using a linear model whereas the other two kernerls' times are estimated by quadratic models. The scheduler then estimates the acceleration time from the latency profiles of the five building blocks and the data transfer bandwidth, both obtained offline. The scheduler triggers the accelerator when the CPU time would be longer than acceleration time.


%% file: eval.tex
\section{Evaluation}
\label{sec:eval}

\no{We first describe our experimental methodology (\Sect{sec:eval:exp}) and show the FPGA resource utilization (\Sect{sec:eval:res}). We demonstrate the effectiveness of \proj (\Sect{sec:eval:sys}) and analyze its behaviors in the frontend (\Sect{sec:eval:fe}) and backend (\Sect{sec:eval:be}). Finally, we evaluate the backend scheduler (\Sect{sec:eval:sched}).}

\subsection{Experimental Setup}
\label{sec:eval:exp}

\paragraph{Hardware Platform} We build an FPGA prototype (\sys{Edx-Car}) and evaluate it on our commercial autonomous vehicle. We use a Xilinx Virtex-7 XC7V690T FPGA board~\cite{vertex} connected to a PC machine, which has a four-core Intel Kaby Lake CPU (1.6 GHz and 9 MB LLC) and 8 GB memory. To show the flexibility and general applicability of our design framework, we also build another prototype targeting drones (\sys{Edx-Drone}), for which we use a Zynq Ultrascale+ ZU9CG board~\cite{zynq}, which integrates a quad-core ARM Cortex-A53 CPU with an FPGA on the same chip.

The actual accelerator implementations on both instances are almost the same except that \sys{Edx-Car} uses a larger matrix multiplication/decomposition unit and larger line-buffers and SPMs to deal with a higher input resolution.

The FPGA is directly interfaced with the cameras and IMU/GPS sensors. The host and the FPGA accelerator communicate three times in each frame: the first time from the FPGA, which transfers the frontend results and the IMU/GPS samples to the host; the second time from the host, which passes the inputs of a backend kernel (e.g., the $\mathbf{H}$, $\mathbf{P}$, and $\mathbf{R}$ matrices to calculate Kalman gains) to the FPGA, and the last time transferring the backend results back to the host.

For \mbox{\sys{Edx-Car}}, FPGA reads data from PC through PCI-e 3.0, with a max bandwidth of 7.9 GB/s. For \mbox{\sys{Edx-Drone}}, the FPGA reads data from DRAM through the AXI4 bus, with a max bandwidth of 1.2 GB/s.

\paragraph{Baselines} To our best knowledge, today's localization systems are mostly implemented on general-purpose CPU platforms. Thus, we compare \sys{Edx-Car} against the software implementation on the PC machine without the FPGA, and compare \sys{Edx-Drone} against the software implementation on the quad-core Arm Cortex-A57 processor on the TX1 platform~\cite{tx1}, a representative mobile platform today. To obtain stronger baselines, the software implementations leverage multi-core and the SIMD capabilities of the CPUs. For a comprehensive evaluation, we will also compare against GPU and DSP implementations in \Sect{sec:eval:unopt}.




\paragraph{Dataset} For \sys{Edx-Drone}, we use EuRoC~\cite{burri2016euroc} (Machine Hall sequences), a widely-used drone dataset. Since EuRoC contains only indoor scenes, we complement EuRoC with our in-house outdoor dataset. The two datasets combined have 50\% outdoor frames, 25\% indoor frames without map, and 25\% indoor frames with map. The input images are sized to the $640 \times 480$ resolution. For \sys{Edx-Car}, we use KITTI Odometry~\cite{geiger2012we} (grayscale sequence), which is a widely-used self-driving car dataset. Similarly, since KITTI contains only outdoor scenes, we complement it with our in-house dataset for indoor scenes. The distribution of the three scenes is the same as above. The input images are uniformly sized to the $1280 \times 720$ resolution. The energy results are averaged across all the evaluated frames.

To evaluate the effect of the runtime scheduler (\Sect{sec:be:rt}), we use 25\% of the frames in the datasets to construct the regression models offline, and evaluate on the rest 75\%.

\subsection{Resource Consumption}
\label{sec:eval:res}

\begin{table}[t]
\centering
\caption{FPGA resource consumption of \sys{Edx-Car} and \sys{Edx-Drone} and their utilizations on the actual FPGA boards. Data is obtained after the design passes post-place \& route timing. The ``N.S.'' columns denote the hypothetical resource usage if we do not share the frontend and the building blocks in the backend across the three modes.}
\renewcommand*{\arraystretch}{1}
\renewcommand*{\tabcolsep}{3pt}
\resizebox{\columnwidth}{!}
{
\begin{tabular}{c|ccc|ccc}
\toprule[0.15em]
\textbf{Resource} & \textbf{Car} & \textbf{Virtex-7} & \textbf{N.S.} & \textbf{Drone}  & \textbf{Zynq} & \textbf{N.S.} \\
\midrule[0.05em]
LUT   & 350671 & 80.9\% & 795604 & 231547 & 84.5\% & 659485   \\ 
Flip-Flop  & 239347 & 27.6\% & 628346 & 171314 & 31.2\% & 459485  \\ 
DSP   & 1284   & 35.6\% & 3628   & 1072   & 42.5\% & 3064   \\ 
BRAM  & 5.0    & 87.5\% & 13.2   & 3.67   & 92.3\% & 10.6  \\ 
\bottomrule[0.15em]
\end{tabular}
}
\label{tbl:fpga}
\end{table}

We show the FPGA resource usage in \Tbl{tbl:fpga}. Overall, \sys{Edx-Car} consumes more resources than \sys{Edx-Drone} as the former uses larger hardware structures to cope with higher input resolutions (\Sect{sec:eval:exp}). To demonstrate the effectiveness of our hardware design that shares the frontend and the various backend building blocks across the three modes (\Tbl{tbl:beoperator}), the ``N.S.'' columns show the hypothetical resource consumption \textit{without} sharing these structures in both instances. Resource consumption of all types would more than double, exceeding the available resources on the FPGA boards.

Frontend dominates the resource consumption. In \sys{Edx-Car}, the frontend uses 83.2\% LUT, 62.2\% Flip-Flop, 80.2\% DSP, and 73.5\% BRAM of the total used resource; the percentages in \sys{Edx-Drone} are similar. In particular, feature extraction consumes over two-thirds of frontend resource, corroborating our design decision to multiplex the feature extraction hardware between left and right camera streams (\Sect{sec:fe:para}).

\paragraph{Other Storage Requirements} The MSCKF window size is 30, which translates to a total storage requirement is 1.2 MB (state vector, covariance matrix, Jacobian matrix, Kalman gain). These data structures initially reside in DRAM and are transferred to the FPGA depending on what matrix operations are offloaded. The dictionary for loop detection is about 60 MB, which initially resides in the DRAM. Only the \textit{Projection} kernel of loop closure is offloaded to FPGA (\mbox{\Tbl{tbl:beoperator}}).




\subsection{Overall Results}
\label{sec:eval:sys}

\begin{figure}[t]
\centering
\subfloat[\small{Latency and variation improvement on \sys{Edx-Car}.}]
{
  \includegraphics[trim=0 0 0 0, clip, width=0.45\columnwidth]{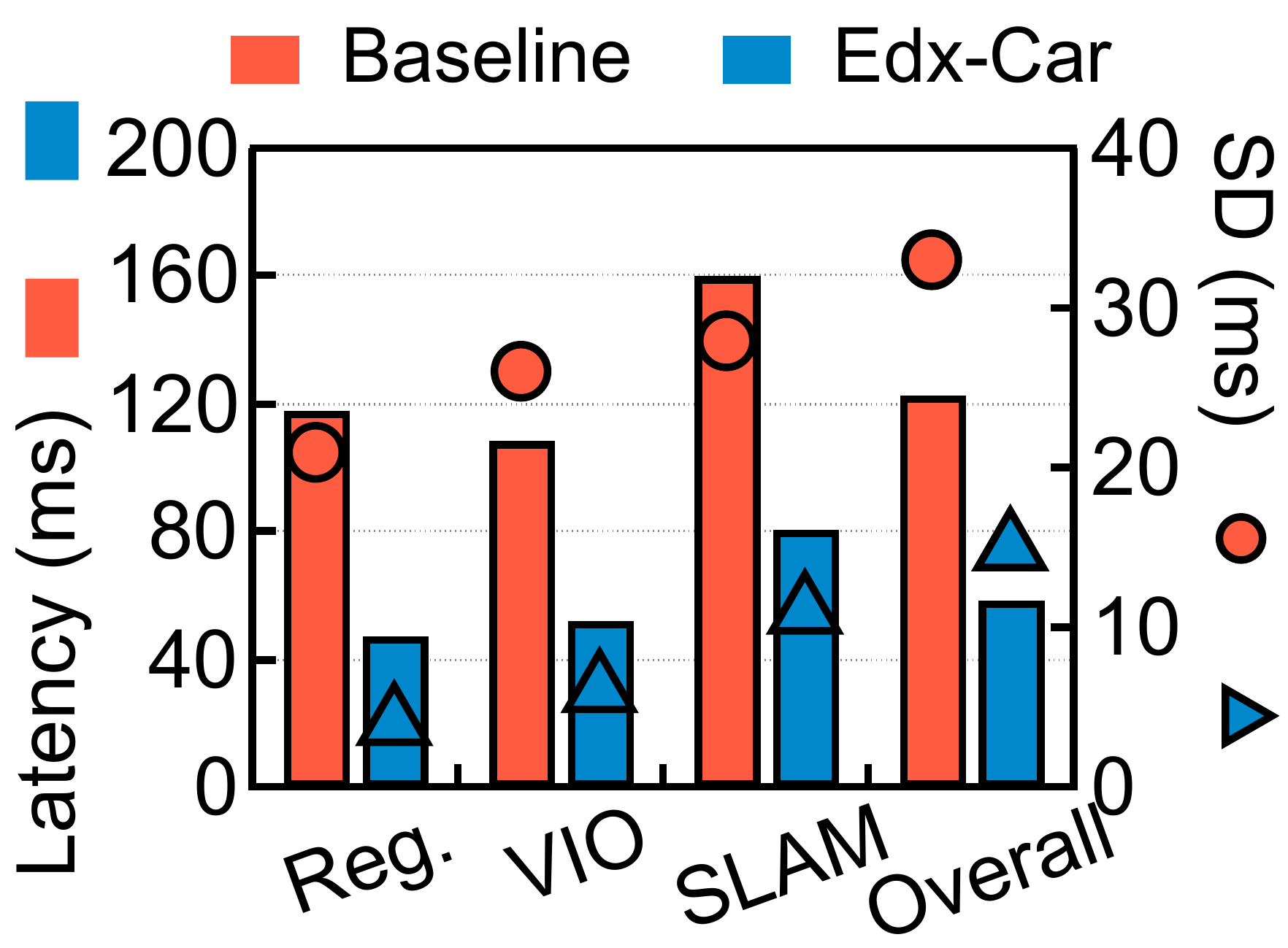}
  \label{fig:sysall_cpu}
}
\hspace{2pt}
\subfloat[\small{Latency and variation improvement on \sys{Edx-Drone}.}]
{
  \includegraphics[trim=0 0 0 0, clip, width=0.45\columnwidth]{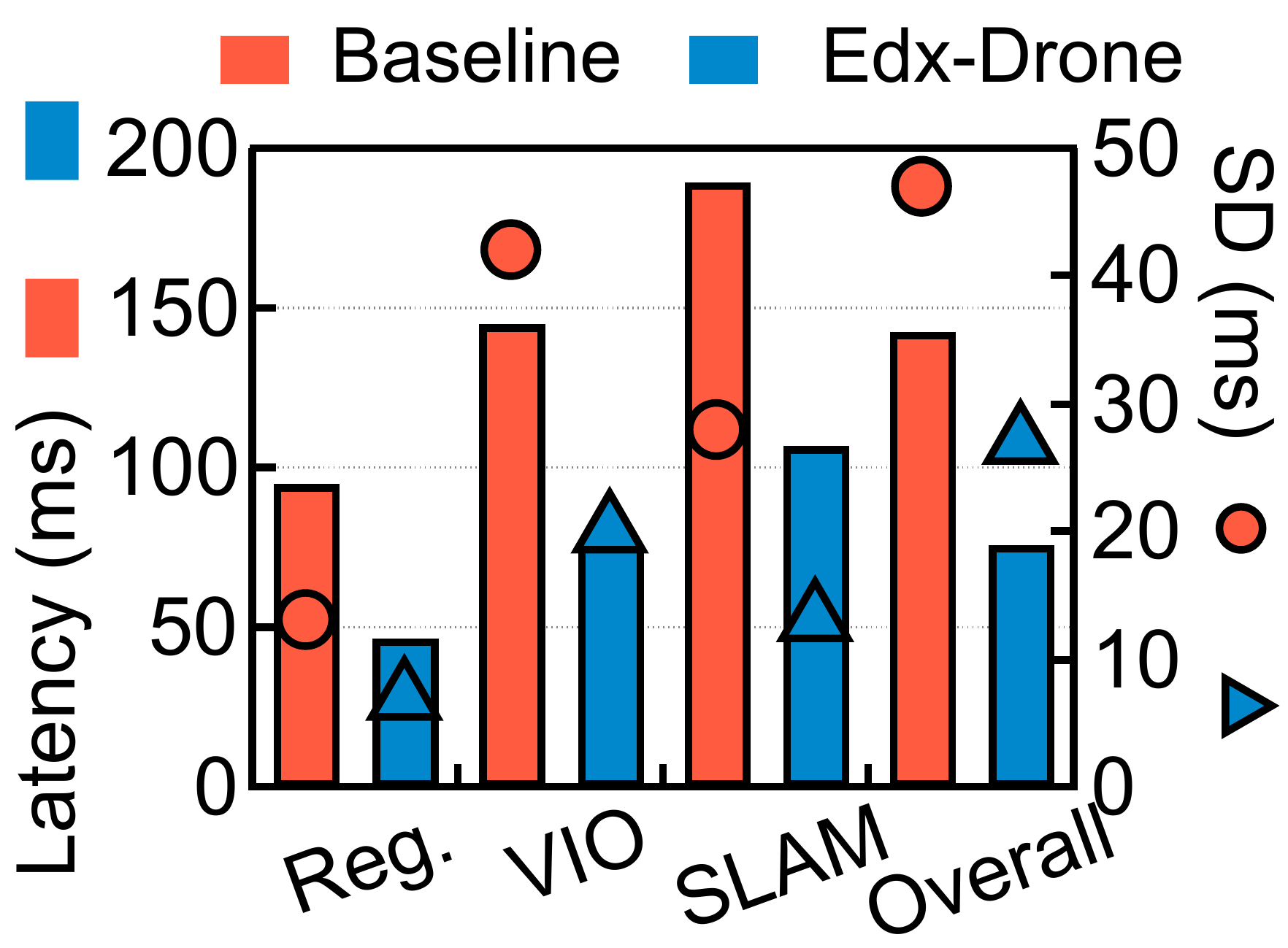}
  \label{fig:sysall_tx1}
}
\caption{Overall system latency and variation reduction.}
\label{fig:eval_sys}
\end{figure}

\begin{figure}[t]
\vspace{5pt}
\centering
\begin{minipage}[t]{0.48\columnwidth}
  \centering
  \includegraphics[width=\columnwidth]{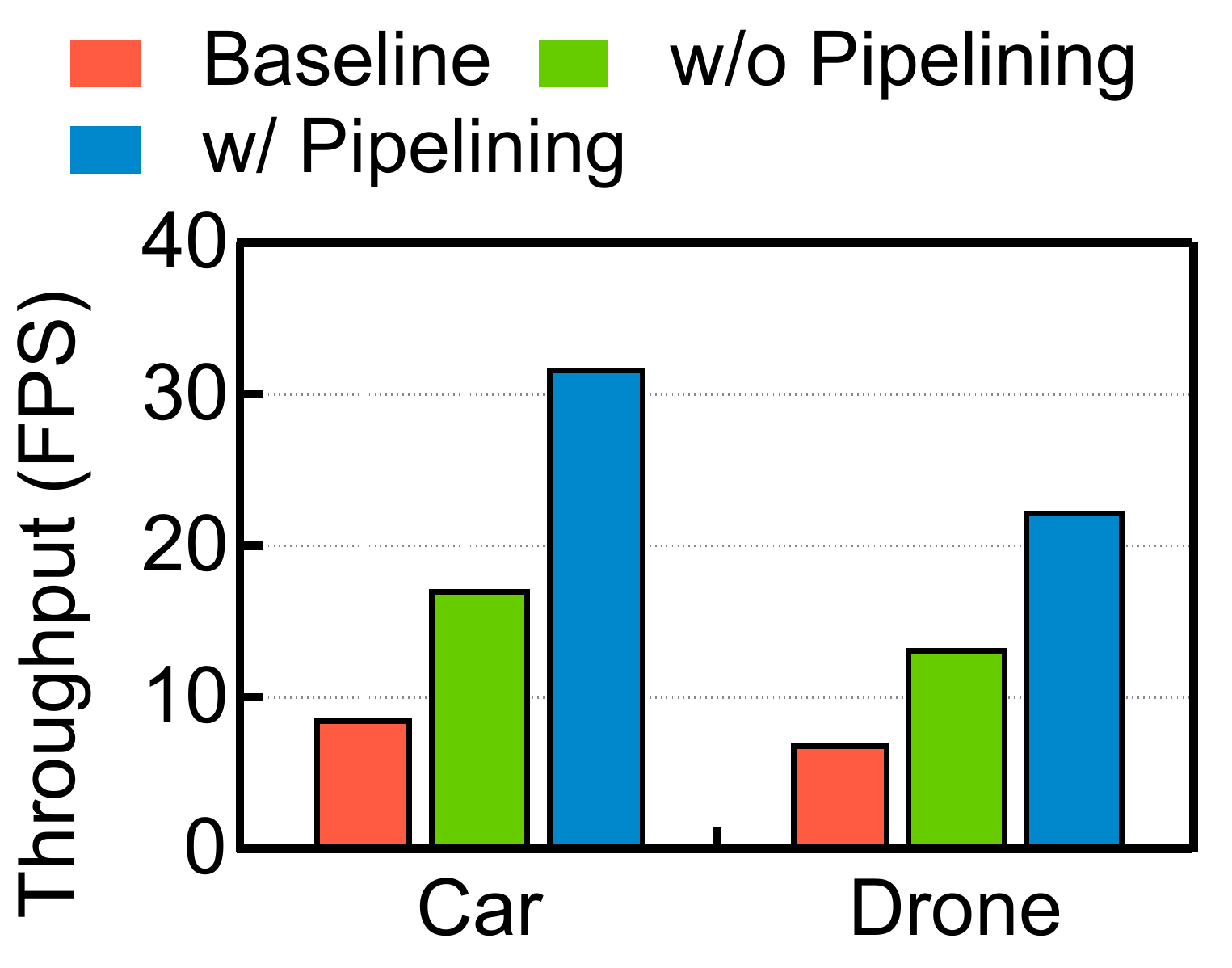}
  \caption{FPS of baseline and \proj with and without frontend/backend pipelining.}
  \label{fig:sys_fps}
\end{minipage}
\hspace{2pt}
\begin{minipage}[t]{0.48\columnwidth}
  \centering
  \includegraphics[width=\columnwidth]{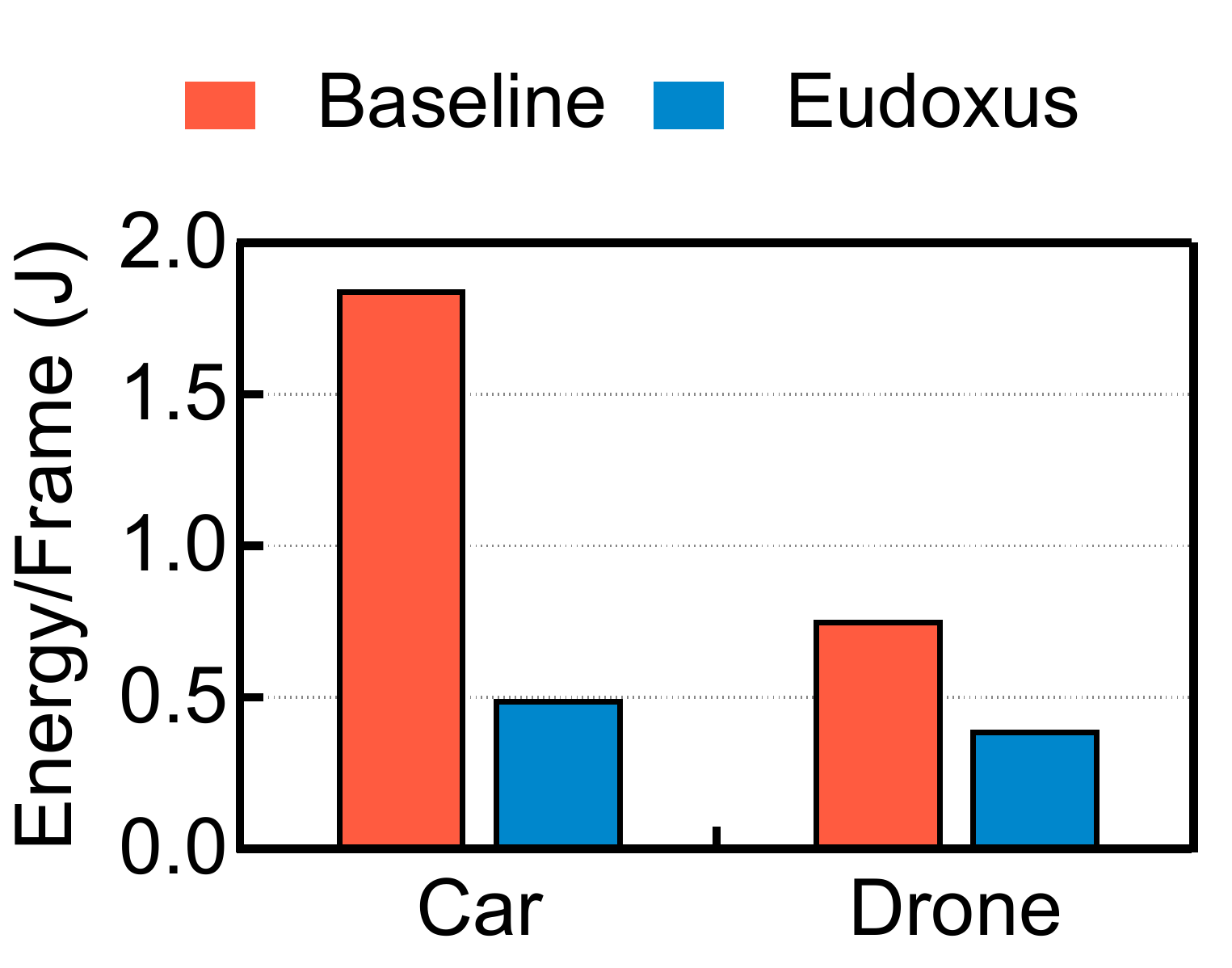}
  \caption{Energy per frame comparison between baseline and \proj.}
  \label{fig:sys_eng}
\end{minipage}
\end{figure}

\paragraph{Car Results} We compare the average frame latency and variation between the baseline and \sys{Edx-Car} in \Fig{fig:sysall_cpu}. We show the overall results as well as the results in the three modes separately. The end-to-end frame latency is reduced by 2.5$\times$, 2.1$\times$, and 2.0$\times$ in registration, VIO, and SLAM mode, respectively, which leads to overall 2.1$\times$ speedup. \sys{Edx-Car} also significantly reduces the variation. The standard deviation (SD) is reduced by 58.4\%. The latency reduction directly translates to higher throughput, which reaches 17.2 FPS from 8.6 FPS, as shown in \Fig{fig:sys_fps}. Further pipelining the frontend with the backend improves the FPS to 31.9.

The energy is also significantly reduced. \Fig{fig:sys_eng} compares the energy per frame between the baseline and \sys{Edx-Car}. With hardware acceleration, \sys{Edx-Car} reduces the average energy by 73.7\%, from 1.9 J to 0.5 J per frame.

\paragraph{Drone Results} We show the results of \sys{Edx-Drone} in \Fig{fig:sysall_tx1}. The frame latency has a speedup of 2.0$\times$, 1.9$\times$, and 1.8$\times$ in the three modes, respectively, which leads to a 1.9$\times$ overall speedup. The overall SD is reduced by 42.7\%. The average throughput is improved from 7.0 FPS to 22.4 FPS. The average energy per frame is reduced by 47.4\% from 0.8 J to 0.4 J per frame. The energy saving is lower than \sys{Edx-Car} because the FPGA static power stands out as \sys{Edx-Drone} reduces the dynamic power.


\subsection{Frontend Results}
\label{sec:eval:fe}

\begin{figure}[t]
\vspace{-5pt}
\centering
\subfloat[\small{Latency comparison.}]
{
  \includegraphics[trim=0 0 0 0, clip, width=0.47\columnwidth]{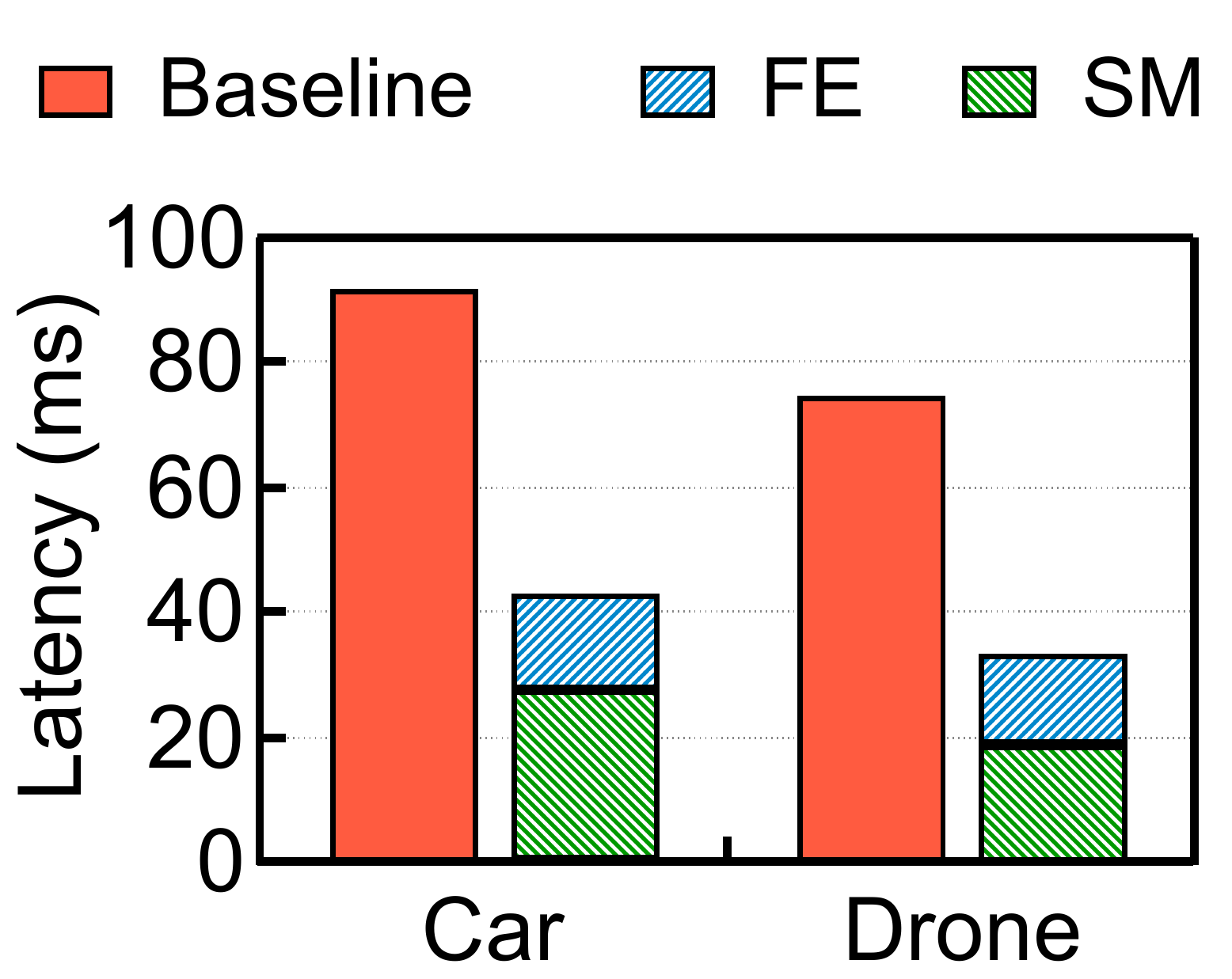}
  \label{fig:fe_acc}
}
\hfill
\subfloat[\small{Throughput comparison.}]
{
  \includegraphics[trim=0 0 0 0, clip, width=0.47\columnwidth]{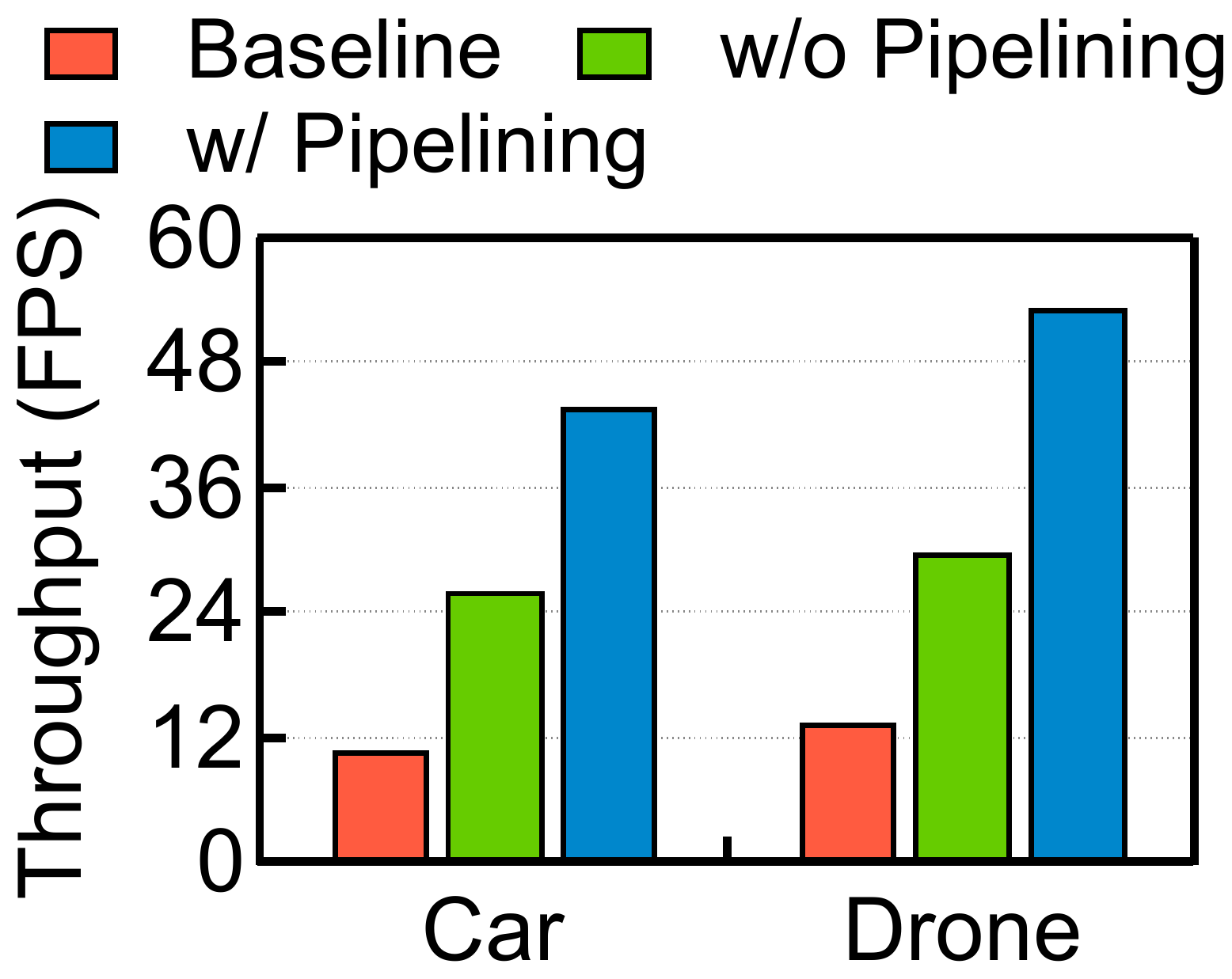}
  \label{fig:fe_fps}
}
\caption{Frontend acceleration results.}
\label{fig:fe_res}
\end{figure}

\paragraph{Car Results} We show the frontend latency results of \sys{Edx-Car} in \Fig{fig:fe_acc}. We break down the frontend latency of \sys{Edx-Car} into two components: feature extraction (FE) and stereo matching (SM). Note that the temporal matching is hidden behind the critical path\no{ (\Fig{fig:tlp})}, and thus does not contribute to the latency. Compared to the baseline, the average frontend latency is reduced from 92.4 $ms$ to 42.7 $ms$, a 2.2$\times$ speedup. The SM dominates the frontend latency, further confirming our design decision to multiplex the FE hardware between left and right camera streams (\Sect{sec:fe:para}).

\Fig{fig:fe_fps} compares the throughput of the baseline with two versions of \sys{Edx-Car}, one that pipelines FE and SM and the other that does not. With pipelining, the frontend throughput is 44.0 FPS, higher than the overall system throughput of 31.9 (\Fig{fig:sys_fps}). Note that the throughput without pipelining is 26.1, lower than the overall throughput. This indicates that without FE/SM pipelining the frontend is the system bottleneck, while pipelining pushes the bottleneck to the backend.



\paragraph{Drone Results} The frontend latency and throughput improvements in \sys{Edx-Drone} are shown in \Fig{fig:fe_acc} and \Fig{fig:fe_fps}, respectively, too. \sys{Edx-Drone} achieves 2.2$\times$ latency speed up and 4.0$\times$ throughput speedup with pipelining. The frontend latency of \sys{Edx-Drone} is slightly faster than that of \sys{Edx-Car}. This is because drones deal with a 3$\times$ lower image resolution than self-driving cars, leading to lower compute workload overall.

\paragraph{On-chip Memory Design} Owing to our SB optimization (\Fig{fig:sb_opt}), the SB size is very small while the SPM by far dominates the memory resources. For instance on \sys{Edx-Car}, SPM consumes about 3.6~MB memory while SB consumes 0.4~MB. Without the SB optimization, the SB size would increase by about 9~MB, far exceeding the FPGA resource provision. This is because a pixel would have to stay in the SB for over 3 million cycles after being consumed by FD/IF and before being consumed by DR (\Sect{sec:fe:data}).

\subsection{Backend Results}
\label{sec:eval:be}

\begin{figure}[t]
\centering
\subfloat[\small{Backend latency and variation improvements of \sys{Edx-Car}.}]
{
  \includegraphics[trim=0 0 0 0, clip, width=0.45\columnwidth]{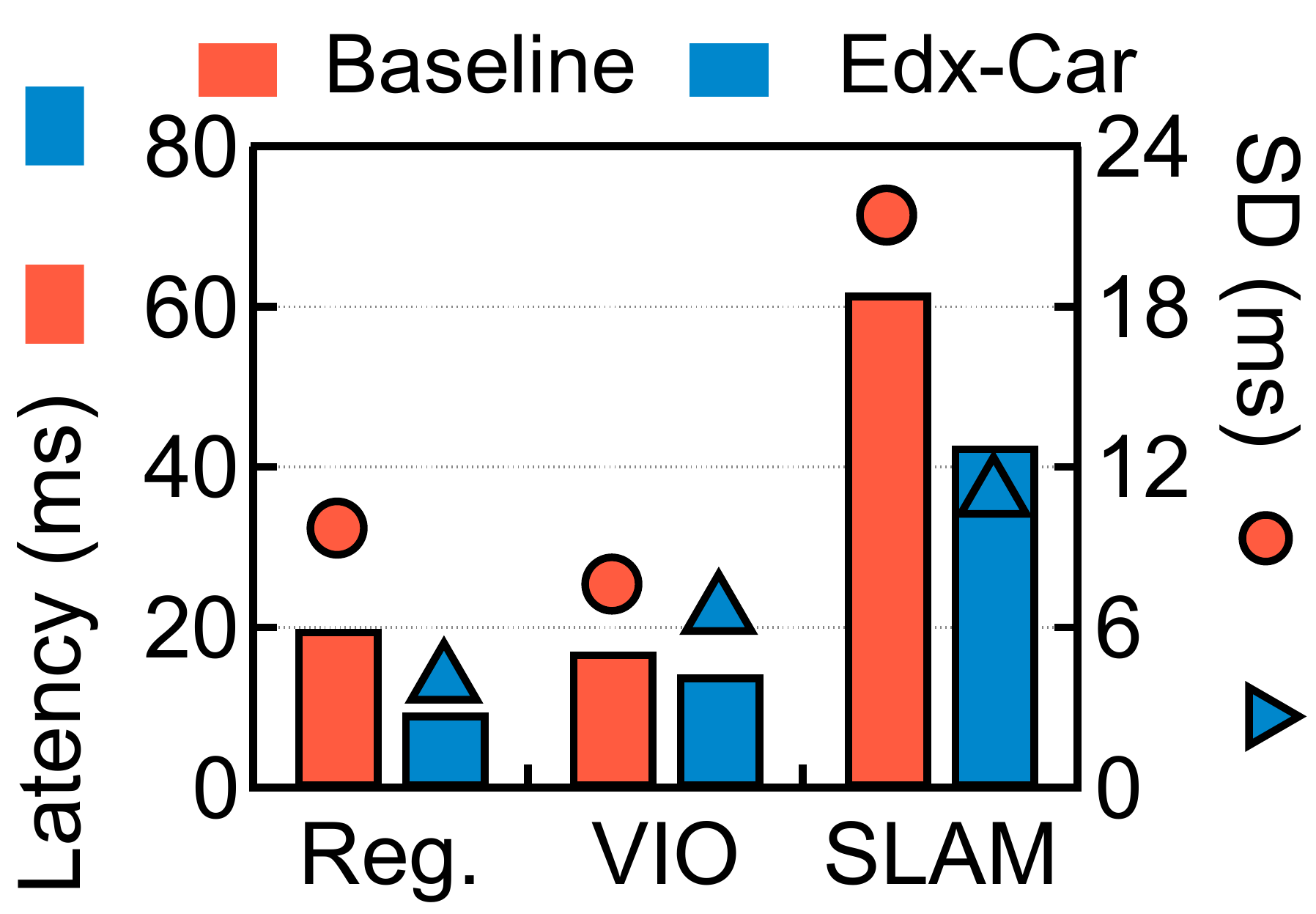}
  \label{fig:be_all_cpu}
}
\hspace{2pt}
\subfloat[\small{Backend latency and variation improvements of \sys{Edx-Drone}.}]
{
  \includegraphics[trim=0 0 0 0, clip, width=0.45\columnwidth]{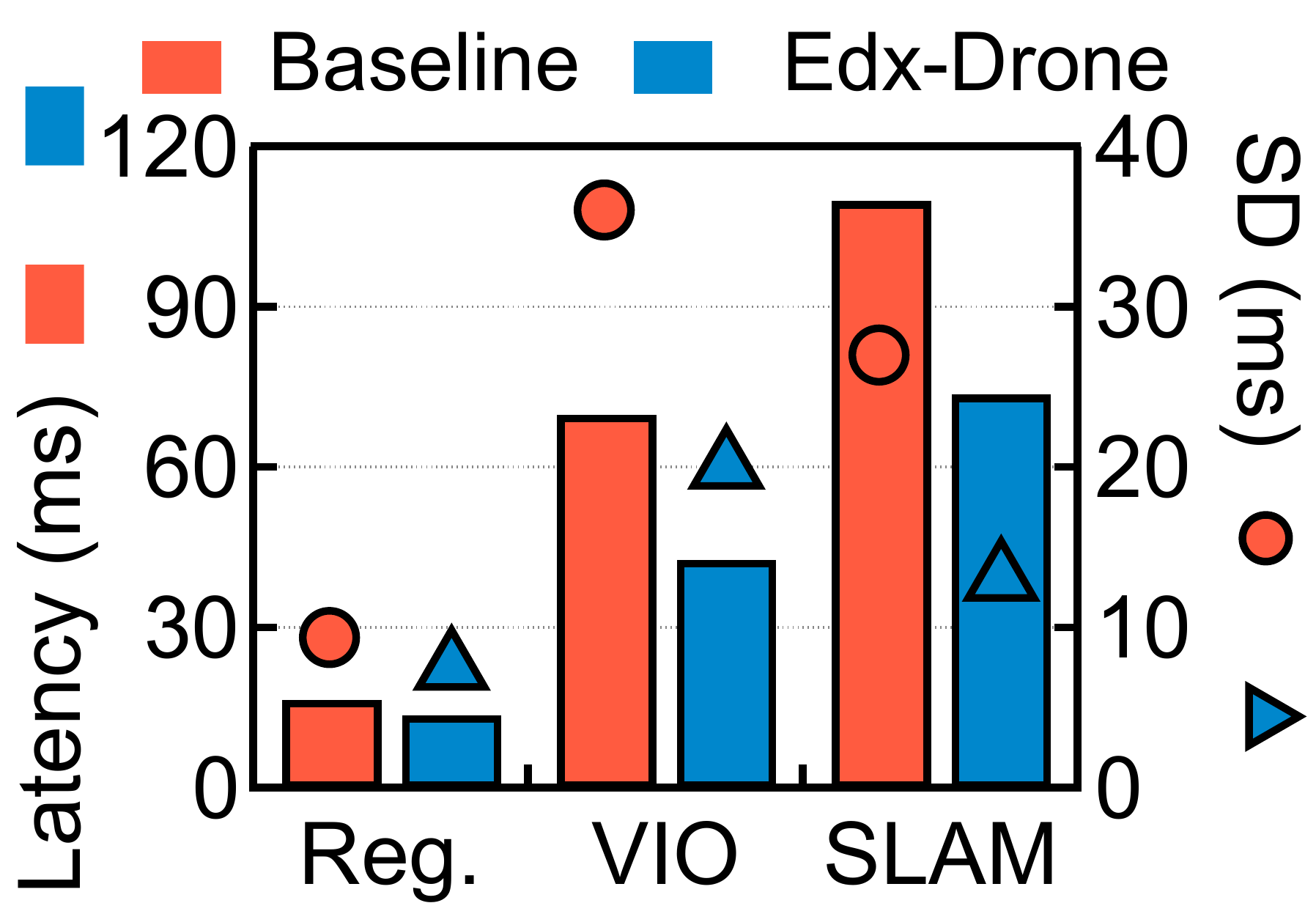}
  \label{fig:be_all_tx1}
}
\label{fig:be_all}
\caption{Backend latency and variation reduction.}
\end{figure}

\paragraph{Car Results} We show the backend performance results of \sys{Edx-Car} in \Fig{fig:be_all_cpu}. The left $y$-axis shows the latency, and the right $y$-axis shows the standard deviation. The average backend latency in the registration model is reduced by 49.4\%. The large latency reduction is a direct result of accelerating the Projection kernel, whose latency is reduced by 95.3\%. Meanwhile, the backend variation is also significantly reduced. The SD is reduced by 58.7\% from 9.6 $ms$ to 4.0 $ms$. \no{The significant drop of both the latency and the latency variation is evident in \Fig{fig:cpu_proj}, which shows the per-frame Projection time on both the baseline and \sys{Edx-Car}, sorted in the ascending order.}


In the VIO mode, the Kalman gain kernel is accelerated by 2.0$\times$, which translates to 16.3\% overall backend latency reduction as Kalman gain contributes to about 33.3\% of the VIO latency (\Fig{fig:vio_be_time}). \no{\Fig{fig:cpu_vio} compares the per-frame latency of Kalman Gain on the baseline and \sys{Edx-Car}. Overall, the SD of the VIO mode is reduced by 13.9\%.} The improvements in the SLAM mode are significant. The Marginalization kernel is accelerated by 2.4$\times$, which translates to 30.2\% overall backend latency reduction. The SD also reduces from 21.4 $ms$ to 10.9 $ms$. \no{The variation reduction is clear in \Fig{fig:cpu_slam}, which compare the per-frame Marginalization latency on the baseline and \sys{Edx-Car}.}

\paragraph{Drone Results} \Fig{fig:be_all_tx1} compares the backend latency and latency variation between \sys{Edx-Drone} and baseline. The backend latency reduction is 16.7\%, 38.9\%, and 32.7\% in the registration, VIO, and SLAM modes, respectively. The SD reductions are 19.4\%, 44.4\%, and 51.9\%, respectively. The per-frame latency comparison figures are omitted due to space limitations. Unlike frontend, the backend is faster on \sys{Edx-Car} than on \sys{Edx-Drone} even though the former has to deal with a larger input resolution. This is because the backend accelerator in \sys{Edx-Car} uses a larger matrix multiplication and decomposition unit.

\no{An interesting observation is that the registration latency and variation reduction is more significant in \sys{Edx-Car} than in \sys{Edx-Drone}, whereas the opposite is true for VIO. This is because the different input resolutions shift the contributions of the accelerated kernels in registration and VIO.}

\subsection{Effectiveness of Backend Scheduling}
\label{sec:eval:sched}

\no{To ensure that offloading backend kernels does not increase the execution time, our runtime scheduler uses regression models constructed offline to offload backend kernels only when it deems that offloading provides speedups (\Sect{sec:be:rt}).}


\no{The per-frame latency data in \Fig{fig:besched} gives an intuition of the scheduler's runtime behaviors. As expected, the scheduler does not offload frames that execute fast on the CPU, which is evident in the figures where frames toward the left $x$-axes tend to have the same execution time as the baseline.}

We find that the regression model has a $R^2$ value (i.e., coefficient of determination~\cite{friedman2001elements}) of 0.83, 0.82, and 0.98 for registration, VIO, and SLAM, respectively, indicating high accuracy. To quantify the effectiveness of the runtime scheduler, we compare it against an oracle scheduler which always correctly schedules the frames. In all three cases, our runtime scheduler results in almost the same speedup compared to the oracle scheduler (less than 0.001\% difference).

Using \mbox{\sys{Edx-Car}} as a case-study, almost all the frames are offloaded to FPGA in the registration and VIO mode. However, only 76.4\% frames are offloaded in SLAM. Always offloading SLAM frames increases latency by 8.3\%.

\no{\begin{figure}[t]
\centering
\subfloat[\small{Projection.}]
{
  \includegraphics[trim=0 0 0 0, clip, width=0.33\columnwidth]{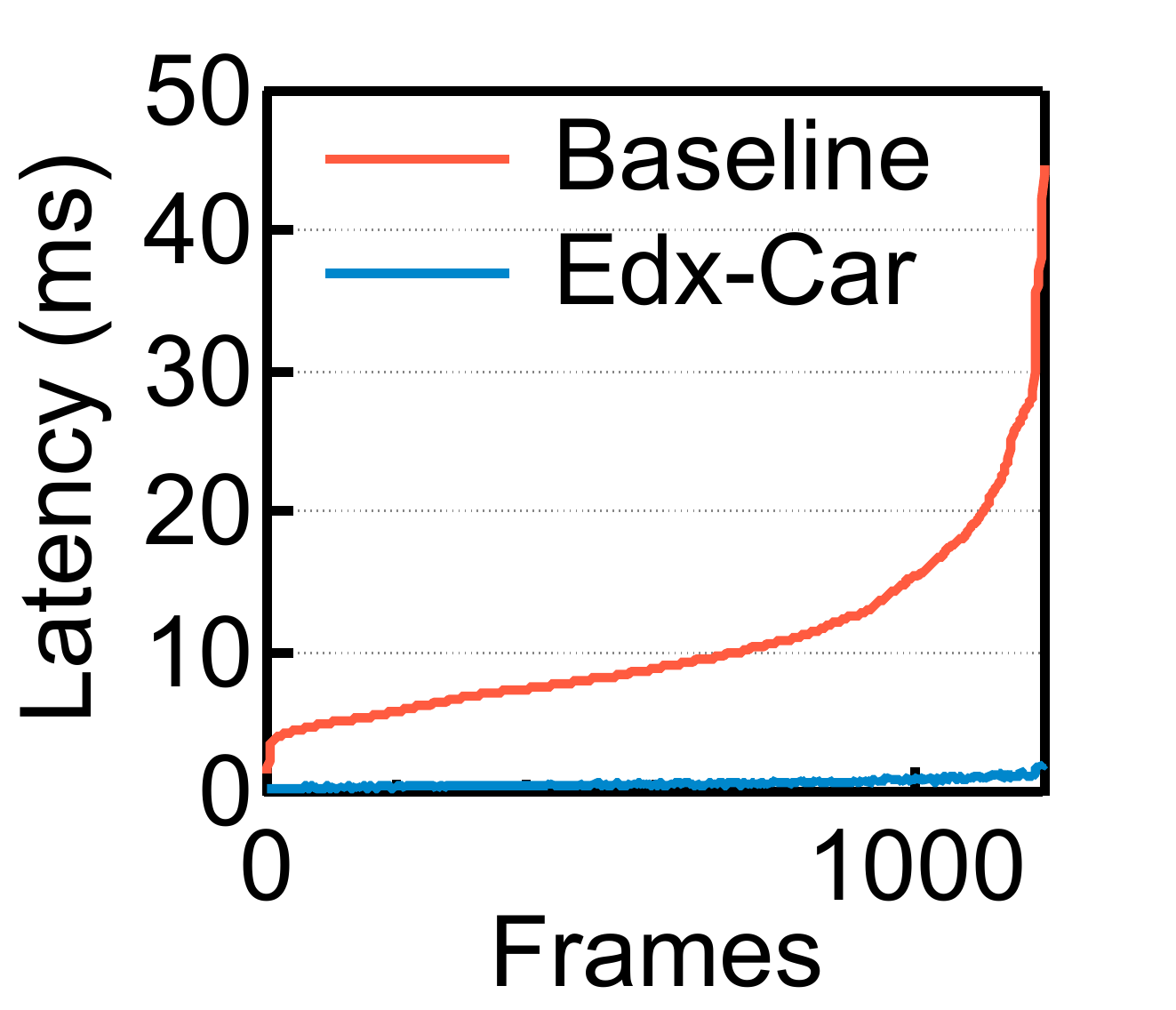}
  \label{fig:cpu_proj}
}
\subfloat[\small{Kalman gain.}]
{
  \includegraphics[trim=0 0 0 0, clip, width=0.33\columnwidth]{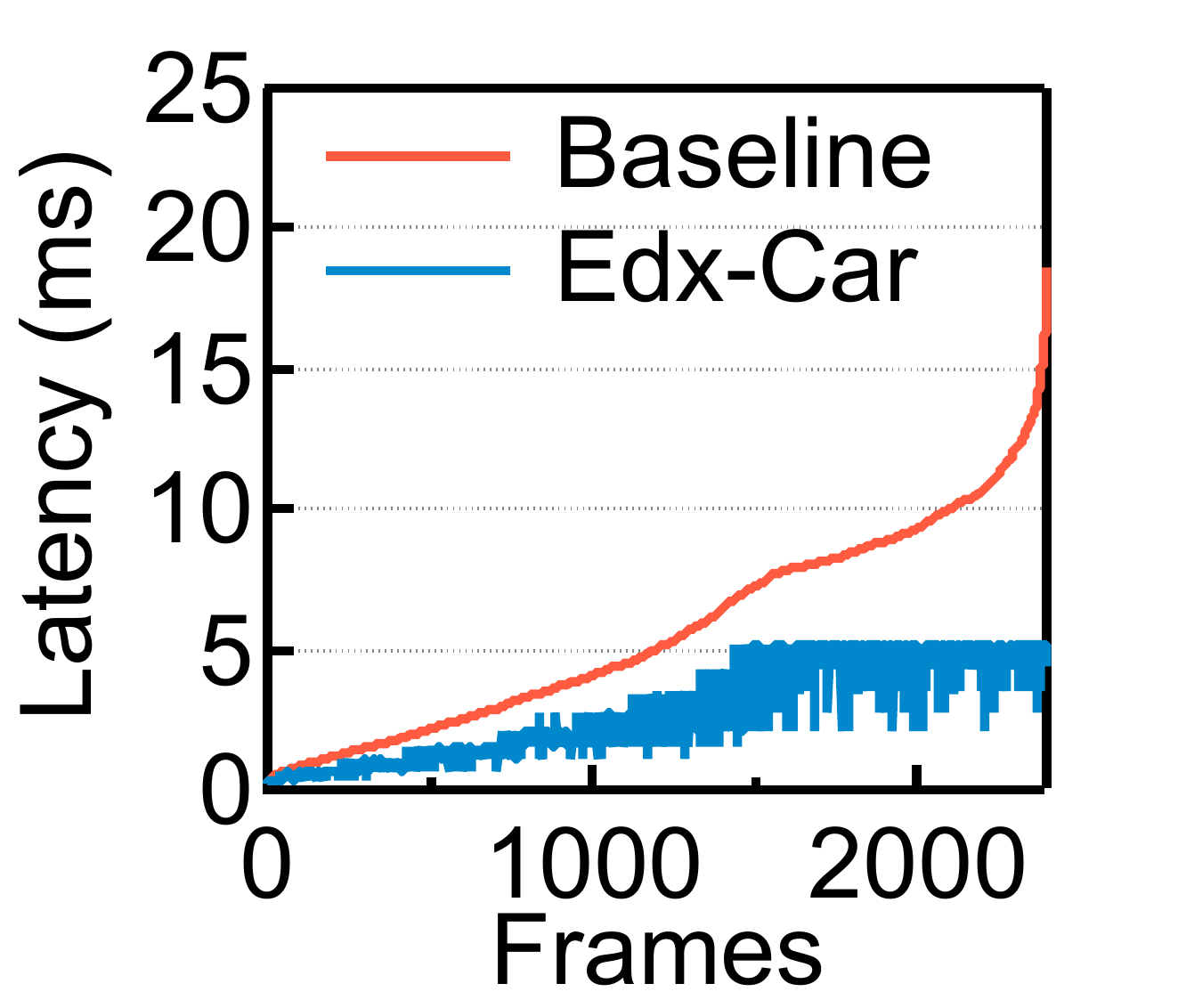}
  \label{fig:cpu_vio}
}
\subfloat[\small{Marginalization.}]
{
  \includegraphics[trim=0 0 0 0, clip, width=0.33\columnwidth]{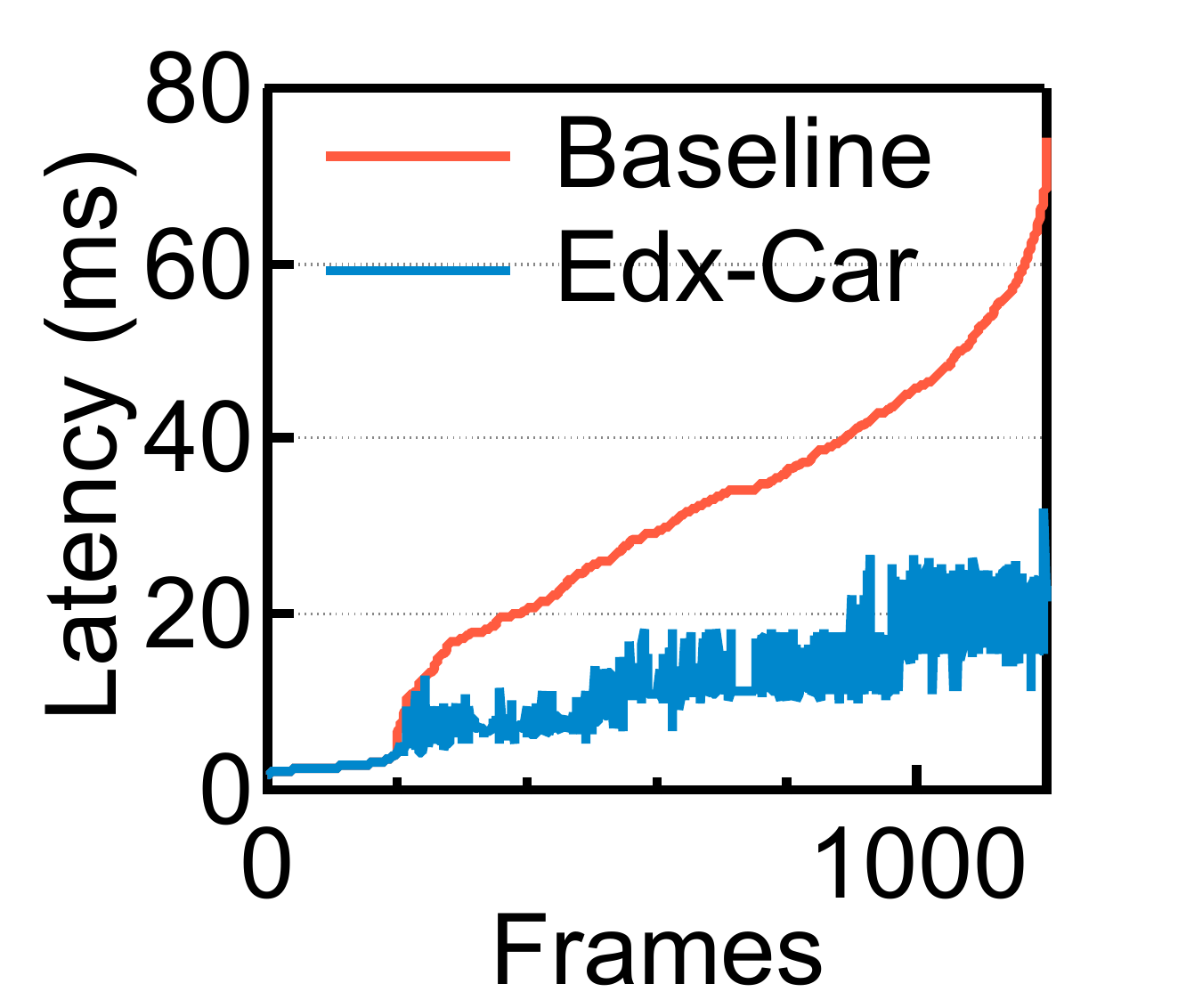}
  \label{fig:cpu_slam}
}
\caption{Per-frame latency comparison between the baseline and \sys{Edx-Car} in the three accelerated backend kernels.}
\label{fig:besched}
\end{figure}}

\subsection{Existing Accelerator Comparison}
\label{sec:eval:comp}

Existing ASIC/FPGA localization accelerators target different algorithmic variants than ours, so a fair comparison is difficult. Below is our best-effort comparison.

Suleiman et. al.~\mbox{\cite{suleiman2019navion}} (ASIC VIO) uses factor-graph optimizations. In comparison, our system has slightly higher error (0.28\%--0.42\% vs. 0.28\%) on EuRoC. The ASIC runs at 28--171 FPS whereas \mbox{\sys{Edx-Drone}} runs at 22.4 FPS on average.

Compared to Li et. al.~\mbox{\cite{li2019879gops}} (ASIC SLAM), our algorithm has a much lower error (0.01\% vs. 2.1\%) on KITTI. The ASIC runs at 30 -- 80 FPS while \mbox{\sys{Edx-Car}} runs at 31.9 FPS.

Compared to Suleiman et. al.~\mbox{\cite{suleiman2018navion}} (FPGA VIO), our algorithm has similar error (0.23 $m$ vs. 0.19 $m$) on EuRoC. They run at 5 -- 20 FPS while \mbox{\sys{Edx-Drone}} runs at 22.4 FPS.

We note that: 1) 30 FPS is considered real-time; our FPS could be higher with more capable FPGA platforms, and 2) \mbox{\proj} unifies different algorithms in one design while prior accelerators focus on one algorithm each.

\subsection{CPU/GPU/DSP Comparison}
\label{sec:eval:unopt}


\Tbl{tbl:unopt} presents a comprehensive performance comparison with various CPU/GPU/DSP baselines. Due to space limit, we show the data of only \sys{Edx-Car} as a case study.

\begin{table}[t]
\centering
\caption{\mbox{\sys{Edx-Car}} speedup over CPU/GPU/DSP.}
\renewcommand*{\arraystretch}{1}
\renewcommand*{\tabcolsep}{14pt}
\resizebox{\columnwidth}{!}
{
\begin{tabular}{lc}
\toprule[0.15em]
\textbf{Baseline} & \textbf{Speedup ($\times$)} \\
\midrule[0.05em]
Single-core w/ ROS & 3.5\\
Single-core w/o ROS & 3.3 \\
Multi-core w/ ROS & 2.2 \\
Multi-core w/o ROS (Our baseline) & 2.1 \\
Adreno 530 mobile GPU + CPU & 4.4 \\
Hexagon 680 DSP + CPU & 2.5 \\
Maxwell mobile GPU + CPU & 2.5 \\
\bottomrule[0.15em]
\end{tabular}
}
\label{tbl:unopt}
\end{table}

\mbox{\sys{Edx-Car}} has the lowest speedup over the baseline used in this paper (i.e., multi-core without ROS dependencies), indicating that our baseline is optimized. The GPU implementations are slower than CPU, mainly because of the launch/setup time (40 $ms$ on Adreno GPU; batching is unavailable) and the inefficiency in dealing with sparse matrices in SLAM/VIO backend. For this reason, we are not aware of widely-used GPU implementations of end-to-end SLAM/VIO system.

%% file: conc.tex
\section{Conclusion and Future Work}
\label{sec:conc}

\proj identifies and addresses a key challenge in deploying autonomous machines in the real world: \textit{efficient and flexible localization under resource constraints}. We summarize key lessons we learned and outline a few future directions.

\paragraph{Establishing Proper Software Target for Acceleration} Using data collected from our commercial deployment and standard benchmarks, we show that no existing localization algorithm (e.g., constrained optimization-based SLAM~\cite{mur2017orb, engel2014lsd, pumarola2017pl}, probabilistic estimation-based VIO~\cite{marchetti2006comparative, mourikis2007multi, jetto1999development, mao2007design}, and registration against the map~\cite{mur2017orb, mur2015orb}) is sufficiently flexible to adapt to different real-world operating scenarios. Therefore, accelerating any single algorithm will unlikely be useful in real products. We propose a general-purpose localization algorithm that integrates key primitives in existing algorithms. The new algorithm retains high accuracy of individual algorithms, simplifies our software stack, and provides a desirable acceleration target.

Our initial exploration suggests that the same ``one algorithm does not fit all'' observation applies to other autonomous machine tasks such as planning and control, suggesting opportunities for software-hardware co-design in the future.

\paragraph{Unified Architecture} The unified algorithm framework enables a unified architecture substrate, a key difference compared to prior localization accelerators on ASIC~\cite{suleiman2019navion, li2019879gops, yoon2010graphics} and FPGA~\cite{zhang2017visual, zhang2018pirvs, liu2019eslam, fang2017fpga, gautier2019fpga, boikos2016semi, tertei2016fpga} that target only specific algorithms and scenarios. We show that simply stacking dedicated accelerators built for each algorithm (the N.S. columns in \mbox{\Tbl{tbl:fpga}}) leads to over 2$\times$ resource waste. Additionally, with the same design principle, \proj can be instantiated differently to target different autonomous machines with varying performance requirements and constraints, e.g., drones vs. self-driving cars, as we demonstrate in \mbox{\Sect{sec:eval}}. The unified methodology and architecture greatly simplify our design and deployment flow, and are instrumental to allow us to expand our product lines going forward.


\paragraph{System-level Optimizations} While many have studied accelerators for individual tasks in localization, such as convolution~\cite{qadeer2013convolution}, stereo matching~\cite{feng2019asv, li20171920}, and optical flow~\cite{gultekin2013fpga, xiang2016hardware}, an end-to-end localization system necessarily integrates different tasks, and requires us to explore optimizations at the system level~\cite{alp}. Our hardware design exploits the parallelism and data communication patterns across different tasks, and judiciously applies pipelining, resource-sharing, on-chip buffer specialization to achieve meaningful system-level gains.

\paragraph{Architectural Support for Autonomous Machines} Architectural and systems support for autonomous machines has received increasing attention lately in academia, ranging from accelerating motion planning and control~\cite{sacks2018robox, murray2016microarchitecture, murray2016robot}, point cloud analytics~\cite{feng2020real, feng2020mesorasi, xu2019tigris}, enabling distributed cognition~\cite{hadidi2018distributed, merck2019characterizing, hadidi2019understanding}, to benchmarking~\cite{weisz2016robobench, boroujerdian2018mavbench, slambench}. We present a case study on localization, which we hope could promote open software stacks and hardware platforms for autonomous machines in the near future, much like today's machine learning domain.




%% file: main.bbl
\begin{thebibliography}{10}
\providecommand{\url}[1]{#1}
\csname url@samestyle\endcsname
\providecommand{\newblock}{\relax}
\providecommand{\bibinfo}[2]{#2}
\providecommand{\BIBentrySTDinterwordspacing}{\spaceskip=0pt\relax}
\providecommand{\BIBentryALTinterwordstretchfactor}{4}
\providecommand{\BIBentryALTinterwordspacing}{\spaceskip=\fontdimen2\font plus
\BIBentryALTinterwordstretchfactor\fontdimen3\font minus
  \fontdimen4\font\relax}
\providecommand{\BIBforeignlanguage}[2]{{%
\expandafter\ifx\csname l@#1\endcsname\relax
\typeout{** WARNING: IEEEtranS.bst: No hyphenation pattern has been}%
\typeout{** loaded for the language `#1'. Using the pattern for}%
\typeout{** the default language instead.}%
\else
\language=\csname l@#1\endcsname
\fi
#2}}
\providecommand{\BIBdecl}{\relax}
\BIBdecl

\bibitem{Cameraprice}
\BIBentryALTinterwordspacing
``Camera price.'' [Online]. Available:
  \url{https://www.amazon.com/s?k=car+camera&ref=nb_sb_noss_1}
\BIBentrySTDinterwordspacing

\bibitem{ceresuse}
\BIBentryALTinterwordspacing
``Ceres users.'' [Online]. Available: \url{http://ceres-solver.org/users.html}
\BIBentrySTDinterwordspacing

\bibitem{dbow2}
\BIBentryALTinterwordspacing
``Dbow2.'' [Online]. Available: \url{https://github.com/dorian3d/DBoW2}
\BIBentrySTDinterwordspacing

\bibitem{GPSprice}
\BIBentryALTinterwordspacing
``Gps price.'' [Online]. Available:
  \url{https://www.amazon.com/s?k=GPS&ref=nb_sb_noss_2}
\BIBentrySTDinterwordspacing

\bibitem{mlhololens}
\BIBentryALTinterwordspacing
``Highly efficient machine learning for hololens.'' [Online]. Available:
  \url{https://www.microsoft.com/en-us/research/uploads/prod/2018/03/Andrew-Fitzgibbon-Fitting-Models-to-Data-Accuracy-Speed-Robustness.pdf}
\BIBentrySTDinterwordspacing

\bibitem{IMUprice}
\BIBentryALTinterwordspacing
``Imu price.'' [Online]. Available:
  \url{https://www.amazon.com/s?k=IMU&ref=nb_sb_noss_2}
\BIBentrySTDinterwordspacing

\bibitem{irobot}
\BIBentryALTinterwordspacing
``irobot brings visual mapping and navigation to the roomba 980.'' [Online].
  Available:
  \url{https://spectrum.ieee.org/automaton/robotics/home-robots/irobot-brings-visual-mapping-and-navigation-to-the-roomba-980}
\BIBentrySTDinterwordspacing

\bibitem{djiviogps}
\BIBentryALTinterwordspacing
``Mark: the world’s first 4k drone positioned by visual inertial odometry.''
  [Online]. Available:
  \url{https://www.provideocoalition.com/mark-the-worlds-first-4k-drone-positioned-by-visual-inertial-odometry/}
\BIBentrySTDinterwordspacing

\bibitem{mipicsi2}
\BIBentryALTinterwordspacing
``{MIPI Camera Serial Interface 2 (MIPI CSI-2)}.'' [Online]. Available:
  \url{https://www.mipi.org/specifications/csi-2}
\BIBentrySTDinterwordspacing

\bibitem{msckfvio}
\BIBentryALTinterwordspacing
``Msckf vio.'' [Online]. Available:
  \url{https://github.com/KumarRobotics/msckf_vio}
\BIBentrySTDinterwordspacing

\bibitem{slamcore}
\BIBentryALTinterwordspacing
``Slamcore.'' [Online]. Available: \url{https://www.slamcore.com/}
\BIBentrySTDinterwordspacing

\bibitem{tx1}
\BIBentryALTinterwordspacing
``Tx1 datasheet.'' [Online]. Available:
  \url{http://images.nvidia.com/content/tegra/embedded-systems/pdf/JTX1-Module-Product-sheet.pdf}
\BIBentrySTDinterwordspacing

\bibitem{vertex}
\BIBentryALTinterwordspacing
``Vertex-7 datasheet.'' [Online]. Available:
  \url{https://www.xilinx.com/support/documentation/data_sheets/ds180_7Series_Overview.pdf}
\BIBentrySTDinterwordspacing

\bibitem{vinsfusion}
\BIBentryALTinterwordspacing
``Vins-fusion.'' [Online]. Available:
  \url{https://github.com/HKUST-Aerial-Robotics/VINS-Fusion}
\BIBentrySTDinterwordspacing

\bibitem{arcoremsckf}
\BIBentryALTinterwordspacing
``Visual inertial fusion.'' [Online]. Available:
  \url{http://rpg.ifi.uzh.ch/docs/teaching/2018/13_visual_inertial_fusion_advanced.pdf#page=33}
\BIBentrySTDinterwordspacing

\bibitem{zynq}
\BIBentryALTinterwordspacing
``Znyq datasheet.'' [Online]. Available:
  \url{https://www.xilinx.com/support/documentation/data_sheets/ds891-zynq-ultrascale-plus-overview.pdf}
\BIBentrySTDinterwordspacing

\bibitem{boikos2016semi}
K.~Boikos and C.-S. Bouganis, ``Semi-dense slam on an fpga soc,'' in \emph{2016
  26th International Conference on Field Programmable Logic and Applications
  (FPL)}.\hskip 1em plus 0.5em minus 0.4em\relax IEEE, 2016, pp. 1--4.

\bibitem{boroujerdian2018mavbench}
B.~Boroujerdian, H.~Genc, S.~Krishnan, W.~Cui, A.~Faust, and V.~Reddi,
  ``Mavbench: Micro aerial vehicle benchmarking,'' in \emph{2018 51st Annual
  IEEE/ACM International Symposium on Microarchitecture (MICRO)}.\hskip 1em
  plus 0.5em minus 0.4em\relax IEEE, 2018, pp. 894--907.

\bibitem{budiyono2013towards}
A.~Budiyono, L.~Chen, S.~Wang, K.~McDonald-Maier, and H.~Hu, ``Towards
  autonomous localization and mapping of auvs: a survey,'' \emph{International
  Journal of Intelligent Unmanned Systems}, 2013.

\bibitem{burri2016euroc}
M.~Burri, J.~Nikolic, P.~Gohl, T.~Schneider, J.~Rehder, S.~Omari, M.~W.
  Achtelik, and R.~Siegwart, ``The euroc micro aerial vehicle datasets,''
  \emph{The International Journal of Robotics Research}, vol.~35, no.~10, pp.
  1157--1163, 2016.

\bibitem{calonder2010brief}
M.~Calonder, V.~Lepetit, C.~Strecha, and P.~Fua, ``Brief: Binary robust
  independent elementary features,'' in \emph{European conference on computer
  vision}.\hskip 1em plus 0.5em minus 0.4em\relax Springer, 2010, pp. 778--792.

\bibitem{chen2018ionet}
C.~Chen, X.~Lu, A.~Markham, and N.~Trigoni, ``Ionet: Learning to cure the curse
  of drift in inertial odometry,'' in \emph{Thirty-Second AAAI Conference on
  Artificial Intelligence}, 2018.

\bibitem{chen2016eyeriss}
Y.-H. Chen, J.~Emer, and V.~Sze, ``Eyeriss: A spatial architecture for
  energy-efficient dataflow for convolutional neural networks,'' in \emph{ACM
  SIGARCH Computer Architecture News}, vol.~44, no.~3.\hskip 1em plus 0.5em
  minus 0.4em\relax IEEE Press, 2016, pp. 367--379.

\bibitem{chen2014dadiannao}
Y.~Chen, T.~Luo, S.~Liu, S.~Zhang, L.~He, J.~Wang, L.~Li, T.~Chen, Z.~Xu,
  N.~Sun \emph{et~al.}, ``Dadiannao: A machine-learning supercomputer,'' in
  \emph{Proceedings of the 47th Annual IEEE/ACM International Symposium on
  Microarchitecture}.\hskip 1em plus 0.5em minus 0.4em\relax IEEE Computer
  Society, 2014, pp. 609--622.

\bibitem{chi2018soda}
Y.~Chi, J.~Cong, P.~Wei, and P.~Zhou, ``Soda: stencil with optimized dataflow
  architecture,'' in \emph{2018 IEEE/ACM International Conference on
  Computer-Aided Design (ICCAD)}.\hskip 1em plus 0.5em minus 0.4em\relax IEEE,
  2018, pp. 1--8.

\bibitem{dudek2010computational}
G.~Dudek and M.~Jenkin, \emph{Computational principles of mobile
  robotics}.\hskip 1em plus 0.5em minus 0.4em\relax Cambridge university press,
  2010.

\bibitem{dusha2012error}
D.~Dusha and L.~Mejias, ``Error analysis and attitude observability of a
  monocular gps/visual odometry integrated navigation filter,'' \emph{The
  International Journal of Robotics Research}, vol.~31, no.~6, pp. 714--737,
  2012.

\bibitem{el2004wavelet}
N.~El-Sheimy, S.~Nassar, and A.~Noureldin, ``Wavelet de-noising for imu
  alignment,'' \emph{IEEE Aerospace and Electronic Systems Magazine}, vol.~19,
  no.~10, pp. 32--39, 2004.

\bibitem{engel2014lsd}
J.~Engel, T.~Sch{\"o}ps, and D.~Cremers, ``Lsd-slam: Large-scale direct
  monocular slam,'' in \emph{European conference on computer vision}.\hskip 1em
  plus 0.5em minus 0.4em\relax Springer, 2014, pp. 834--849.

\bibitem{fang2017fpga}
W.~Fang, Y.~Zhang, B.~Yu, and S.~Liu, ``Fpga-based orb feature extraction for
  real-time visual slam,'' in \emph{2017 International Conference on Field
  Programmable Technology (ICFPT)}.\hskip 1em plus 0.5em minus 0.4em\relax
  IEEE, 2017, pp. 275--278.

\bibitem{feng2020real}
Y.~Feng, S.~Liu, and Y.~Zhu, ``Real-time spatio-temporal lidar point cloud
  compression,'' in \emph{2020 IEEE/RSJ international conference on intelligent
  robots and systems (IROS)}, 2020.

\bibitem{feng2020mesorasi}
Y.~Feng, B.~Tian, T.~Xu, P.~Whatmough, and Y.~Zhu, ``Mesorasi: Architecture
  support for point cloud analytics via delayed-aggregation,'' in \emph{2020
  53rd Annual IEEE/ACM International Symposium on Microarchitecture
  (MICRO)}.\hskip 1em plus 0.5em minus 0.4em\relax IEEE, 2020, pp. 1037--1050.

\bibitem{feng2019asv}
Y.~Feng, P.~Whatmough, and Y.~Zhu, ``Asv: Accelerated stereo vision system,''
  in \emph{Proceedings of the 52nd Annual IEEE/ACM International Symposium on
  Microarchitecture}.\hskip 1em plus 0.5em minus 0.4em\relax ACM, 2019, pp.
  643--656.

\bibitem{fraundorfer2012visual}
F.~Fraundorfer and D.~Scaramuzza, ``Visual odometry: Part ii: Matching,
  robustness, optimization, and applications,'' \emph{IEEE Robotics \&
  Automation Magazine}, vol.~19, no.~2, pp. 78--90, 2012.

\bibitem{friedman2001elements}
J.~Friedman, T.~Hastie, and R.~Tibshirani, \emph{The elements of statistical
  learning}.\hskip 1em plus 0.5em minus 0.4em\relax Springer series in
  statistics New York, 2001, vol.~1, no.~10.

\bibitem{galvez2012bags}
D.~G{\'a}lvez-L{\'o}pez and J.~D. Tardos, ``Bags of binary words for fast place
  recognition in image sequences,'' \emph{IEEE Transactions on Robotics},
  vol.~28, no.~5, pp. 1188--1197, 2012.

\bibitem{gautier2019fpga}
Q.~Gautier, A.~Althoff, and R.~Kastner, ``Fpga architectures for real-time
  dense slam,'' in \emph{2019 IEEE 30th International Conference on
  Application-specific Systems, Architectures and Processors (ASAP)}, vol.
  2160.\hskip 1em plus 0.5em minus 0.4em\relax IEEE, 2019, pp. 83--90.

\bibitem{geiger2012we}
A.~Geiger, P.~Lenz, and R.~Urtasun, ``Are we ready for autonomous driving? the
  kitti vision benchmark suite,'' in \emph{2012 IEEE Conference on Computer
  Vision and Pattern Recognition}.\hskip 1em plus 0.5em minus 0.4em\relax IEEE,
  2012, pp. 3354--3361.

\bibitem{slambench}
N.~S. Ghalehshahi, R.~Hadidi, and H.~Kim, ``Slam performance on embedded
  robots,'' in \emph{Student Research Competition at Embedded System Week (SRC
  ESWEEK)}, 2019.

\bibitem{gultekin2013fpga}
G.~K. Gultekin and A.~Saranli, ``An fpga based high performance optical flow
  hardware design for computer vision applications,'' \emph{Microprocessors and
  Microsystems}, vol.~37, no.~3, pp. 270--286, 2013.

\bibitem{hadidi2019understanding}
R.~Hadidi, J.~Cao, M.~Merck, A.~Siqueira, Q.~Huang, A.~Saraha, C.~Jia, B.~Wang,
  D.~Lim, L.~Liu \emph{et~al.}, ``Understanding the power consumption of
  executing deep neural networks on a distributed robot system,'' in
  \emph{Algorithms and Architectures for Learning in-the-Loop Systems in
  Autonomous Flight, International Conference on Robotics and Automation
  (ICRA)}, vol. 2019, 2019.

\bibitem{hadidi2018distributed}
R.~Hadidi, J.~Cao, M.~Woodward, M.~S. Ryoo, and H.~Kim, ``Distributed
  perception by collaborative robots,'' \emph{IEEE Robotics and Automation
  Letters}, vol.~3, no.~4, pp. 3709--3716, 2018.

\bibitem{hartley2003multiple}
R.~Hartley and A.~Zisserman, \emph{Multiple view geometry in computer
  vision}.\hskip 1em plus 0.5em minus 0.4em\relax Cambridge university press,
  2003.

\bibitem{Hegarty2014darkroom}
J.~Hegarty, J.~Brunhaver, Z.~DeVito, J.~Ragan-Kelley, N.~Cohen, S.~Bell,
  A.~Vasilyev, M.~Horowitz, and P.~Hanrahan, ``Darkroom: Compiling high-level
  image processing code into hardware pipelines,'' 2014.

\bibitem{hegarty2016rigel}
J.~Hegarty, R.~Daly, Z.~DeVito, J.~Ragan-Kelley, M.~Horowitz, and P.~Hanrahan,
  ``Rigel: Flexible multi-rate image processing hardware,'' \emph{ACM
  Transactions on Graphics (TOG)}, vol.~35, no.~4, pp. 1--11, 2016.

\bibitem{hennessy2017computer}
J.~L. Hennessy and D.~A. Patterson, \emph{Computer architecture: a quantitative
  approach}, 6th~ed.\hskip 1em plus 0.5em minus 0.4em\relax Elsevier, 2017.

\bibitem{alp}
M.~D. Hill and V.~J. Reddi, ``Accelerator level parallelism,'' \emph{arXiv
  preprint arXiv:1907.02064}, 2019.

\bibitem{jakubowski2013block}
M.~Jakubowski and G.~Pastuszak, ``Block-based motion estimation algorithms—a
  survey,'' \emph{Opto-Electronics Review}, vol.~21, no.~1, pp. 86--102, 2013.

\bibitem{jetto1999development}
L.~Jetto, S.~Longhi, and G.~Venturini, ``Development and experimental
  validation of an adaptive extended kalman filter for the localization of
  mobile robots,'' \emph{IEEE Transactions on Robotics and Automation},
  vol.~15, no.~2, pp. 219--229, 1999.

\bibitem{jouppi2017datacenter}
N.~P. Jouppi, C.~Young, N.~Patil, D.~Patterson, G.~Agrawal, R.~Bajwa, S.~Bates,
  S.~Bhatia, N.~Boden, A.~Borchers \emph{et~al.}, ``In-datacenter performance
  analysis of a tensor processing unit,'' in \emph{2017 ACM/IEEE 44th Annual
  International Symposium on Computer Architecture (ISCA)}.\hskip 1em plus
  0.5em minus 0.4em\relax IEEE, 2017, pp. 1--12.

\bibitem{julier2004unscented}
S.~J. Julier and J.~K. Uhlmann, ``Unscented filtering and nonlinear
  estimation,'' \emph{Proceedings of the IEEE}, vol.~92, no.~3, pp. 401--422,
  2004.

\bibitem{kelly2013mobile}
A.~Kelly, \emph{Mobile robotics: mathematics, models, and methods}.\hskip 1em
  plus 0.5em minus 0.4em\relax Cambridge University Press, 2013.

\bibitem{kos2010effects}
T.~Kos, I.~Markezic, and J.~Pokrajcic, ``Effects of multipath reception on gps
  positioning performance,'' in \emph{Proceedings ELMAR-2010}.\hskip 1em plus
  0.5em minus 0.4em\relax IEEE, 2010, pp. 399--402.

\bibitem{li2012improving}
M.~Li and A.~I. Mourikis, ``Improving the accuracy of ekf-based visual-inertial
  odometry,'' in \emph{2012 IEEE International Conference on Robotics and
  Automation}.\hskip 1em plus 0.5em minus 0.4em\relax IEEE, 2012, pp. 828--835.

\bibitem{li2013high}
M.~Li and A.~I. Mourikis, ``High-precision, consistent ekf-based
  visual-inertial odometry,'' \emph{The International Journal of Robotics
  Research}, vol.~32, no.~6, pp. 690--711, 2013.

\bibitem{li2019879gops}
Z.~Li, Y.~Chen, L.~Gong, L.~Liu, D.~Sylvester, D.~Blaauw, and H.-S. Kim, ``An
  879gops 243mw 80fps vga fully visual cnn-slam processor for wide-range
  autonomous exploration,'' in \emph{2019 IEEE International Solid-State
  Circuits Conference-(ISSCC)}.\hskip 1em plus 0.5em minus 0.4em\relax IEEE,
  2019, pp. 134--136.

\bibitem{li20171920}
Z.~Li, Q.~Dong, M.~Saligane, B.~Kempke, L.~Gong, Z.~Zhang, R.~Dreslinski,
  D.~Sylvester, D.~Blaauw, and H.-S. Kim, ``A 1920$\times $1080 30-frames/s 2.3
  tops/w stereo-depth processor for energy-efficient autonomous navigation of
  micro aerial vehicles,'' \emph{IEEE Journal of Solid-State Circuits},
  vol.~53, no.~1, pp. 76--90, 2017.

\bibitem{liu2019eslam}
R.~Liu, J.~Yang, Y.~Chen, and W.~Zhao, ``Eslam: An energy-efficient accelerator
  for real-time orb-slam on fpga platform,'' in \emph{Proceedings of the 56th
  Annual Design Automation Conference 2019}, 2019, pp. 1--6.

\bibitem{lucas1981iterative}
B.~D. Lucas and T.~Kanade, ``{An iterative image registration technique with an
  application to stereo vision},'' in \emph{Proceedings of the 7th
  International Joint Conference on Artificial Intelligence}, 1981.

\bibitem{mao2007design}
G.~Mao, S.~Drake, and B.~D. Anderson, ``Design of an extended kalman filter for
  uav localization,'' in \emph{2007 Information, Decision and Control}.\hskip
  1em plus 0.5em minus 0.4em\relax IEEE, 2007, pp. 224--229.

\bibitem{marchetti2006comparative}
L.~Marchetti, G.~Grisetti, and L.~Iocchi, ``A comparative analysis of particle
  filter based localization methods,'' in \emph{Robot Soccer World Cup}.\hskip
  1em plus 0.5em minus 0.4em\relax Springer, 2006, pp. 442--449.

\bibitem{merck2019characterizing}
M.~L. Merck, B.~Wang, L.~Liu, C.~Jia, A.~Siqueira, Q.~Huang, A.~Saraha, D.~Lim,
  J.~Cao, R.~Hadidi \emph{et~al.}, ``Characterizing the execution of deep
  neural networks on collaborative robots and edge devices,'' in
  \emph{Proceedings of the Practice and Experience in Advanced Research
  Computing on Rise of the Machines (learning)}, 2019, pp. 1--6.

\bibitem{more1978levenberg}
J.~J. Mor{\'e}, ``The levenberg-marquardt algorithm: implementation and
  theory,'' in \emph{Numerical analysis}.\hskip 1em plus 0.5em minus
  0.4em\relax Springer, 1978, pp. 105--116.

\bibitem{mourikis2007multi}
A.~I. Mourikis and S.~I. Roumeliotis, ``A multi-state constraint kalman filter
  for vision-aided inertial navigation,'' in \emph{Proceedings 2007 IEEE
  International Conference on Robotics and Automation}.\hskip 1em plus 0.5em
  minus 0.4em\relax IEEE, 2007, pp. 3565--3572.

\bibitem{mur2015orb}
R.~Mur-Artal, J.~M.~M. Montiel, and J.~D. Tardos, ``Orb-slam: a versatile and
  accurate monocular slam system,'' \emph{IEEE transactions on robotics},
  vol.~31, no.~5, pp. 1147--1163, 2015.

\bibitem{mur2014fast}
R.~Mur-Artal and J.~D. Tard{\'o}s, ``Fast relocalisation and loop closing in
  keyframe-based slam,'' in \emph{2014 IEEE International Conference on
  Robotics and Automation (ICRA)}.\hskip 1em plus 0.5em minus 0.4em\relax IEEE,
  2014, pp. 846--853.

\bibitem{mur2017orb}
R.~Mur-Artal and J.~D. Tard{\'o}s, ``Orb-slam2: An open-source slam system for
  monocular, stereo, and rgb-d cameras,'' \emph{IEEE Transactions on Robotics},
  vol.~33, no.~5, pp. 1255--1262, 2017.

\bibitem{murray2016robot}
S.~Murray, W.~Floyd-Jones, Y.~Qi, D.~J. Sorin, and G.~Konidaris, ``Robot motion
  planning on a chip.'' in \emph{Robotics: Science and Systems}, 2016.

\bibitem{murray2016microarchitecture}
S.~Murray, W.~Floyd-Jones, Y.~Qi, G.~Konidaris, and D.~J. Sorin, ``The
  microarchitecture of a real-time robot motion planning accelerator,'' in
  \emph{The 49th Annual IEEE/ACM International Symposium on
  Microarchitecture}.\hskip 1em plus 0.5em minus 0.4em\relax IEEE Press, 2016,
  p.~45.

\bibitem{profanter2019opc}
S.~Profanter, A.~Tekat, K.~Dorofeev, M.~Rickert, and A.~Knoll, ``Opc ua versus
  ros, dds, and mqtt: performance evaluation of industry 4.0 protocols,'' in
  \emph{Proceedings of the IEEE International Conference on Industrial
  Technology (ICIT)}, 2019.

\bibitem{pumarola2017pl}
A.~Pumarola, A.~Vakhitov, A.~Agudo, A.~Sanfeliu, and F.~Moreno-Noguer,
  ``Pl-slam: Real-time monocular visual slam with points and lines,'' in
  \emph{2017 IEEE international conference on robotics and automation
  (ICRA)}.\hskip 1em plus 0.5em minus 0.4em\relax IEEE, 2017, pp. 4503--4508.

\bibitem{qadeer2013convolution}
W.~Qadeer, R.~Hameed, O.~Shacham, P.~Venkatesan, C.~Kozyrakis, and M.~A.
  Horowitz, ``Convolution engine: balancing efficiency \& flexibility in
  specialized computing,'' in \emph{Proceedings of the 40th IEEE Annual
  International Symposium on Computer Architecture}, 2013.

\bibitem{qin2017vins}
T.~Qin, P.~Li, and S.~Shen, ``Vins-mono: A robust and versatile monocular
  visual-inertial state estimator,'' \emph{IEEE Transactions on Robotics},
  vol.~34, no.~4, pp. 1004--1020, 2018.

\bibitem{rosten2006machine}
E.~Rosten and T.~Drummond, ``Machine learning for high-speed corner
  detection,'' in \emph{European conference on computer vision}.\hskip 1em plus
  0.5em minus 0.4em\relax Springer, 2006, pp. 430--443.

\bibitem{rublee2011orb}
E.~Rublee, V.~Rabaud, K.~Konolige, and G.~Bradski, ``Orb: An efficient
  alternative to sift or surf,'' in \emph{2011 International conference on
  computer vision}.\hskip 1em plus 0.5em minus 0.4em\relax Ieee, 2011, pp.
  2564--2571.

\bibitem{sacks2018robox}
J.~Sacks, D.~Mahajan, R.~C. Lawson, and H.~Esmaeilzadeh, ``Robox: an end-to-end
  solution to accelerate autonomous control in robotics,'' in \emph{Proceedings
  of the 45th Annual International Symposium on Computer Architecture}.\hskip
  1em plus 0.5em minus 0.4em\relax IEEE Press, 2018, pp. 479--490.

\bibitem{suleiman2018navion}
A.~Suleiman, Z.~Zhang, L.~Carlone, S.~Karaman, and V.~Sze, ``Navion: a fully
  integrated energy-efficient visual-inertial odometry accelerator for
  autonomous navigation of nano drones,'' in \emph{2018 IEEE Symposium on VLSI
  Circuits}.\hskip 1em plus 0.5em minus 0.4em\relax IEEE, 2018, pp. 133--134.

\bibitem{suleiman2019navion}
A.~Suleiman, Z.~Zhang, L.~Carlone, S.~Karaman, and V.~Sze, ``Navion: A 2-mw
  fully integrated real-time visual-inertial odometry accelerator for
  autonomous navigation of nano drones,'' \emph{IEEE Journal of Solid-State
  Circuits}, vol.~54, no.~4, pp. 1106--1119, 2019.

\bibitem{sun2018robust}
K.~Sun, K.~Mohta, B.~Pfrommer, M.~Watterson, S.~Liu, Y.~Mulgaonkar, C.~J.
  Taylor, and V.~Kumar, ``Robust stereo visual inertial odometry for fast
  autonomous flight,'' \emph{IEEE Robotics and Automation Letters}, vol.~3,
  no.~2, pp. 965--972, 2018.

\bibitem{tertei2016fpga}
D.~T. Tertei, J.~Piat, and M.~Devy, ``Fpga design of ekf block accelerator for
  3d visual slam,'' \emph{Computers \& Electrical Engineering}, vol.~55, pp.
  123--137, 2016.

\bibitem{wei2016rt}
H.~Wei, Z.~Shao, Z.~Huang, R.~Chen, Y.~Guan, J.~Tan, and Z.~Shao, ``Rt-ros: A
  real-time ros architecture on multi-core processors,'' \emph{Future
  Generation Computer Systems}, vol.~56, pp. 171--178, 2016.

\bibitem{weisz2016robobench}
J.~Weisz, Y.~Huang, F.~Lier, S.~Sethumadhavan, and P.~Allen, ``Robobench:
  Towards sustainable robotics system benchmarking,'' in \emph{2016 IEEE
  International Conference on Robotics and Automation (ICRA)}.\hskip 1em plus
  0.5em minus 0.4em\relax IEEE, 2016, pp. 3383--3389.

\bibitem{whatmough2019fixynn}
P.~N. Whatmough, C.~Zhou, P.~Hansen, S.~K. Venkataramanaiah, J.-s. Seo, and
  M.~Mattina, ``Fixynn: Efficient hardware for mobile computer vision via
  transfer learning,'' \emph{arXiv preprint arXiv:1902.11128}, 2019.

\bibitem{xiang2016hardware}
J.~Xiang, Z.~Li, H.~S. Kim, and C.~Chakrabarti, ``Hardware-efficient
  neighbor-guided sgm optical flow for low power vision applications,'' in
  \emph{2016 IEEE International Workshop on Signal Processing Systems
  (SiPS)}.\hskip 1em plus 0.5em minus 0.4em\relax IEEE, 2016, pp. 1--6.

\bibitem{xu2019tigris}
T.~Xu, B.~Tian, and Y.~Zhu, ``Tigris: Architecture and algorithms for 3d
  perception in point clouds,'' in \emph{Proceedings of the 52nd Annual
  IEEE/ACM International Symposium on Microarchitecture}.\hskip 1em plus 0.5em
  minus 0.4em\relax ACM, 2019, pp. 629--642.

\bibitem{yoon2010graphics}
J.-S. Yoon, J.-H. Kim, H.-E. Kim, W.-Y. Lee, S.-H. Kim, K.~Chung, J.-S. Park,
  and L.-S. Kim, ``A graphics and vision unified processor with 0.89 $\mu$w/fps
  pose estimation engine for augmented reality,'' in \emph{2010 IEEE
  International Solid-State Circuits Conference-(ISSCC)}.\hskip 1em plus 0.5em
  minus 0.4em\relax IEEE, 2010, pp. 336--337.

\bibitem{yu2020building}
B.~Yu, W.~Hu, L.~Xu, J.~Tang, S.~Liu, and Y.~Zhu, ``Building the computing
  system for autonomous micromobility vehicles: Design constraints and
  architectural optimizations,'' in \emph{2020 53rd Annual IEEE/ACM
  International Symposium on Microarchitecture (MICRO)}.\hskip 1em plus 0.5em
  minus 0.4em\relax IEEE, 2020, pp. 1067--1081.

\bibitem{zhang2018pirvs}
Z.~Zhang, S.~Liu, G.~Tsai, H.~Hu, C.-C. Chu, and F.~Zheng, ``Pirvs: An advanced
  visual-inertial slam system with flexible sensor fusion and hardware
  co-design,'' in \emph{2018 IEEE International Conference on Robotics and
  Automation (ICRA)}.\hskip 1em plus 0.5em minus 0.4em\relax IEEE, 2018, pp.
  1--7.

\bibitem{zhang2017visual}
Z.~Zhang, A.~A. Suleiman, L.~Carlone, V.~Sze, and S.~Karaman, ``Visual-inertial
  odometry on chip: An algorithm-and-hardware co-design approach,'' in
  \emph{Proceedings of Robotics Science and Systems (RSS)}, 2017.

\end{thebibliography}
